\documentclass[iop]{emulateapj}
\usepackage{apjfonts}
\usepackage{amssymb}
\usepackage{natbib}
\usepackage{amsmath}
\usepackage{color}
\shorttitle{Stellar and Nebular Spectra at $z\sim 2.4$}

\newcommand{\Ha}{\ensuremath{\rm H\alpha}}
\newcommand{\Hb}{\ensuremath{\rm H\beta}}
\newcommand{\lya}{\ensuremath{\rm Ly\alpha}}

\newcommand{\kms}{\rm km~s\ensuremath{^{-1}\,}}

\newcommand{\nhi}{\ensuremath{N_{\rm HI}}}

\newcommand{\secpoint}{\mbox{$''\mskip-7.6mu.\,$}}

\def\ltsima{$\; \buildrel < \over \sim \;$}
\def\simlt{\lower.5ex\hbox{\ltsima}}
\def\gtsima{$\; \buildrel > \over \sim \;$}
\def\simgt{\lower.5ex\hbox{\gtsima}}

\begin{document}

\title{Reconciling the Stellar and Nebular Spectra of High Redshift Galaxies\altaffilmark{*}}

\slugcomment{Accepted for publication in ApJ}

\author{\sc Charles C. Steidel\altaffilmark{1}, Allison L. Strom\altaffilmark{1}, Max Pettini\altaffilmark{2}, Gwen C. Rudie\altaffilmark{3}, \\ 
Naveen A. Reddy\altaffilmark{4,5}, \& Ryan F. Trainor\altaffilmark{6,7}}

\altaffiltext{*}{Based on data obtained at the
W.M. Keck Observatory, which
is operated as a scientific partnership among the California Institute of
Technology, the
University of California, and NASA, and was made possible by the generous
financial
support of the W.M. Keck Foundation.
}
\altaffiltext{1}{Cahill Center for Astronomy and Astrophysics, California Institute of Technology, MS 249-17,
Pasadena, CA 91105, USA}
\altaffiltext{2}{Institute of Astronomy, Madingley Road, Cambridge CB3 OHA, UK}
\altaffiltext{3}{Carnegie Observatories, 813 Santa Barbara Street, Pasadena, CA 91101, USA} 
\altaffiltext{4}{Department of Physics and Astronomy, University of California, Riverside, 900 University Avenue, Riverside, CA 92521, USA} 
\altaffiltext{5}{Alfred P. Sloan Foundation Research Fellow}
\altaffiltext{6}{Department of Astronomy, University of California, Berkeley, Campbell Hall, Berkeley, CA 94720 USA}
\altaffiltext{7}{Miller Fellow}

\begin{abstract}

We present a combined analysis of rest-frame far-UV (1000-2000 \AA) and rest-frame optical (3600-7000 \AA) 
composite spectra formed from very deep Keck/LRIS and Keck/MOSFIRE observations of a sample of 30 star-forming 
galaxies with $z=2.40\pm0.11$, selected to be representative of the full KBSS-MOSFIRE spectroscopic survey.  
Since the same massive stars are responsible for the observed FUV continuum {\it and} for the excitation of the observed
nebular emission in the FUV and optical, a self-consistent stellar population synthesis model must {\it simultaneously} 
match the details of the far-UV stellar+nebular continuum, including stellar photospheric and wind features, 
{\it and}-- when inserted as the excitation source in photoionization models-- account for all observed nebular emission 
line ratios over the full rest-frame FUV to rest-frame optical range.  We find that only models including massive 
star binaries, having low {\it stellar} metallicity ($Z_{\ast}/Z_{\odot} \simeq 0.1$) but relatively high ionized 
gas-phase oxygen abundances ($Z_{\rm neb}/Z_{\odot} \simeq 0.5$), can successfully match all of the observational constraints.
We argue that this apparent discrepancy is naturally explained by highly super-solar O/Fe [${\rm \simeq 4-5~(O/Fe)_{\odot}}$], 
expected for gas whose enrichment is dominated by the products of core-collapse supernovae. While O dominates 
the physics of the ionized gas (and thus the nebular emission lines), Fe dominates the EUV and FUV opacity and 
controls the mass loss rate from massive stars, resulting in dramatic effects on the evolution of massive stars in binary systems. 
When O/Fe is high, stellar spectra appear ``metal-poor'' in the far-UV and ``hard'' in the ionizing EUV 
even when oxygen abundances approach solar metallicity. That high nebular excitation is much more common at high redshift (at fixed O/H) 
compared to low redshift is a direct consequence of systematic differences in star formation history: galaxies whose chemical enrichment
is dominated by the products of stars formed over the previous $< 1$ Gyr are the rule at high redshift, but the exception at low redshift. 
Once the correct ionizing stellar spectrum is identified, photoionization models constrained only by ratios of strong emission lines 
reproduce the direct $T_{\rm e}$ measurement of O/H and accurate gas-phase abundance ratios of N/O and C/O -- both of 
which are significantly sub-solar but, like the inferred O/Fe, are in remarkable agreement with abundance patterns 
observed in Galactic thick disk, bulge, and halo stars with similar O/H. 
\end{abstract}


\section{Introduction}
\label{sec:intro}

The formation and evolution of massive stars impact most -- if not all-- observable properties of the most
distant galaxies. 
Their winds, radiation pressure, and supernova explosions are the principal sources of feedback that ultimately control
the galaxy formation process
(e.g., \citealt{murray05,fauch13,hopkins14,muratov15}), and they dominate the bolometric luminosity of 
rapidly forming galaxies, whether most of the luminosity emerges in the UV or is re-radiated in the far-IR. 
Massive stars are responsible for
producing-- and, arguably, dispersing -- most of the metals that have enriched galaxies and the circumgalactic (CGM) and intergalactic (IGM) medium
over cosmic time, and were
also the most likely sources responsible 
for producing the photons that reionized the universe at $z\simgt 7$ (e.g., \citealt{robertson15}).  

In spite of the crucial role played by massive stars in all of these baryonic processes, 
we know surprisingly little about them. Contributing to our ignorance is the fact that massive stars
are naturally short-lived, and therefore any accounting must be done {\it in situ} at the epoch of their formation.
Short lifetimes also means that massive stars 
die close to their birthplace, and these sites are often heavily obscured by dense
ISM and dust. When not obscured, massive stars emit most of their energy in the UV, requiring observations from space for nearby examples. 
Although they produce most of the H-ionizing photons within galaxies and, at $z \simgt 2$, in the intergalactic medium as well,
no direct observation of the ionizing spectrum of an O-star has ever been made, at any redshift or metallicity.   

On top of observational challenges, massive stars are also more difficult to model than their less-massive counterparts.
The ionizing spectra, main sequence lifetimes, evolutionary stages, end states, and chemical yields from massive stars all depend on 
the details of mass loss (e.g., \citealt{smith14}), which itself depends on 
the initial chemical composition (e.g., \citealt{puls08,kudritzki+puls2000,cantiello07,eldridge11,brott11,langer12}). 
The
last decade has witnessed a rapidly-changing paradigm for massive star winds due to the appreciation that the wind
material is clumpy, reducing the mass loss rates by factors of $\sim 2-3$ compared to what was believed previously (\citealt{smith14}, and
references therein). In parallel, it has also been established that {\it most} massive stars do not evolve as single stars,
but as part of binary systems \citep{crowther07,sana12}.  Multiplicity affects nearly all aspects of the expected evolutionary path of 
massive stars, 
and can also lead to phenomena that could not occur for single massive
stars, particularly in low-metallicity environments.  
For example, binary massive stars  
with metallicities $Z_{\ast} < 0.004$ ($Z_{\ast}/Z_{\odot} < 0.35$) can lead to ``quasi-homogeneous evolution'' (QHE; \citealt{yoon05,cantiello07,eldridge11,brott11,eldridge12}) 
in which the more massive star over-flows its Roche lobe and dumps mass onto the secondary, which (because of weak winds inherent to lower
stellar metallicity) accretes rapidly and spins up to high rotational velocity, 
causing it to become fully mixed. Such stars can burn all of their H to He, 
last up to 3 times longer on the main sequence than a single star of the same mass, would  
not produce stellar winds, and  
become progressively {\it hotter} as they evolve away from the zero-age main sequence. QHE could produce sources of extreme-UV 
radiation that would mimic classical Wolf-Rayet
stars, but would be much longer-lived, would require less massive progenitors, and may be absent emission features of C or N. 

In the context of ionizing photon production, 
since the more massive donor star reaches its end state first, the hot Wolf-Rayet-like
companion may receive a kick capable of perturbing it well out of its birth cloud-- potentially resolving 
the puzzle of how 
stellar ionizing photons manage to escape their immediate surroundings to contribute to the ionization of the IGM. 
As we discuss below, there may be significant observational
evidence that QHE is important, particularly at high redshift.   

Most importantly, the uncertainties associated with massive star populations have profound effects on the interpretation of
observations of the high redshift universe. 
The recent sea-changes in our understanding of massive stars -- lower mass loss, the dominance of binary evolution -- have been slow to 
percolate to the galaxy formation community, and most investigations that make use of massive star population synthesis  
use models based on key assumptions now known to be incorrect, or at least incomplete.  
Quantities that may be subject to revision include estimates of SFR, stellar mass (${\rm M_{\ast}}$), stellar population ages from SED fitting, 
the ratio of ionizing to non-ionizing UV photons produced by young stellar populations, the rate of core-collapse supernovae and associated chemical yields, 
and the measurement of gas-phase metallicities from observations of nebular emission lines. These uncertainties
propagate through to affect our understanding of  
stellar and supernova feedback, gas accretion and outflows, and the chemical evolution of galaxies-- in short, all of the details relevant
to baryonic processes in forming galaxies.  

The goal of this paper is to seek a deeper understanding of these processes by combining two complementary observational signatures
of massive stars in forming galaxies-- the
rest-frame far-UV continuum ($1000 \simlt \lambda_0 \simlt 2000$ \AA) and nebular emission lines from \ion{H}{2} regions-- in
the same galaxies. Clearly the two are complementary as well as causally-connected: the far-UV continuum provides direct access to the light
produced by OB stars, encoding information on stellar metallicity, initial mass function (IMF), population age, star formation rate, continuum
attenuation, etc. The nebular spectrum is the response of the galaxy ISM
to the ionizing UV continuum ($200 \simlt \lambda_0 \simlt 900$ \AA) -- which cannot be observed directly-- from the same OB stars, and
is also sensitive to the physical conditions (metallicity, electron density $n_{\rm e}$) in the ionized gas.     

Because many diagnostic nebular emission lines for \ion{H}{2} regions are found in the rest wavelength range $3500 \simlt \lambda_0 \simlt 9000$,
easily accessible from the ground for nearby galaxies, there is a long history of nebular spectroscopy and its application to measuring abundances and other
gas-phase physical parameters. Large optical spectroscopic databases such as the Sloan Digital Sky Survey (SDSS) 
are now playing a major role establishing a statistical baseline in the $z \sim 0$ universe.  
However, complementary far-UV spectroscopy of low-redshift galaxies requires observations from
space, where the absence of multiplexing capability, limitations in sensitivity and wavelength coverage, and competition for scarce
{\it Hubble Space Telescope} resources has kept the samples small\footnote{Joint analyses of the UV stellar and optical nebular spectra
of massive star systems in the local universe have been undertaken (e.g., \citealt{gonzalez2000,gonzalez02}), primarily for individual
young massive star clusters or \ion{H}{2} regions, rather than entire star-forming galaxies. The massive star populations in such
systems are extremely sensitive to age, and are typically modeled as instantaneous bursts with ages of only a few Myr. As a result,
(and as we explain below) many of the modeling issues for the high redshift galaxies are significantly different, since they
apply to a system of $\sim 10^{6-7}$ O stars with dynamical times much greater than a few Myr.}   
 
The situation is nearly the reverse in the high-redshift universe, where a large fraction of what we know about 
star-forming galaxies at $z \simgt 1.5$ comes from rest-UV spectra obtained using multiplexed optical spectrometers
on large ground-based telescopes. Rest-frame optical spectroscopy has been slower to develop, awaiting 
sensitive multiplexed near-IR spectroscopic capabilities. Nevertheless, it has long been appreciated that, once such
capabilities arrive, there are specific redshift ranges which maximize access to diagnostic
nebular emission lines due to fortuitous placement with respect to the atmospheric transmission windows:
$1.4 \simlt z \simlt 1.7$, $2.0 \simlt z \simlt 2.6$, and $3 \simlt z \simlt 3.8$.  Of these intervals, the first
is too low to provide full access (from the ground) to the rest-frame far-UV, and the last is
too high to permit observations of \Ha\ due to prohibitively-bright thermal backgrounds. 

The redshift range $z=2-2.6$, however, is optimal for ground-based observations: it provides
access in the J, H, and K band atmospheric windows to a suite of nebular emission lines collectively
sensitive to electron density ($n_{\rm e}$), star formation
rate (SFR), ionization parameter ($U$), metallicity [${\rm 12+log(O/H)}$], extinction, and the  
stellar ionizing radiation field in the 1-4 Ryd (13.6-54.4 eV) range. 
At the same time, $2 \simlt z \simlt 2.6$ brings the rest-frame far-UV (1000-2000 \AA) portion of galaxy spectra
above the atmospheric cutoff near 3100 \AA, and into a region which enjoys unparalleled    
sensitivity for direct observation of  
the integrated far-UV light from massive OB stars {\it in the same galaxies} 
(see, e.g., \citealt{pettini02,shapley03,steidel03,steidel04,rix04,steidel2010}).    
The information content of the far-UV spectra is extremely high: in addition to the integrated
stellar UV continuum, with thousands of stellar photospheric absorption features and several
strong P-Cygni lines from massive star winds, the $\lambda_0 = 1000-2000$ \AA\ wavelength range includes 
nebular emission lines that complement those available in the rest-frame optical. These include 
lines of \ion{O}{3}, \ion{Si}{3}, \ion{N}{3}, and \ion{C}{3}-- in each case the dominant ionization stage in high-excitation
\ion{H}{2} regions --
as well as the \ion{He}{2}$\lambda 1640$ emission feature, which can be observed as a nebular recombination line and/or as a
broad stellar emission feature in the spectra of very hot stars.

It was clear from the initial KBSS-MOSFIRE sample presented by \citet{steidel14} (S14) that standard population
synthesis models of massive stars could not easily account for the high excitation nebular emission observed
in most of the high redshift galaxies.
Recognizing the  
large uncertainties in massive star population synthesis models, S14 used single temperature  
blackbody spectra to approximate the integrated stellar ionizing spectrum in the 1-4 Ryd energy range. It was found  
that the observed ratios of strong nebular emission lines could be reproduced if the energy distribution of the 1-4 Ryd 
radiation field resembles that of a 
55,000-60,000 K blackbody in the {\it steady state}. Such high ionizing radiation field temperatures can 
be produced only over very short durations in single-star population synthesis models, over timescales
much shorter than the (central) dynamical times of $t_{\rm d} \simgt 20-30$~Myr typical
of the $z\sim2.3$ KBSS-MOSFIRE galaxies (e.g., \citealt{erb+06b,reddy08a}). The fact that these galaxies are both
relatively massive ($\langle {\rm log(M_{\ast}/M_{\odot}) \rangle \sim 10}$) and very common
at high redshift suggests a high duty cycle for the prevailing ionization conditions. 
S14 speculated further that massive star population synthesis models that include rapidly 
rotating massive stars (e.g., \citealt{brott11}) and/or that account for 
massive star binary evolution (\citealt{eldridge12}) might mitigate the problem, for the reasons discussed
above. 

In any case, it seems likely that the use of strong nebular lines to infer gas-phase metallicity-- a technique now employed {\it en masse}
at all redshifts-- could be strongly affected by the degeneracy between gas-phase physical conditions 
and the nature of the ionizing stars. It appears that uncertainties in the EUV stellar ionizing spectrum are at least as significant--
if not more so-- than uncertainties in gas-phase abundances and physical conditions when it comes to modeling the nebular spectra of high redshift galaxies.  
Fortunately, it is now feasible to make considerable headway in constraining the {\it combination} of massive star properties 
and nebular physical conditions that can explain the rapidly-improving observations.  

In this paper, we present the results of a pilot program intended to test our ability to construct a self-consistent model 
that accounts for all observational constraints
provided by joint rest-frame far-UV and rest-optical spectra of a representative sample of star-forming galaxies
at $z\sim 2.4$.  In our view, a successful population synthesis+photoionization model must {\it simultaneously} match the overall
far-UV stellar continuum shape, the observed nebular line intensity ratios, and the detailed stellar spectral features (including
photospheric line blanketing and stellar wind features) of the UV stellar spectrum. The foundation of the project is a set 
of high-quality near-IR spectra obtained as part of KBSS-MOSFIRE (S14; Strom et al 2016), to which we have added new, very
deep optical spectra, providing complementary rest-frame far-UV spectra of comparable quality, using Keck/LRIS.    

The paper is organized as follows: \S\ref{sec:observations} details the observations and reductions of 
the optical and near-IR spectra used for this study;  section~\ref{sec:stellar_pops} describes the stellar 
population synthesis models, and the procedure used to compare them to the  
observed rest-frame far-UV spectrum, while \S\ref{sec:PSMs} makes more detailed comparisons of the observed
and model spectra. Section~\ref{sec:measurements} describes the measurement of nebular emission lines in the
rest-frame far-UV and rest-frame optical; 
\S\ref{sec:inferences} derives physical parameters from the nebular line ratios that
are useful as input to photoionization models of the nebulae. The photoionization models -- using the population synthesis models as input -- 
and the detailed comparison of the model grids
with the observed nebular emission measurements, are covered in \S\ref{sec:cloudy_models}; ionized
gas-phase abundances and abundance ratios are derived in \S\ref{sec:metallicity}. Finally, \S\ref{sec:discussion} discusses
the results and their significance to interpreting far-UV and nebular spectra of high redshift galaxies.  

Throughout this paper, where required we have assumed a cosmology with $\Omega_m=0.3$, $\Omega_{\Lambda}=0.7$, $h=0.7$, and a
\citet{chabrier03} stellar initial mass function (IMF).  All wavelengths of spectral lines are referred to using
their vacuum values. 
For definiteness, we generally refer to metallicity using
the metal fraction by mass, $Z$; when conversion between $Z$ and abundances relative to solar are needed, we
have assumed that $Z_{\odot} = 0.0142$, with oxygen abundance ${\rm 12+log(O/H)=8.69}$ (\citealt{asplund09}). Unless
indicated otherwise, solar abundance ratios are assumed.

\begin{deluxetable}{lcc}
\tabletypesize{\scriptsize}
\tablewidth{0pc}
\tablecaption{KBSS-LM1 Sample Statistics}
\tablehead{
\colhead{Property} & \colhead{Median\tablenotemark{a}} & \colhead{Range} }  
\startdata
Redshift    & $2.396\pm0.111\tablenotemark{b}$       & $[2.113, 2.572]$     \\  
${\rm log~(M_{\ast}/M_{\odot})}$ & $~~~9.8\pm0.3$  & $[9.0, 10.8]$    \\
${\rm SFR_{UV}/(M_{\odot}~yr^{-1})}$ & $~32.9 \pm 26.7$ & $[8,102]$     \\
${\rm SFR_{\Ha}/(M_{\odot}~ yr^{-1})}$ & $~29.2 \pm 17.6$ & $[2,330]$    \\
${\rm log~(sSFR/Gyr^{-1})}$ & $0.54\pm0.33$ & $[-0.30,+1.45]$   \\
${\rm E(B-V)_{sed}}$ & $0.16\pm0.07$ & $[0.00, 0.32]$   \\
${\cal R}$        & $24.44\pm0.46$ & $[23.49, 25.43]$    \\
$ K_{\rm s}$  & $23.35 \pm 0.68$ & $[22.12, 24.55]$            \\
$ U_n-G $      &  $0.90\pm0.38$  & $[0.37, 2.49]$  \\
$ G-{\cal R} $      &  $0.10\pm0.04$   & $[0.07, 0.18]$  \\
${\cal R}-K_{\rm s}$ & $0.69\pm0.45$ & $[-0.21,+1.99]$  
\enddata
\tablenotetext{a}{Error bars on median values are the inter-quartile range.}
\tablenotetext{b}{Mean and rms, rather than median. }
\label{tab:sample}
\end{deluxetable}

\section{Observations and Data} 
\label{sec:observations}
We refer to the sample of galaxies analyzed in this paper as KBSS-LM1: Table~\ref{tab:sample} summarizes its most
important properties. 
The data presented in this paper are drawn from the first two Keck/LRIS \citep{oke95,steidel04} slitmasks observed as part of a larger
program to obtain high-quality rest-UV spectra for a sample of $\simeq 100$ galaxies with $\langle z \rangle \simeq 2.4$ drawn
from the deepest sub-sample of the KBSS-MOSFIRE 
survey (S14; Strom et al 2016). The main criteria used to select targets
for LRIS observations are: 1) the redshift lies in the range $2.10 \le z_{\rm gal} \le 2.58$ so that a full set
of strong rest-frame optical nebular emission lines is accessible in the near-IR atmospheric windows (see e.g. \citealt{steidel14}) and 
2) the galaxy observations with MOSFIRE have already been completed in J, H, and K bands, and have yielded emission line measurements
or sensitive limits on the fluxes of the following emission lines: [OII]$\lambda\lambda3727,3729$, [NeIII]$\lambda 3870$ (J band), 
\Hb\ and [OIII]$\lambda\lambda4960,5008$ (H band), and \Ha, [NII]$\lambda\lambda 6549,6585$, and [SII]$\lambda\lambda6718,6732$ (K band).   
Subsequent to applying these criteria, we attempted to ensure that the observed sample spans the full range of properties in the KBSS-MOSFIRE sample, in terms
of inferred SFR, M$_{\ast}$, and metallicity.  Any remaining space on the slitmasks was assigned to galaxies drawn from the
parent KBSS photometric sample with no prior rest-UV spectra. 
The choice of KBSS field was then dictated by the number of high-priority galaxies accommodated on a single slit mask,
and by the time of scheduled observations. One mask each was observed in the KBSS1442+295 and KBSS2343+125 fields, as summarized
in Table~\ref{tab:observations}; Table~\ref{tab:observations} also summarizes the available MOSFIRE observations of 
the high priority targets.  Each of the two slitmasks discussed here includes 15 high priority KBSS-MOSFIRE galaxies, for a total
of 30. 

\begin{deluxetable*}{lccccc}
\tabletypesize{\scriptsize}
\tablewidth{0pc}
\tablecaption{Summary of Observations}
\tablehead{
\colhead{Field} & \colhead{Telescope} & \colhead{Instrument} & \colhead{Wavelength Range} & \colhead{Integration Time} & \colhead{Dates} \\
\colhead{}  & \colhead{} & \colhead{} & \colhead{(\AA)}   & \colhead{(s)} & \colhead{} }  
\startdata
KBSS2343+125      &  Keck 1 10m   &  LRIS-B 600/4000 & $3400-5100$ & 37800 & 2014 Sep 19,20,21 \\
                &  Keck 1 10m   &  LRIS-R 600/5000 & $4900-7650$ & 35880 & 2014 Sep 19,20,21 \\
                &  Keck 1 10m   &  MOSFIRE J       & $11500-13500$ & 14420\tablenotemark{a} & 2012-2014 Sep/Oct \\
                &  Keck 1 10m   &  MOSFIRE H       & $14300-18100$ & 17500\tablenotemark{a} & 2012-2014 Sep/Oct \\
                &  Keck 1 10m   &  MOSFIRE K       & $19400-24060$ & 22150\tablenotemark{a} & 2012-2014 Sep/Oct \\
\hline
KBSS1442+295   &  Keck 1 10m   &  LRIS-B 600/4000 & $3400-5700$  & 36000 & 2015 May 17,18,19 \\
                &  Keck 1 10m   &  LRIS-R 600/5000 & $5500-8000$  & 37100 & 2015 May 17,18,19 \\
                &  Keck 1 10m   &  MOSFIRE J       & $11500-13500$ & ~9260\tablenotemark{a}   & 2013-2014 May/Jun \\
                &  Keck 1 10m   &  MOSFIRE H       & $14300-18100$ & 18900\tablenotemark{a} & 2013-2014 May/Jun \\
                &  Keck 1 10m   &  MOSFIRE K       & $19400-24060$ & ~9260\tablenotemark{a}  & 2013-2014 May/Jun     
\enddata
\tablenotetext{a}{Average total integration time for the galaxies included in this study.}
\label{tab:observations}
\end{deluxetable*}

\subsection{MOSFIRE Observations}

Observations and data reduction for the KBSS-MOSFIRE survey are described in detail elsewhere (\citealt{steidel14}, Strom et al 2016.) Here,
we briefly address the issue of ``cross-band'' calibration, i.e., ensuring that emission line fluxes measured in each atmospheric band
are corrected for slit losses so that ratios of emission lines falling in different bands are accurate. 
Considerable care was used to establish the cross-band calibration for KBSS-MOSFIRE observations, described
in detail by Strom et al (2016). We use a combination of ``slit star'' observations -- the placement of a star with good
broad-band photometric measurements (bright enough to ensure a continuum spectrum of high S/N) on each observed mask--
and multiple observations of a given target on independent masks, to correct for slit losses on an object-by-object and band-by-band basis. 
Each slit-loss-corrected spectrum was assigned a quality flag based on the internal consistency of measurements on different 
masks after applying the nominal correction from the slit star measurements.  
All spectra in the current KBSS-LM1 sample were assigned the highest quality flag, meaning that
the relative flux calibrations have estimated systematic errors of $< 10$\%.   

\begin{figure}[htb]
\centerline{\includegraphics[width=8.5cm]{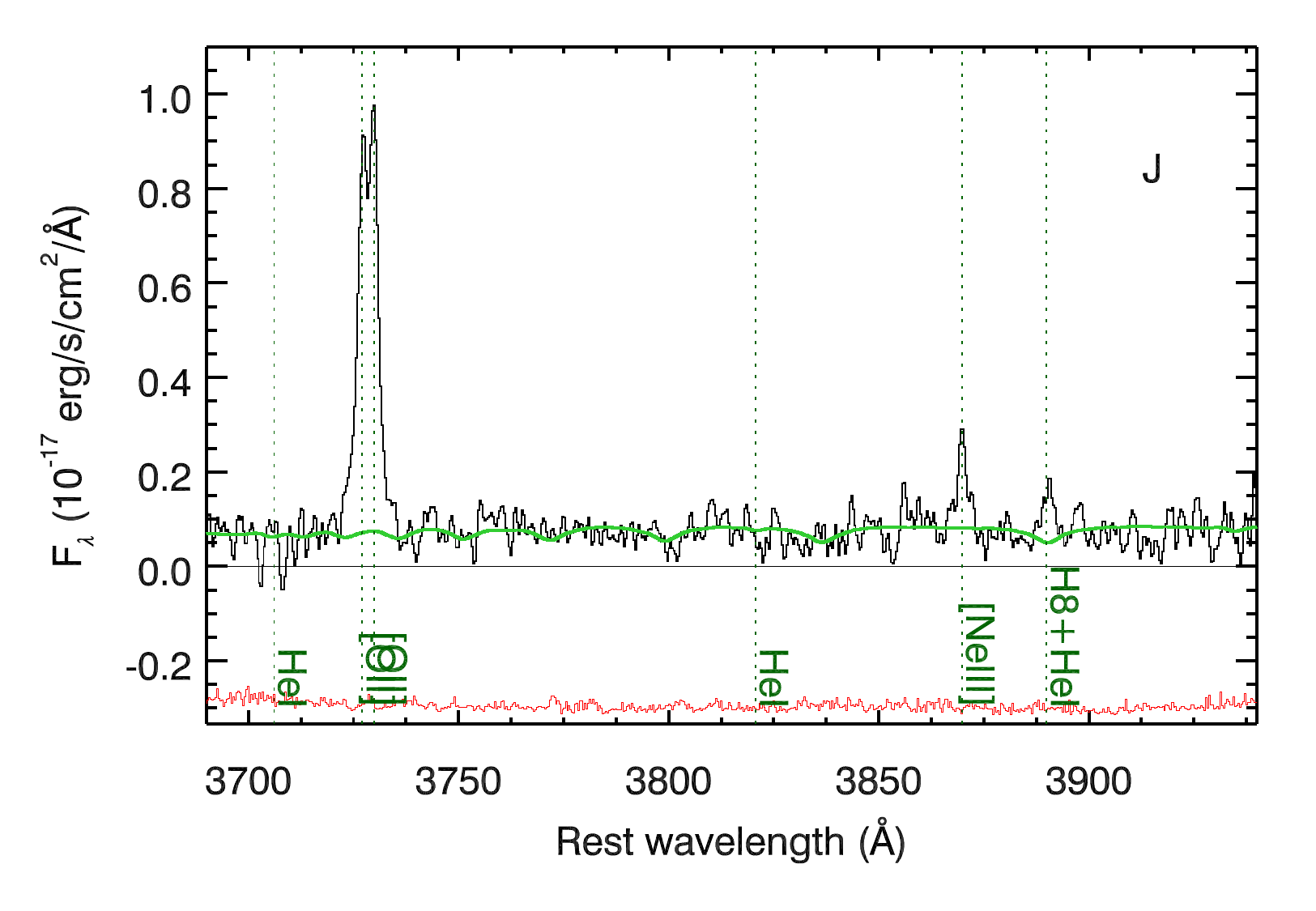}}
\centerline{\includegraphics[width=8.5cm]{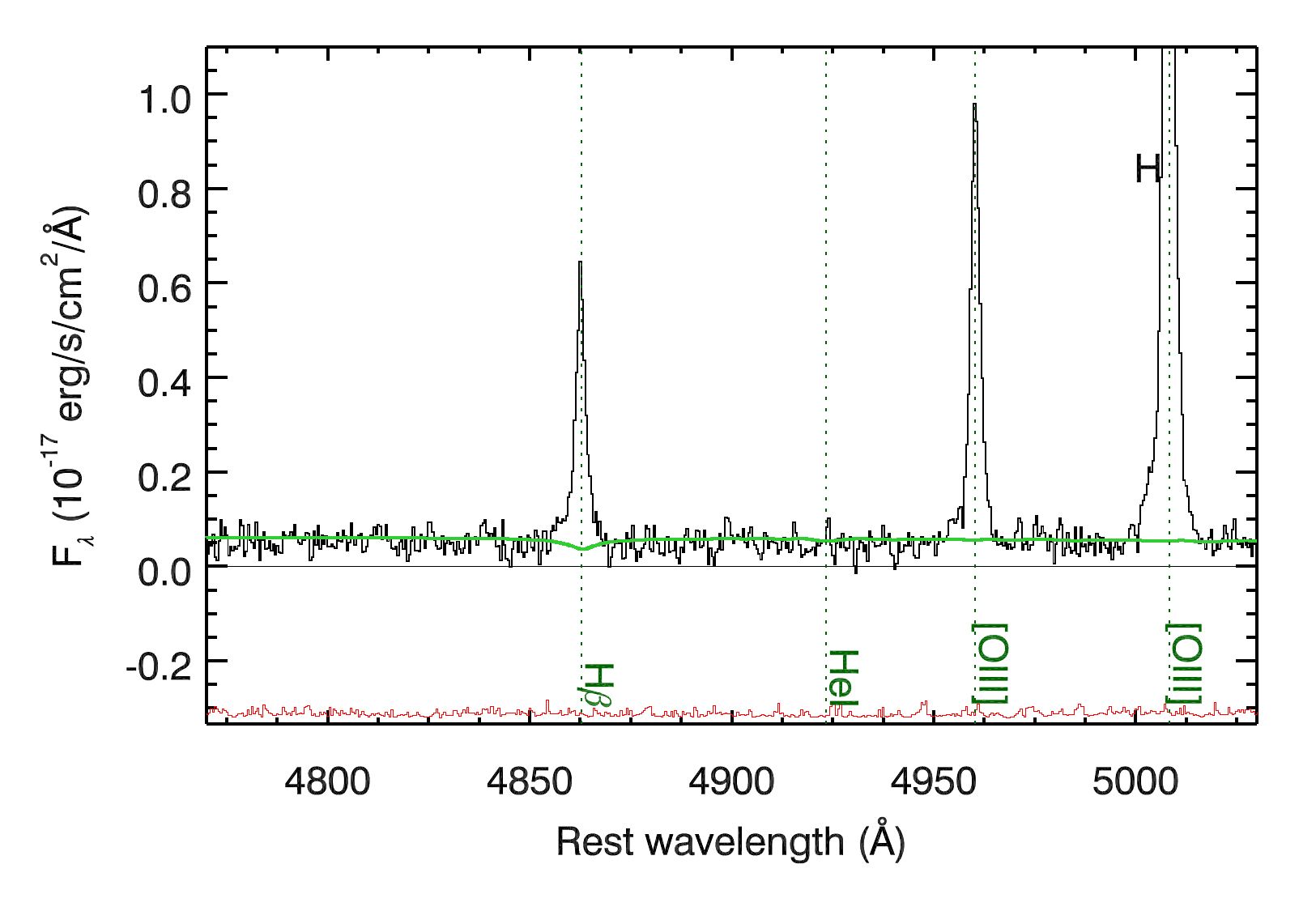}}
\centerline{\includegraphics[width=8.5cm]{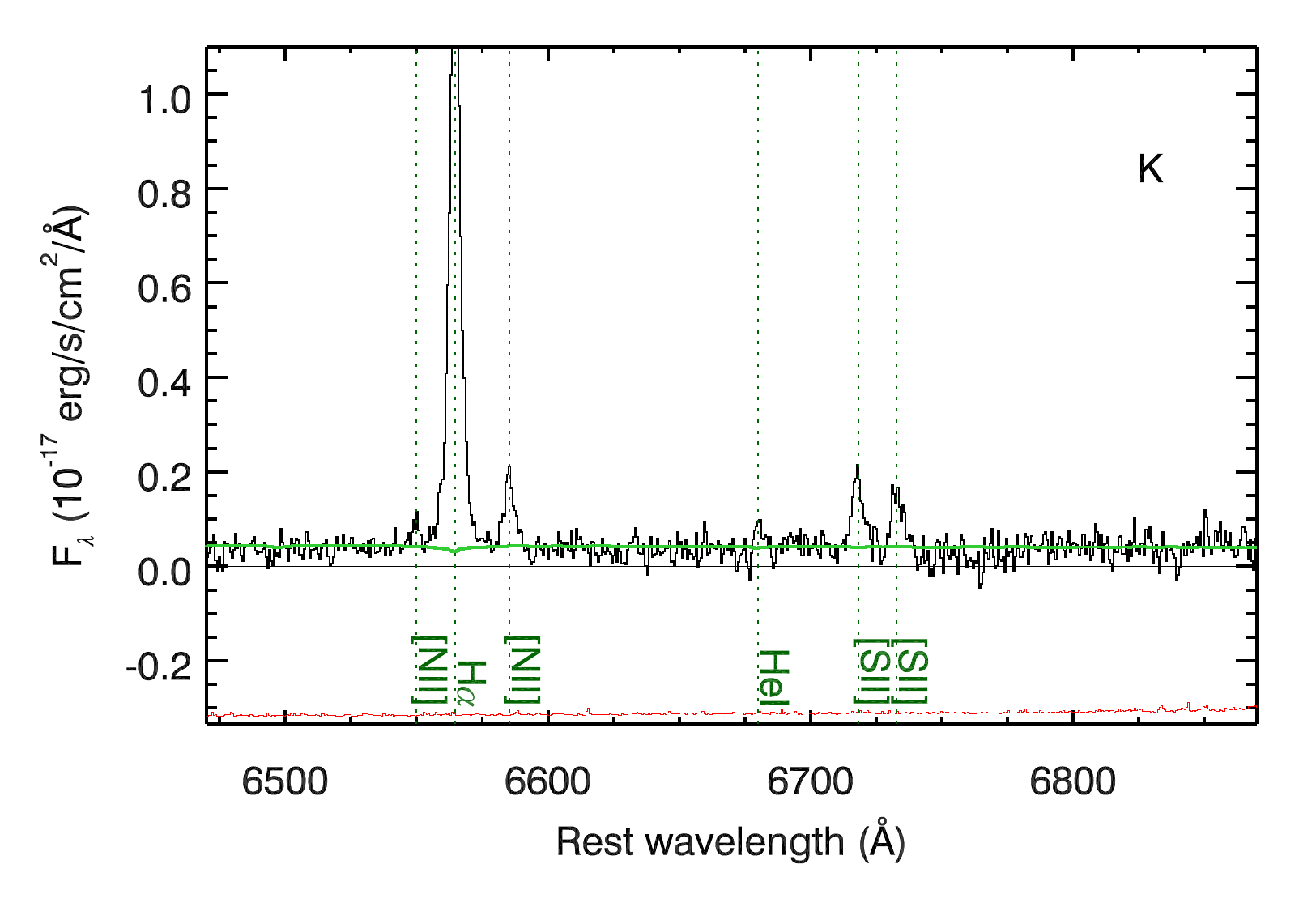}}
\caption{Portions of the MOSFIRE KBSS-LM1 composite spectra (comprised of the same 30 galaxies included in the composite
LRIS spectrum in Figures~\ref{fig:all_deep_full} in the J, H, and K bands (top, middle, and bottom
panels, respectively).  Each panel also shows (red) the 1$\sigma$ error spectrum for the composite. The locations of various strong emission lines are indicated
in each panel; the green curve is the adopted stellar continuum fit.  The ordinate in each panel is the average
flux density that would be {\it observed} if all objects were at the mean redshift, $z=2.396$, while the wavelength scale is shifted to the rest frame. }
\label{fig:mosfire_spectra}
\end{figure}

The slit-loss-corrected J, H, and K band spectra of the 30 galaxies comprising the KBSS-LM1 sample were shifted into
the rest-frame according to the measured nebular redshift $z_{\rm neb}$\footnote{The redshifts were measured independently
in each band, with rms difference $\sigma(\Delta v)=18$ \kms. The average of the H-band and K-band redshifts was
adopted as $z_{\rm neb}$}; the flux scale of each spectrum was adjusted to account
for the small differences in redshift among the sample so that a line of the same luminosity and rest-wavelength
would map to the same line flux in 
each spectrum after resampling 
to a common rest-frame wavelength scale. In practice, the relative scale factors are small ($\simlt 10$\%) 
and make little difference to the final results so long as they are applied in the same way to the spectra in each band. 
The final stacked spectrum in each band was formed by averaging all unmasked pixels at each dispersion point; unless
otherwise noted, the composite spectra are used for measurements only where the wavelength of a spectral feature was observed for all 
30 objects. The 1$\sigma$ error
spectra for each object were propagated to produce a statistical 1$\sigma$ error estimate at each dispersion point in the composite spectrum
within each atmospheric band (J, H, and K). 
Portions of the resulting 
KBSS-LM1 composite spectra, together with their 1$\sigma$ error spectra, are shown in Figure~\ref{fig:mosfire_spectra}; all 30
individual objects contributed to the composite spectra over the entire wavelength regions shown. 

\begin{figure*}[htb]
\centerline{\includegraphics[width=18cm]{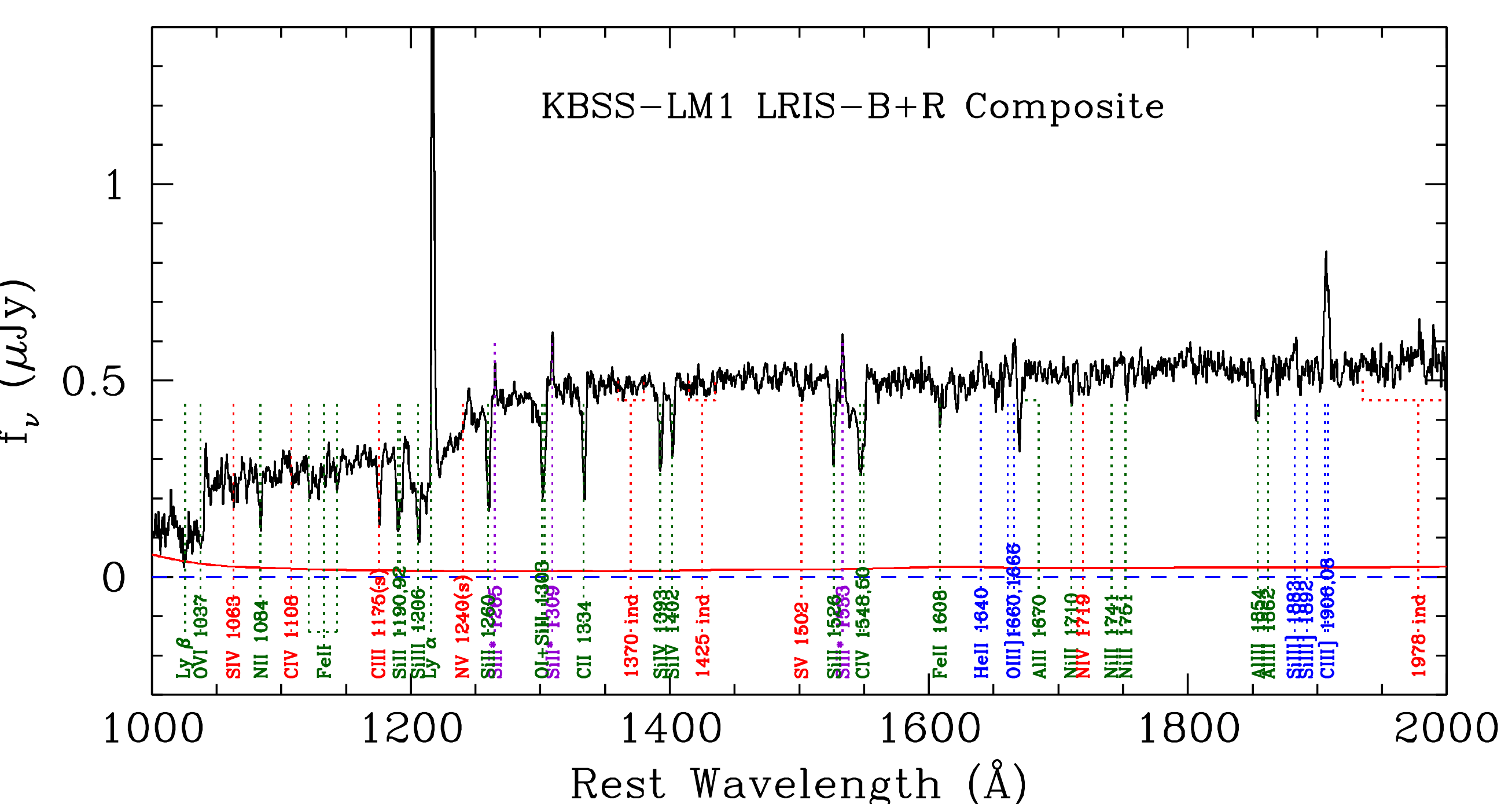}}
\caption{Stacked composite rest-frame UV spectrum of 30 galaxies in the initial KBSS-LM1 sample, with $\langle z \rangle = 2.396\pm0.111$ 
(black histogram). 
Some prominent emission and absorption features are identified, with color-coded labels: stellar absorption features (red), interstellar 
absorption features (dark green), nebular emission lines (blue), and excited fine structure emission lines (dark violet). The emission line spectrum
is discussed in \S\ref{sec:measurements}.}
\label{fig:all_deep_full}
\end{figure*}

\subsection{LRIS-B+R Observations}   

The LRIS spectrograph configuration was chosen to optimize the sensitivity and spectral resolution
over the observed wavelength range $3400 \simlt \lambda \simlt 7300$ \AA, corresponding to a rest wavelength
range of $1000 \simlt \lambda_0 \simlt 2150$ \AA\ at the mean redshift of the sample, $\langle z \rangle = 2.396\pm0.111$.
We used a 600 line/mm grism blazed at 4000 \AA\ in the blue channel (hereinafter LRIS-B), and a 600 line/mm grating blazed
at 5000\AA\ in the red channel (hereinafter LRIS-R); see Table~\ref{tab:observations}. 
For the observations of mask KBSS2343-LM1, we used the d500 dichroic beamsplitter, which
sends wavelengths shortward of $\simeq 5000$\AA\ to LRIS-B and longer
wavelengths to LRIS-R; a configuration using the d560 (with crossover wavelength near
5600 \AA) was used for mask KBSS1442-LM1. Both configurations achieve a spectral resolving power of
$R\sim 1300-1500$ over the full wavelength range when combined with 1\secpoint2 entrance slits and typical seeing of 0\secpoint7 (see, e.g.,
\citealt{steidel2010}).   

The LRIS instrument is located at the Cassegrain focus of the Keck 1 10m telescope, behind the Cassegrain atmospheric dispersion corrector (ADC). 
The ADC allows slitmask position angles to be chosen to optimize the targets on the mask without concern about operating far from
the parallactic angle during the course of the observations. The LRIS-B data were obtained as a series of typically 1800s individual
integrations, with the 4k$\times$4k E2V CCD mosaic binned $1 \times 2$ (binning in the dispersion direction). The LRIS-R exposures, obtained
contemporaneously, were generally 1200s, with $1\times2$ binning of the 4k$\times$4k deep depletion detector as on the blue side. The shorter
individual exposure times in the red were used to mitigate the much higher rate of cosmic ray events owing to the thick CCD substrate
of the LBNL deep depletion CCDs \citep{rockosi10}. 
Wavelength calibration was achieved for both LRIS-B and LRIS-R using a combination of Hg, Cd, Zn, and Ne arc lamps obtained
during afternoon and morning calibrations; the lamp solutions were adjusted to account for any small illumination or flexure-induced
shifts using night sky emission lines observed in the individual science frames. Typical residuals to the wavelength solutions were
$\simlt 0.10$ \AA\ for both LRIS-B and LRIS-R spectra.  

The data were reduced using a suite of custom IRAF and IDL routines, described in detail elsewhere (e.g., \citealt{steidel03,steidel2010}). 
In brief, each slit is edge-traced and rectified to remove curvature in the spatial direction (i.e., to produce rectilinear regions
for which the traces of dispersed object spectra follow straight lines). The 2d spectrograms for each observation are flat-fielded, 
cosmic-ray rejected, and background subtracted. A stacked average 2d spectrogram is then formed after shifting each exposure into
spatial and spectral registration and averaging all unmasked pixels.  The final 1-d spectra were extracted from the 2d spectrograms based
on the observed profiles of the 2d traces of each galaxy. 
Night sky spectra were extracted using the same profile used for the object spectra, from frames processed
identically to the science spectra except for the background subtraction. 
These were used to determine small shifts to the wavelength solutions relative to the arc lamp spectra, and to construct
error spectra based on a noise model that includes the contributions of Poisson uncertainties associated with the object+sky background 
counting statistics plus that of the detector read noise. 
The final 1d spectra were corrected
to vacuum, heliocentric wavelengths and resampled to a log-linear dispersion of 65 \kms\ pix$^{-1}$ ($\simeq 1/3$ of a spectral resolution
element, and close to the original pixel sampling) for both LRIS-B and LRIS-R. 

Flux calibration was achieved using observations of spectrophotometric standard stars from the list of \cite{massey88}; for some
purposes, including those addressed in this paper, it is important to tie the flux calibration of LRIS-B and LRIS-R together,
and to tie both to line flux measurements observed in the near-IR using MOSFIRE. The former is made possible by the contemporaneous
measurement of the standard star (thus both red and blue channels are subject to identical slit losses and atmospheric
conditions), and using as a cross-check the wavelength region of overlap between blue and red, within $\simeq \pm 100$ \AA\ of the nominal dichroic 
cutoff. In general, it was not necessary to apply any scaling between
the flux calibrated spectra in the two different wavelength channels to produce spectra with good relative spectrophotometry. Absolute
spectrophotometry was then achieved by comparing the flux-calibrated LRIS-B+R optical spectra to existing broad-band photometry
of the same objects; such corrections were typically $\le 0.3$ magnitudes. 

The final, stacked, $z=2.4$ KBSS-LM1 rest-UV spectrum was produced in a manner identical to that used
for the MOSFIRE spectra discussed above: the data were shifted to the rest frame using $z_{\rm neb}$ measured
from the MOSFIRE spectra, the flux scales were adjusted to the values they would have at $z=2.396$, 
and averaged without additional weighting. Outlier pixels due to sky subtraction residuals near bright night
sky emission lines were excluded, but otherwise all 30 spectra contributed to the stack over the full range
in rest wavelength $1000-2150$ \AA. 
The final stacked spectrum is shown in Figure~\ref{fig:all_deep_full}, together with the 1$\sigma$ error spectrum 
propagated from the error spectra of the individual objects. The signal-to-noise ratio (S/N) per pixel in the composite
spectrum varies, with
$5 \simlt {\rm S/N} \simlt 25$ over the wavelength range shown in Figure~\ref{fig:all_deep_full}, with median
${\rm S/N \simeq 15}$; the S/N per spectral resolution element ($\simeq 3$ pixels) is higher by a factor of $\sim 1.7$.  
For the purpose of measuring observed emission line intensities, the stacked rest-frame composite was shifted
back to the observed frame using the mean redshift of the sample, $\langle z \rangle = 2.396$, and converted to an $f_{\lambda}$ 
flux density scale.

\section{Stellar Population Synthesis Models}
\label{sec:stellar_pops}

In \citet{steidel14} we were concerned primarily with constraining the {\it shape} of the ionizing radiation field
responsible for producing the observed nebular emission lines in the KBSS-MOSFIRE $z \sim 2.3$ sample; 
as a first approximation, we parametrized the ionizing spectrum using a single-temperature
blackbody with temperature $T_{\rm eff}$\footnote{
$T_{\rm eff}$ should not be confused with a stellar effective temperature that reproduces the bolometric luminosity of
a star given its physical size-- in our usage, $T_{\rm eff}$ refers only to the overall spectral shape of the radiation field 
over the range 1-4 Ryd that is most relevant to the observed emission lines.}. 
In this paper, we are interested in a more detailed 
comparison of the observed far-UV spectrum-- in both its overall spectral shape and its photospheric and stellar wind features-- with state of the art population synthesis models having a range of
stellar metallicities and including additional physics deemed potentially important for massive star populations.  

As discussed in \S\ref{sec:intro}, advances in understanding massive stars
suggest that a realistic population synthesis of the far-UV spectrum of star-forming galaxies should include 
the effects of lower mass-loss rates, the preponderance of binary evolution, and the related phenomenon of rapid
stellar rotation and QHE (e.g., \citealt{brott11,leitherer14}), all of which may be essential to correctly predicting the stellar UV spectrum
and its evolution with time. 
As in \citet{steidel14}, we deliberately refrain from assuming that the stellar metallicity of a population synthesis model
need necessarily match the ionized gas-phase abundance in the nebulae; in part this is motivated by the desire to include the
widest range of plausible stellar energy distributions, but also because the {\it stellar} spectra are primarily
dependent on the total opacity, which is dominated by Fe, and relatively insensitive to O/H (see, e.g., \citealt{rix04}). To date, essentially all
population synthesis models of the UV spectra of star-forming galaxies assume solar abundance ratios\footnote{An exception is 
\citealt{eldridge12}, who considered models with depletion of C relative to O; C/O is discussed in more detail in \ref{sec:co}}. However, it
seems possible -- or perhaps even likely, given the young inferred ages for most high-redshift star-forming galaxies-- that  
O/Fe may be enhanced relative to solar as expected for an ISM that is enriched primarily by core-collapse supernovae. 
If this is the case, one might find that the best-matching stellar model is one having a metallicity several times {\it lower} than
the nebular oxygen abundance. We return to a discussion of this issue in \S\ref{sec:metallicity} and \S\ref{sec:discussion} below.   

\subsection{Model Details}

\label{sec:model_details}

The models we considered include the most recent revision of Starburst99 \citep{leitherer14}, using the new weaker-wind Geneva tracks 
without stellar rotation\footnote{We also considered the corresponding models with rotation as implemented in Starburst99 using 
the Geneva tracks with rotation, but found that they were indistinguishable (in terms of the observational constraints) 
from the non-rotating models.} 
(referred to as ``S99-v00-[$Z_{\ast}$]'', where $Z_{\ast}$ is the assumed mass fraction of metals). 
These are all ``single star'' models, and thus do not include binary evolution. We considered
models with the default \citet{kroupa01} stellar IMF, with a power law index of $-2.30$ over the mass range
${\rm 0.5 \le M_{\ast}/M_{\odot} \le 100}$, as well as models with ``flatter'' IMF slopes of $-2.00$ and $-1.70$ (referred to
as S00-v00-[$Z_{\ast}$]-IMF2.0 and S99-v00-[$Z_{\ast}$]-IMF1.7).  We considered each of these for metallicities $Z_{\ast}=\left(0.001,0.002,0.008,0.014\right)$,
dictated by the available Geneva tracks. 

We have also included a suite of models from the new release of ``Binary Population
and Spectral Synthesis'' (BPASSv2\footnote{http://bpass.auckland.ac.nz}; \citealt{eldridge16,stanway16}) 
all with an IMF index of $-2.35$. We refer to 3 sets of models according to the stellar metallicity $Z_{\ast}$, the upper
mass cutoff of the assumed IMF (in units of M$_{\odot}$), and whether or not binary evolution is included:  BPASSv2-[$Z_{\ast}$]-100bin models 
include binary evolution and an IMF with
upper mass cutoff of 100 M$_{\odot}$, BPASSv2-[$Z_{\ast}$]-300bin include binary evolution with upper mass cutoff of 300 M$_{\odot}$, 
and BPASSv2-[$Z_{\ast}$]-300 are ``single star'' models with cutoff mass 300 M$_{\odot}$.
The models considered have stellar metallicities 
$Z_{\ast}=\left(0.001,0.002,0.003,0.004,0.006,0.008,0.010,0.014\right)$. 
 
All models (both S99 and BPASSv2) assume continuous star formation with a duration of $10^8$ yr.~\footnote{The models considered 
reach equilibrium in the far-UV after a few $\times 10^7$ yrs, after which the UV ionizing spectral energy distributions are time-independent. The typical
ages inferred from SED fits over the full UV-mid-IR range are $\simeq 300$ Myr (e.g., \citealt{reddy12}).}  

For direct comparison to the KBSS-LM1 far-UV spectrum, we used the high-resolution WM-Basic spectra covering
the wavelength range 900-3000 \AA\ spectral resolution of 0.4 \AA\   
provided by Starburst99, which we smoothed and re-sampled
to match the wavelength scale and spectral resolution of the 
LM1 composite, 65 \kms\ pix$^{-1}$ and FWHM$\simeq 210$ \kms, respectively. 
The BPASSv2 model spectra are provided with 1~\AA\ pix$^{-1}$ sampling
in the UV; these were spline-interpolated onto the observed pixel scale.  

\begin{figure}[htbp]
\centerline{\includegraphics[width=8.5cm]{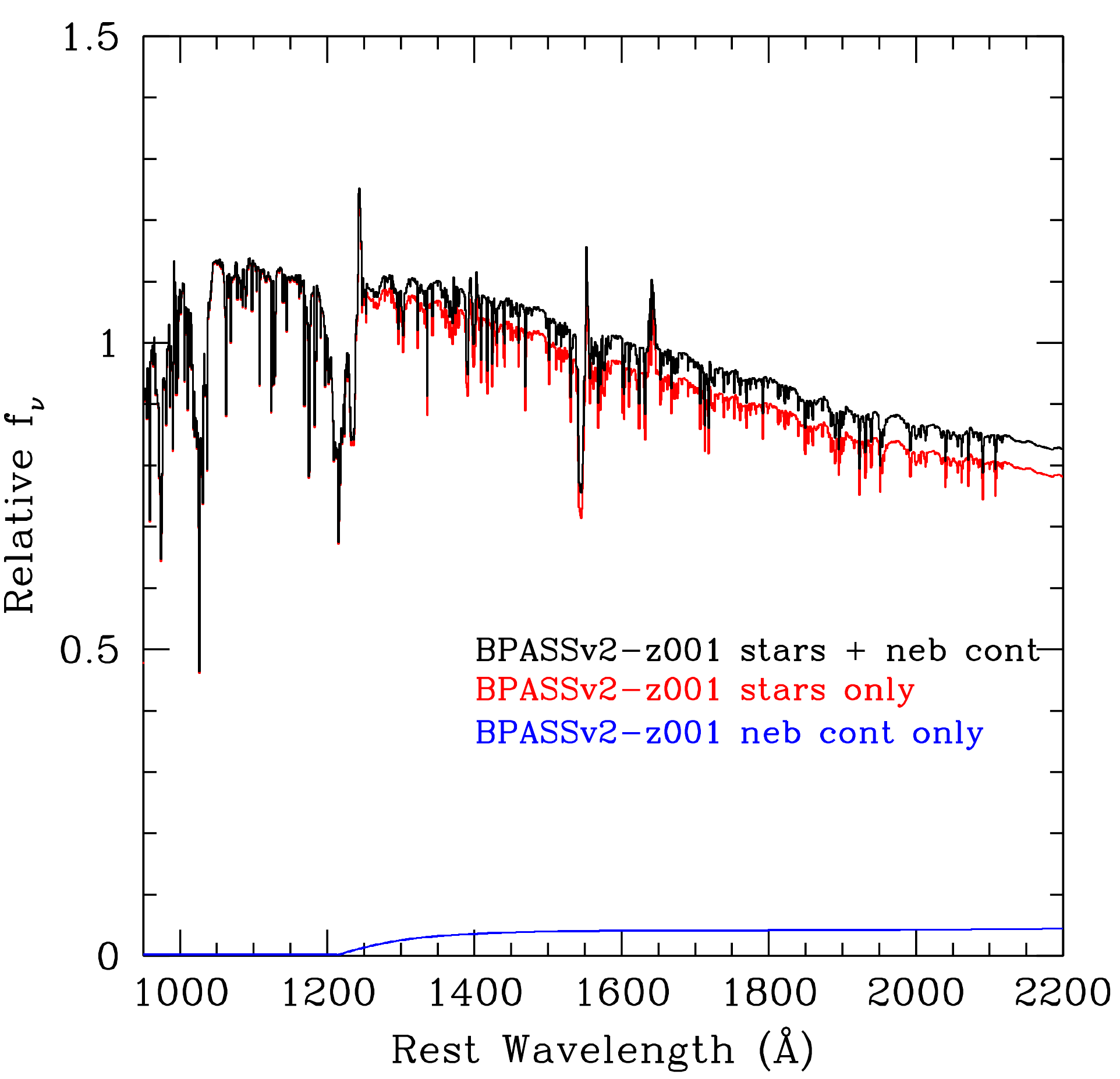}}
\caption{An example illustrating the calculated contribution of nebular continuum emission (blue) to a BPASSv2-z001-100bin stellar population
synthesis model assuming continuous star formation over $10^8$ years. The red spectrum is that of the stars only, while the black
spectrum is the sum of the stellar and nebular continuum that would be used for comparison to the observed spectrum. The spectra
are scaled such that the purely stellar spectrum is normalized to $f_{\nu} = 1$ at 1500 \AA. For this particular model,
the contribution of the nebular continuum
emission to the total ranges from $\simeq 3.5$\% at 1300~\AA\ to $\simeq 6.5$\% at 2200~\AA.
}
\label{fig:nebcont}
\end{figure}

The high spectral resolution models for both S99 and BPASSv2 do not include the contribution of the nebular
continuum to the far-UV spectra; to include this contribution self-consistently,
we used the photoionization models (\S\ref{sec:cloudy_models}) that produced the best matches to 
the observed nebular emission spectrum  
to calculate 
the emergent nebular continuum contribution relative to the stellar continuum. This contribution
was then added to the stellar spectra provided by S99 and BPASSv2 models prior to performing the comparison to the data. For
the assumed star formation ages, the nebular continuum contributes $\simeq 3-7$\% to the total far-UV continuum flux density for rest wavelengths
in the range $1200 \simlt \lambda_0 \simlt 2150$ \AA\ (see the example in Figure~\ref{fig:nebcont}.) 
Including the nebular continuum has the effect of making the total continuum slightly redder 
than the purely stellar continuum over this range-- fits using the stellar continuum only for the same population synthesis models 
require slightly larger values of $E(B-V)_{\rm cont}$ ($\simeq 0.21$) to match the observed spectrum. 

\begin{figure*}[htb]
\centerline{\includegraphics[width=18cm]{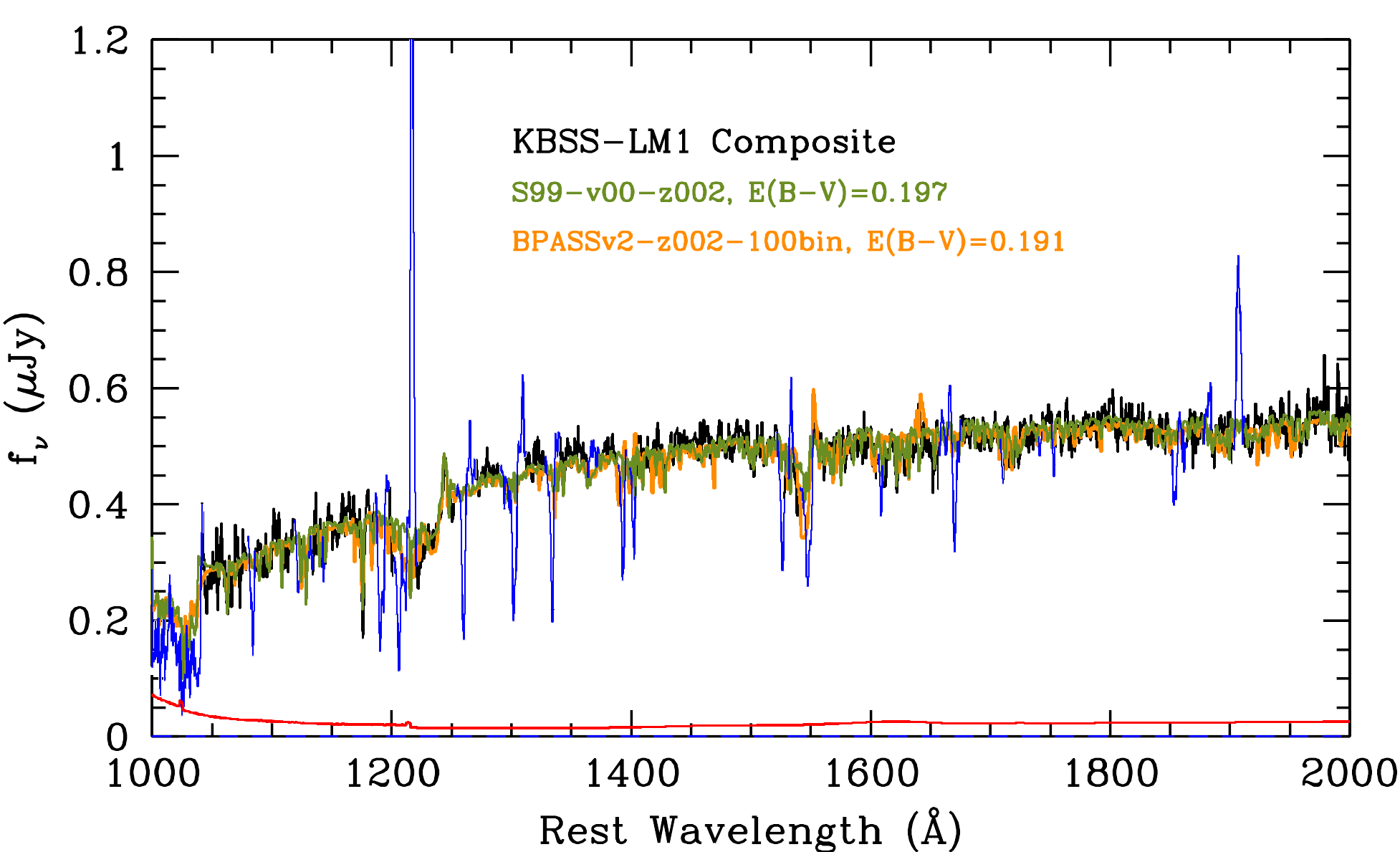}}
\caption{As in Fig.~\ref{fig:all_deep_full}, the KBSS-LM1 composite FUV spectrum is plotted as a black histogram. 
The observed spectrum has been corrected for the mean intergalactic+circumgalactic attenuation appropriate at $z\simeq 2.40$ (\citealt{rudie13}).
Superposed are two of the best-fitting population synthesis models (see Table~\ref{tab:chi2} and \S\ref{sec:model_details}).
Both model spectra include the predicted contribution of the nebular continuum emission, calculated
using the photoionization models described in \S\ref{sec:PSMs}.  
The model spectra were reddened assuming
the \citet{calzetti00} starburst attenuation law, where $E(B-V)$ was adjusted to minimize $\chi^2$ with respect to the observed 
spectrum after masking spectral regions containing strong interstellar absorption or nebular emission lines (violet). 
}
\label{fig:all_deep_full_wmods}
\end{figure*}

In preparation for fitting models to the observations, 
the KBSS-LM1 spectrum shown in Figure~\ref{fig:all_deep_full_wmods} has been corrected for the mean intergalactic (IGM) and circumgalactic CGM)
opacity due to neutral hydrogen (\ion{H}{1}) along the line of sight, for a source with $z=2.40$. This was accomplished using Monte Carlo sampling of
the statistical distribution function of neutral hydrogen column density $f(N_{\rm HI},z)$ measured by \citet{rudie13} from high resolution spectra of the
KBSS QSOs. Including the CGM component accounts for the average additional opacity experienced  
by a galaxy in the KBSS spectroscopic sample, over and above a source located at a random position within the IGM. 
The IGM/CGM correction, which affects the spectrum only shortward of $\lambda_0 = 1216$ \AA\ and
amounts to an average upward correction to the observed continuum of $\sim 20$\%, was
applied to the {\it observations} so that the resulting spectrum represents, as closely as possible,
the intrinsic far-UV spectrum of the stellar populations. 

We then created masks which exclude wavelength pixels
where the IGM-corrected LM1 spectrum (Figure~\ref{fig:all_deep_full_wmods}) 
is affected by features unrelated to the intrinsic stellar spectrum-- i.e., the positions of nebular
emission lines and interstellar absorption lines, as illustrated in Figure~\ref{fig:all_deep_full_wmods}. The parameters of 2 different
masks are given in Table~\ref{tab:masks}-- Mask 1 is intended for global far-UV fitting, and includes pixels over the
full wavelength range 1040-2000 \AA\ as well as the vicinity of strong P-Cygni features from stellar winds. Mask 2 is intended
to emphasize spectral regions that are more sensitive to stellar photospheric line blanketing, for reasons discussed in \S\ref{sec:photospheric} below.  
\begin{deluxetable}{lccc}
\tabletypesize{\scriptsize}
\tablewidth{0pc}
\tablecaption{Rest Wavelength Regions Used for $\chi^2$ Fits\tablenotemark{a}}
\tablehead{
\colhead{$\lambda_{\rm min}$} & \colhead{$\lambda_{\rm max}$} & \colhead{Mask 1} & \colhead{Mask 2} \\
\colhead{(\AA)} & \colhead{(\AA)} & \colhead{($N_{\rm pix}=2771$)} & \colhead{($N_{\rm pix}=1710$)} }
\startdata
1042 & 1080 & 1 & 0 \\
1086 & 1119 & 1 & 0 \\
1123 & 1131 & 1 & 0 \\
1134 & 1141 & 1 & 0 \\
1144 & 1186 & 1 & 0 \\
1198 & 1202 & 1 & 0 \\
1221 & 1254 & 1 & 0 \\
1270 & 1291 & 1 & 0 \\
1312 & 1328 & 1 & 0 \\
1340 & 1363 & 1 & 0 \\
1373 & 1389 & 1 & 0 \\
1396 & 1398 & 1 & 0 \\
1404 & 1521 & 1 & 1 \\
1528 & 1531 & 1 & 0 \\
1536 & 1541 & 1 & 0 \\
1552 & 1606 & 1 & 1 \\
1610 & 1657 & 1 & 1 \\
1675 & 1708 & 1 & 1 \\
1711 & 1740 & 1 & 1 \\
1743 & 1751 & 1 & 1 \\
1754 & 1845 & 1 & 1 \\
1864 & 1878 & 1 & 0 \\
1885 & 1903 & 1 & 0 \\
1920 & 2000 & 1 & 1 
\enddata
\tablenotetext{a}{``1'' indicates that the wavelength region is included in the mask; ``0'' indicates that the region is not included.}
\label{tab:masks}
\end{deluxetable}

The spectrum generated by each model was compared directly to the observed spectrum, varying only the degree of continuum reddening
by dust and the overall normalization of the reddened model spectrum relative to the observations.
The model spectra were reddened using the \citet{calzetti00} continuum attenuation 
relation\footnote{We also evaluated a grid of models using an SMC extinction
curve as parametrized by \citet{gordon03}, which can reproduce the observed spectrum as well as \citet{calzetti00} attenuation,
albeit with smaller values of the best-fit $E_{\rm B-V}$. We adopt \citet{calzetti00} to maintain consistency with the attenuation
relation assumed in the SED fitting.} parametrized by the continuum color excess, $E(B-V)_{\rm cont}$, which was varied over
the range 
$0.000 \le E(B-V)_{\rm cont} \le 0.600$, in steps of 0.001.
Using Mask 1 (Table~\ref{tab:masks}), 
the pixels in each reddened model spectrum were compared to the same pixels in the LM1 composite after multiplying the 
model by a normalization constant
that forced the median intensity of included pixels to be equal to that in the data. 
The total $\chi^2$ was then computed,
\begin{equation}
\chi^2 = \sum_i \left(\frac{{\rm model}(i)-{\rm data}(i)}{\sigma(i)}\right)^2 
\label{eqn:chi2}
\end{equation}
where $\sigma(i)$ is the value of the error spectrum at pixel $i$ and the summation is over all included pixels.
The value of $E(B-V)_{\rm cont}$ that minimized
$\chi^2$ for each population synthesis model, and the corresponding value of $\chi^2/\nu$, where $\nu$ 
is the number of spectral pixels in the applied mask, are recorded in the top portion
of Table~\ref{tab:chi2}.  For reasons of space,  we have omitted from Table~\ref{tab:chi2} the results for metallicities $Z_{\ast} > 0.008$  
since the fits were increasingly poor with increasing $Z_{\ast}$.  

Also listed in Table~\ref{tab:chi2} are two additional parameters: $\Delta \chi^2/\nu$ is the
difference in $\chi^2/\nu$ between that of each model and the best-fitting model in the ensemble. 
The last column, $\Delta\sigma$, is a measure of the significance of the difference between the goodness
of fit of each model and the best-fitting model of the same type (i.e., BPASSv2 or S99), in terms
of the number of $\sigma$ (assuming normally distributed errors) calculated for a $\chi^2$ distribution with $\nu$ degrees of freedom.
With these assumptions, 
$\Delta \sigma=1$ corresponds to $\Delta \chi^2/\nu \approx 0.031$ for $\nu=2771$ and $\Delta \chi^2/\nu \approx 0.039$ for
$\nu=1710$.  

\begin{deluxetable}{lccccc}
\tabletypesize{\scriptsize}
\tablewidth{0pc}
\tablecaption{Results of $\chi^2$ Minimization for Stellar Models}
\tablehead{
\colhead{Model} & \colhead{$Z_{\ast}/Z_{\odot}$\tablenotemark{a}} & \colhead{E(B-V)\tablenotemark{b}} & \colhead{$\chi^2/\nu$} & 
\colhead{$\Delta\chi^2/\nu$\tablenotemark{c}} & \colhead{$\Delta \sigma$\tablenotemark{d}}}
\startdata
\cutinhead{Mask1 (Global) fit ($\nu=2771$)} 
BPASSv2-z001-100bin & 0.07 & 0.185  &      1.098 &  0.026  & ~0.84 \\
BPASSv2-z002-100bin & 0.14 & 0.191  &      1.090 &  0.018  & ~0.58 \\
BPASSv2-z003-100bin & 0.21 & 0.192  &      1.134 &  0.062 &  ~2.00 \\
BPASSv2-z004-100bin & 0.28 & 0.196  &      1.236 &  0.164 &  ~5.29 \\
BPASSv2-z006-100bin & 0.42 & 0.187  &      1.385 &  0.313 &  10.09 \\
BPASSv2-z008-100bin & 0.56 & 0.181  &      1.582 &  0.510 &  16.45 \\
\hline
BPASSv2-z001-300bin &   0.07 &  0.187   &      1.186 &  0.114 & ~3.68 \\
BPASSv2-z002-300bin &   0.14 &  0.190   &      1.159 &  0.087 & ~2.81 \\
BPASSv2-z003-300bin &   0.21 &  0.192   &      1.201 &  0.129 & ~4.16 \\
BPASSv2-z004-300bin &   0.28 &  0.192   &      1.284 &  0.212 & ~6.84 \\
BPASSv2-z006-300bin &   0.42 &  0.186   &      1.369 &  0.297 & ~9.58 \\
BPASSv2-z008-300bin &   0.56 &  0.183   &      1.560 &  0.488 & 15.74 \\
\hline
BPASSv2-z001-300 & 0.07 &  0.173   &      1.072 &  0.000 & ~0.00 \\  
BPASSv2-z002-300 & 0.14 &  0.181   &      1.075 &  0.003 & ~0.10 \\ 
BPASSv2-z003-300 & 0.21 &  0.185   &      1.137 &  0.065 & ~2.10 \\
BPASSv2-z004-300 & 0.28 &  0.188   &      1.253 &  0.181 & ~5.84 \\
BPASSv2-z006-300 & 0.42 &  0.181   &      1.384 &  0.312 & 10.06 \\
BPASSv2-z008-300 & 0.56 &  0.176   &      1.585 &  0.513 & 16.54 \\ 
\hline
S99-v00-z001 &       0.07 &  0.193  &   1.111  & 0.096 & ~3.15 \\
S99-v00-z002 &       0.14 &  0.197  &   1.015 &  0.000 & ~0.00 \\
S99-v00-z008 &       0.56 &  0.200  &   1.530 &  0.515 & 16.90 \\
\hline
S99-v00-z001-IMF2.0 & 0.07 &  0.188 &   1.144 &  0.129 & ~4.24 \\
S99-v00-z002-IMF2.0 & 0.14 &  0.181 &   1.061 &  0.046 & ~1.52 \\
S99-v00-z008-IMF2.0 & 0.56 &  0.197 &   1.467 &  0.452 & 14.83 \\
\hline
S99-v00-z001-IMF1.7 & 0.07 &  0.174 &   1.185 &  0.170 & ~5.57 \\
S99-v00-z002-IMF1.7 & 0.14 &  0.174 &   1.150 &  0.135 & ~4.42 \\
S99-v00-z008-IMF1.7 & 0.56 &  0.187 &   1.456 &  0.441 & 14.45  \\
\cutinhead{Mask 2 fit ($\nu =1710$)} 
BPASSv2-z001-100bin & 0.07 & 0.203 &   1.089 & 0.038 & ~0.97 \\
BPASSv2-z002-100bin & 0.14 & 0.201 &   1.126 & 0.075 & ~1.92 \\
BPASSv2-z003-100bin & 0.21 & 0.203 &   1.181 & 0.130 & ~3.33 \\
BPASSv2-z004-100bin & 0.28 & 0.202 &   1.270 & 0.219 & ~5.62 \\
BPASSv2-z006-100bin & 0.42 & 0.183 &   1.393 & 0.342 & ~8.77 \\
BPASSv2-z008-100bin & 0.56 & 0.177 &   1.533 & 0.482 & 12.36 \\
\hline
BPASSv2-z001-300bin &   0.07 &   0.205 &    1.107 & 0.056 & ~1.43  \\
BPASSv2-z002-300bin &   0.14 &   0.205  &   1.140 & 0.089 & ~2.28 \\
BPASSv2-z003-300bin &   0.21 &   0.204  &   1.187 & 0.136 & ~3.49 \\
BPASSv2-z004-300bin &   0.28 &   0.200  &   1.261 & 0.210 & ~5.39 \\
BPASSv2-z006-300bin &   0.42 &   0.187  &   1.367 & 0.316 & ~8.10 \\
BPASSv2-z008-300bin &   0.56 &   0.183  &   1.508 & 0.457 & 11.72 \\
\hline
BPASSv2-z001-300 & 0.07 &   0.186  &   1.051 & 0.000 & ~0.00 \\ 
BPASSv2-z002-300 & 0.14 &   0.187  &   1.093 & 0.042 & ~1.08 \\
BPASSv2-z003-300 & 0.21 &   0.190  &   1.166 & 0.115 & ~2.95 \\ 
BPASSv2-z004-300 & 0.28 &   0.189  &   1.264 & 0.213 & ~5.46 \\
BPASSv2-z006-300 & 0.42 &   0.177  &   1.383 & 0.332 & ~8.51 \\
BPASSv2-z008-300 & 0.56 &   0.176  &   1.529 & 0.478 & 12.26 \\
\hline 
S99-v00-z001 &     0.07 &   0.204 &      0.986 & 0.000 & ~0.00  \\
S99-v00-z002 &     0.14 &   0.193 &      1.024 & 0.038 & ~0.98  \\
S99-v00-z008 &     0.56 &   0.196 &      1.502 & 0.516 & 13.41 \\
\hline
S99-v00-z001-IMF2.0 &   0.07  &  0.200 &   0.994 & 0.008 & ~0.20  \\
S99-v00-z002-IMF2.0 &   0.14  &  0.188 &   1.004 & 0.018 & ~0.46 \\
S99-v00-z008-IMF2.0 &   0.56  &  0.187 &   1.439 & 0.453 & 11.77 \\
\hline
S99-v00-z001-IMF1.7 &   0.07  &  0.192 &   1.005 & 0.019 & ~0.50 \\
S99-v00-z002-IMF1.7 &   0.14  &  0.180 &   1.004 & 0.018 & ~0.47 \\
S99-v00-z008-IMF1.7 &   0.56  &  0.182 &   1.360 & 0.374 & ~9.71 
\enddata
\tablenotetext{a}{Stellar metallicity relative to solar, where $Z_{\odot} = 0.0142$ (\citealt{asplund09}).}
\tablenotetext{b}{Value of E(B-V)$_{\rm cont}$ that minimizes $\chi^2$.}
\tablenotetext{c}{Difference in $\chi^2/\nu$ compared to best-fitting model within group (BPASSv2 or S99).}
\tablenotetext{d}{Number of $\sigma$ deviation from best-fitting models within group (BPASSv2 or S99).}
\label{tab:chi2}
\end{deluxetable}

As can be seen in Table~\ref{tab:chi2}, the best-fit values of E(B-V)$_{\rm cont}$ are similar across all of the
model spectra (they vary between 0.172-0.200), reflecting the fact that the unreddened FUV model 
spectral shape depends only modestly on stellar abundance and IMF. However, the fits are quite sensitive to the degree
of photospheric line blanketing, which is a relatively strong function of stellar metallicity. 
For the global fits using Mask 1, the best-fitting population synthesis models 
are BPASSv2-z001-300 and BPASSv2-z002-100bin (which differ by $\Delta\sigma < 1$) and S99-v00-z002. Figure~\ref{fig:all_deep_full_wmods} 
shows the two of these models (BPASSv2-z002-100bin and S99-v00-z002) after applying the reddening values 
that minimized $\chi^2$, E(B-V)$_{\rm cont}= 0.191$ and 0.197, respectively.
Table~\ref{tab:chi2} shows that all of the models within $\Delta\sigma< 2$ of the best fitting model have
$Z_{\ast} \le 0.002$,  
and that all models with $Z_{\ast} > 0.003$ ($Z_{\ast}/Z_{\odot} > 0.21$) are
strongly dis-favored in comparison.  
We return to a more detailed comparison of the model versus observed far-UV spectrum in 
\S\ref{sec:PSMs} below.  

The values of $E(B-V)_{\rm cont}$ for the two best-fitting models are 
slightly higher than the median value from the SED-fits of the individual galaxies in the LM1 sample, $E(B-V)_{\rm sed} =0.16$ (Table~\ref{tab:sample}).
The difference is likely due to the fact that the SED fits used a different grid of stellar models (the solar metallicity
models of \citealt{bc03}, which are systematically somewhat redder in the far-UV, and thus require smaller values of $E(B-V)$ to
match the same observed spectrum\footnote{The best-fit continuum color excess for the continuous star formation
BC03 model with an age of $10^8$ yrs   
is $E_{\rm B-V} = 0.156$ using the fitting method described above.}). We compare the inferred continuum attenuation, which implies
extinction $A_{\lambda}\approx 2.0$ mag at $\lambda =1500$ \AA, to that inferred from the nebular emission lines
in \S\ref{sec:nebular_ext} below.  

\section{Detailed Comparison of the Far-UV Spectra}
\label{sec:PSMs}

Whereas the global far-UV spectral shape over the fitted range 1000 \AA$ \le \lambda_0 \le$ 2000 \AA\ 
is easily matched by the two families of population synthesis models after applying 
a standard starburst attenuation relation (the only free parameters are $E(B-V)$ and the overall normalization), the KBSS-LM1 composite
spectrum is of high enough spectral resolution and S/N to merit a more detailed comparison of the observed
and predicted spectral features. 

\begin{figure*}[htbp]
\centerline{\includegraphics[width=8.5cm]{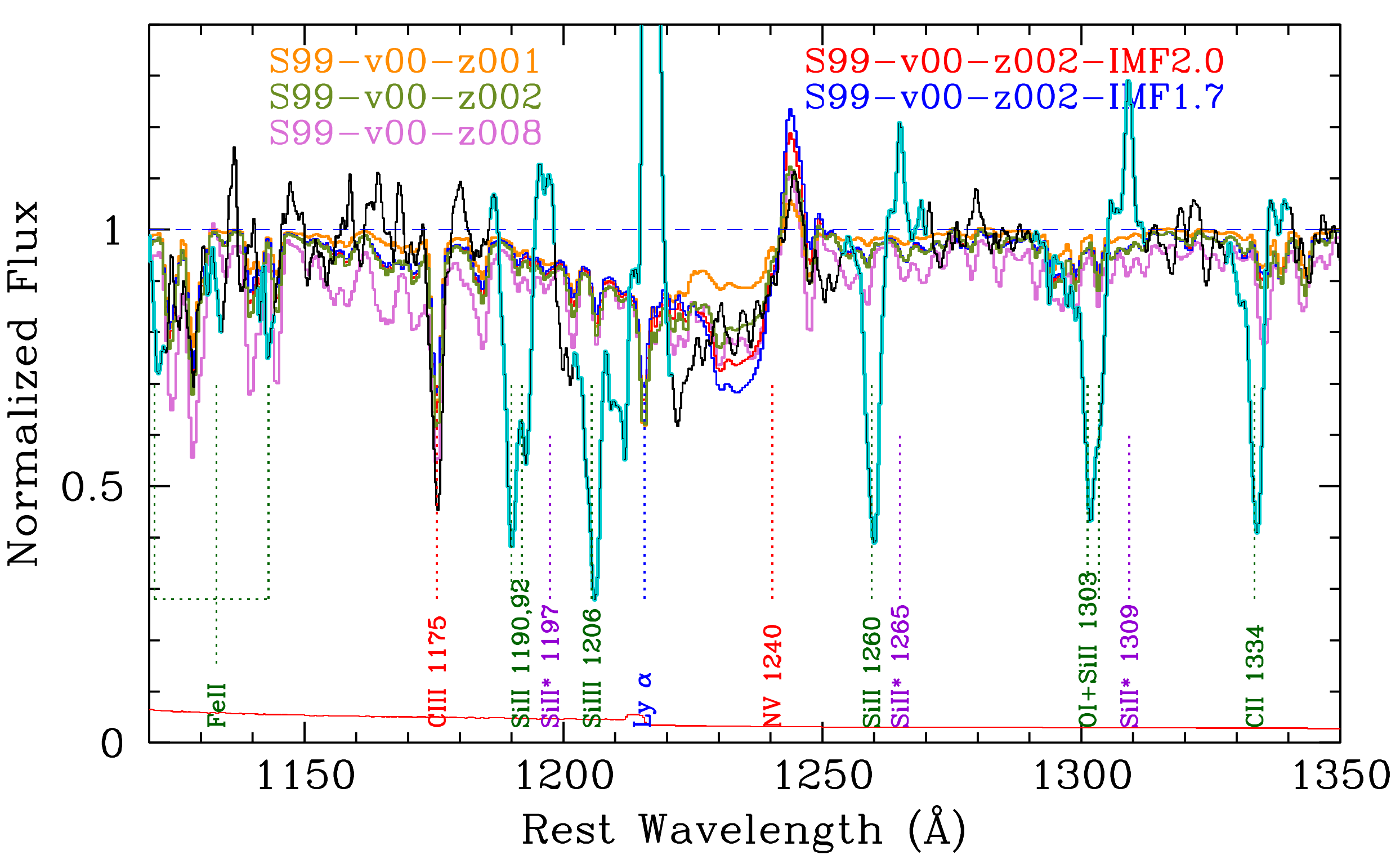}\includegraphics[width=8.5cm]{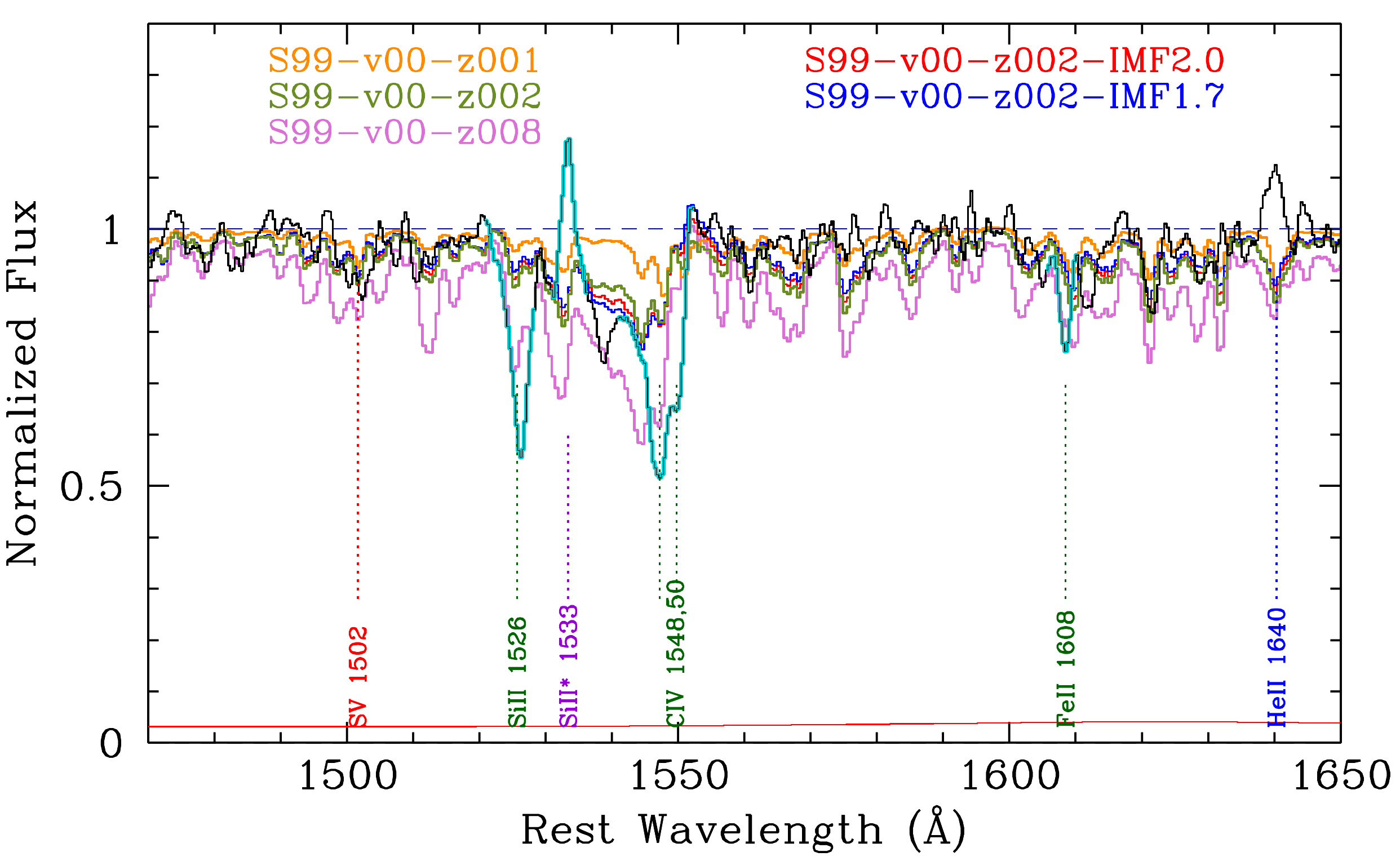}}
\centerline{\includegraphics[width=8.5cm]{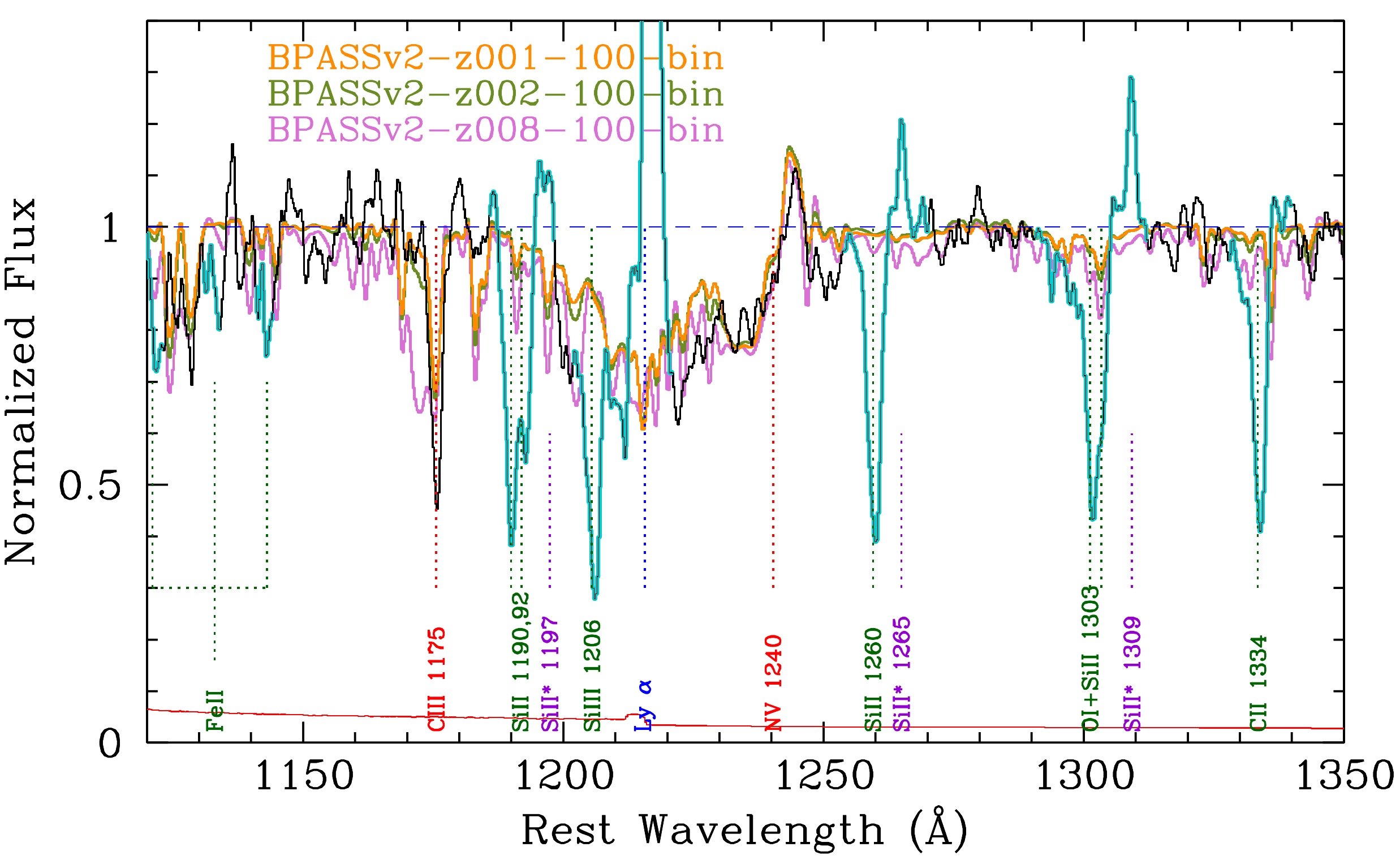}\includegraphics[width=8.5cm]{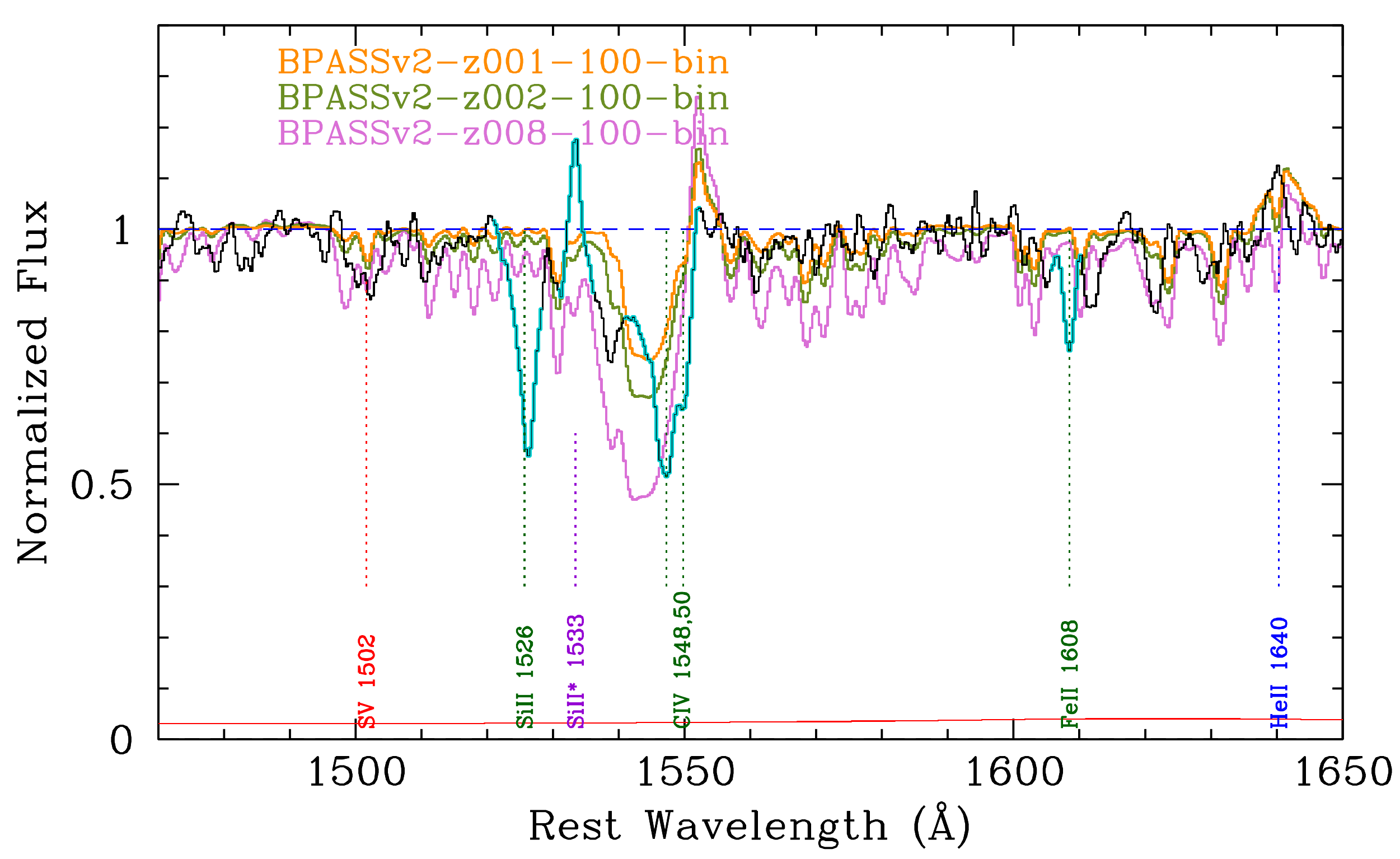}}
\caption{Zoom in of the spectral regions near the \ion{N}{5} (left panels) and \ion{C}{4} (right panels) wind lines, 
comparing various S99 (top) and BPASSv2 (bottom) models with the KBSS-LM1 spectrum. 
The top panels include three different IMFs for the S99-z002 models to illustrate the IMF dependence of the P-Cygni profiles.  
As in Figure~\ref{fig:all_deep_full_wmods}, regions of the spectrum that were 
excluded from the global fits (see \S\ref{sec:PSMs}) are shaded in cyan. The color-coding of the line identifications is
the same as in Fig.~\ref{fig:all_deep_full}.    
}
\label{fig:civ_zoom}
\end{figure*}

\begin{figure*}[htbp]
\centerline{\includegraphics[width=8.5cm]{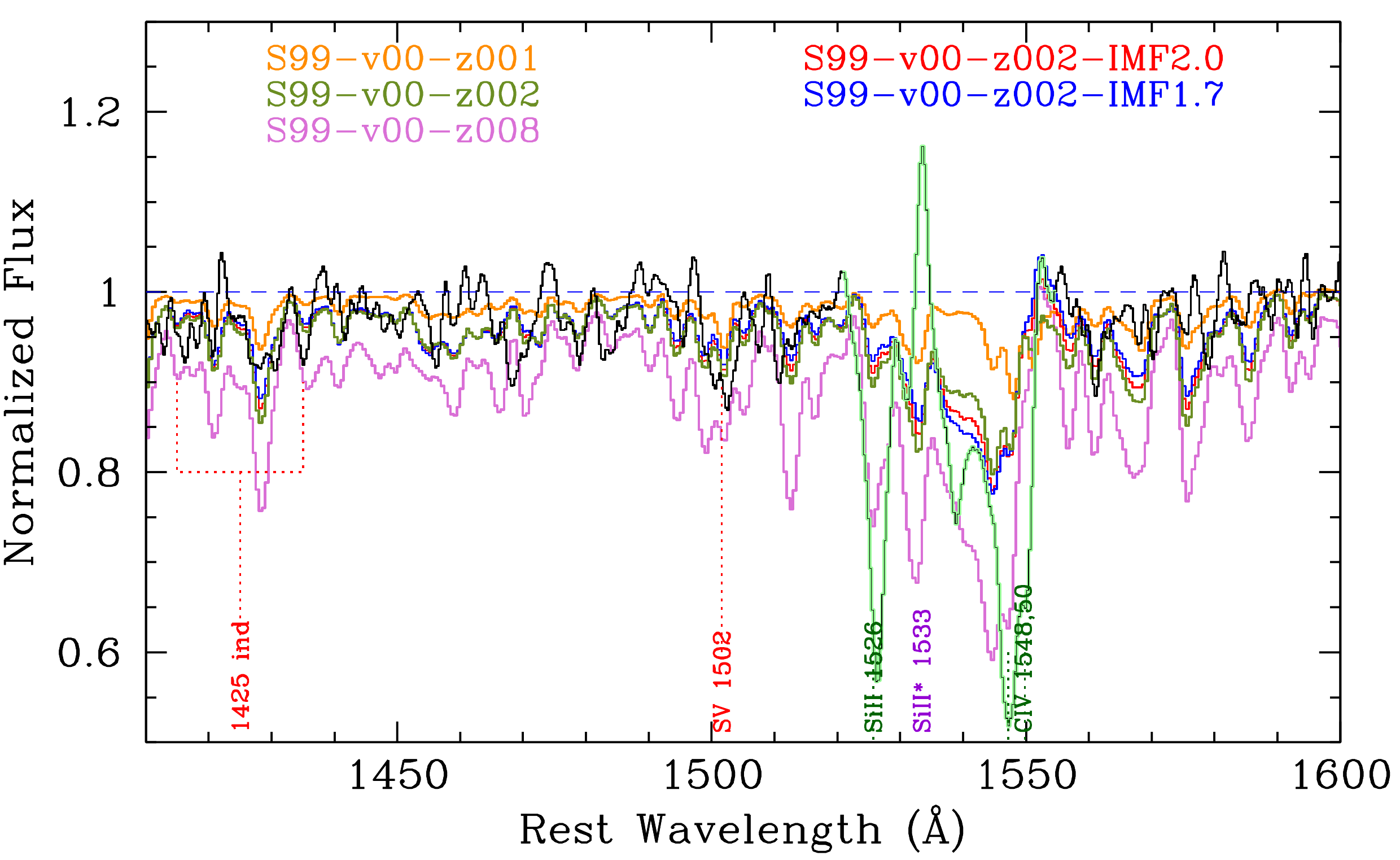}\includegraphics[width=8.5cm]{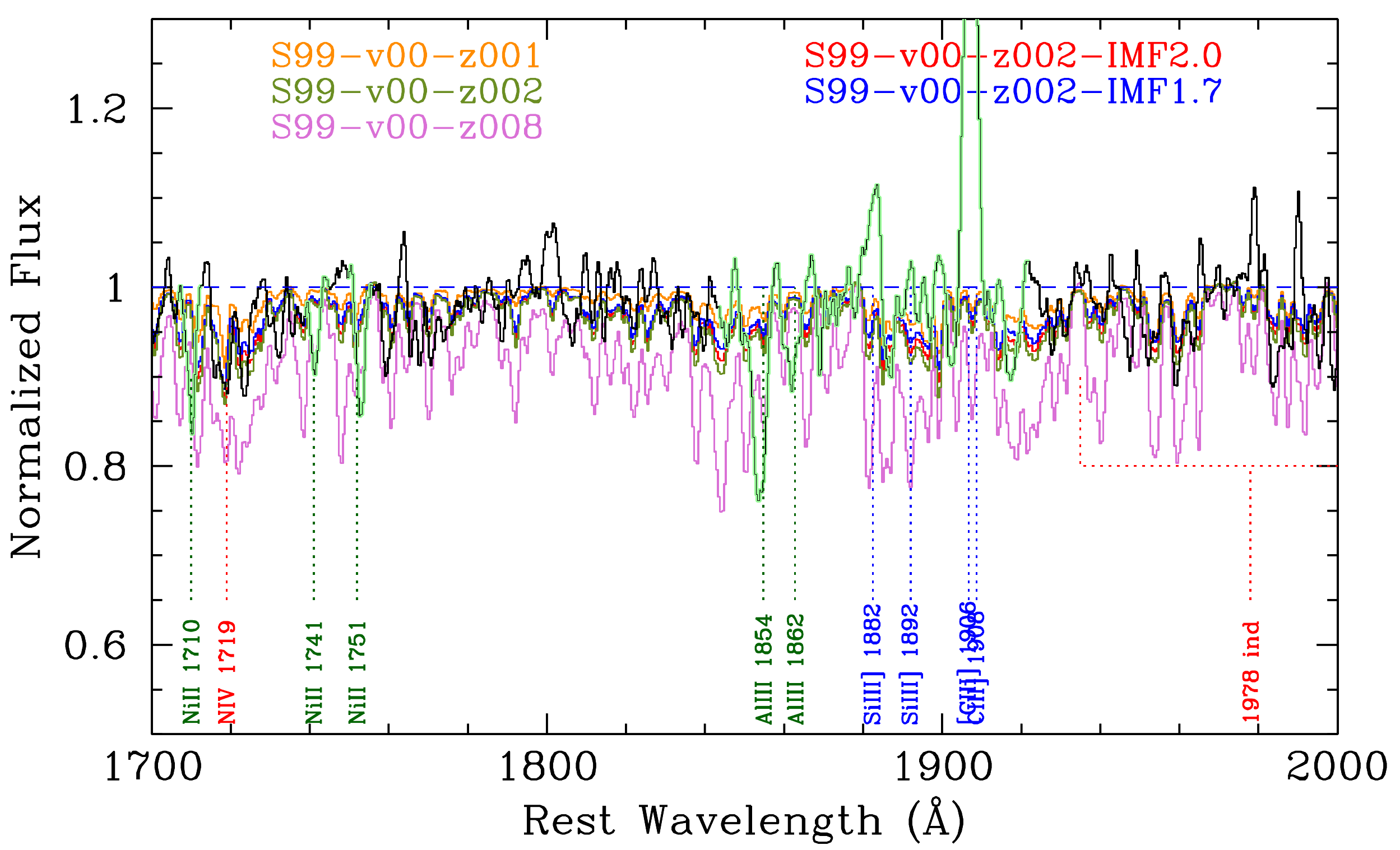}}
\centerline{\includegraphics[width=8.5cm]{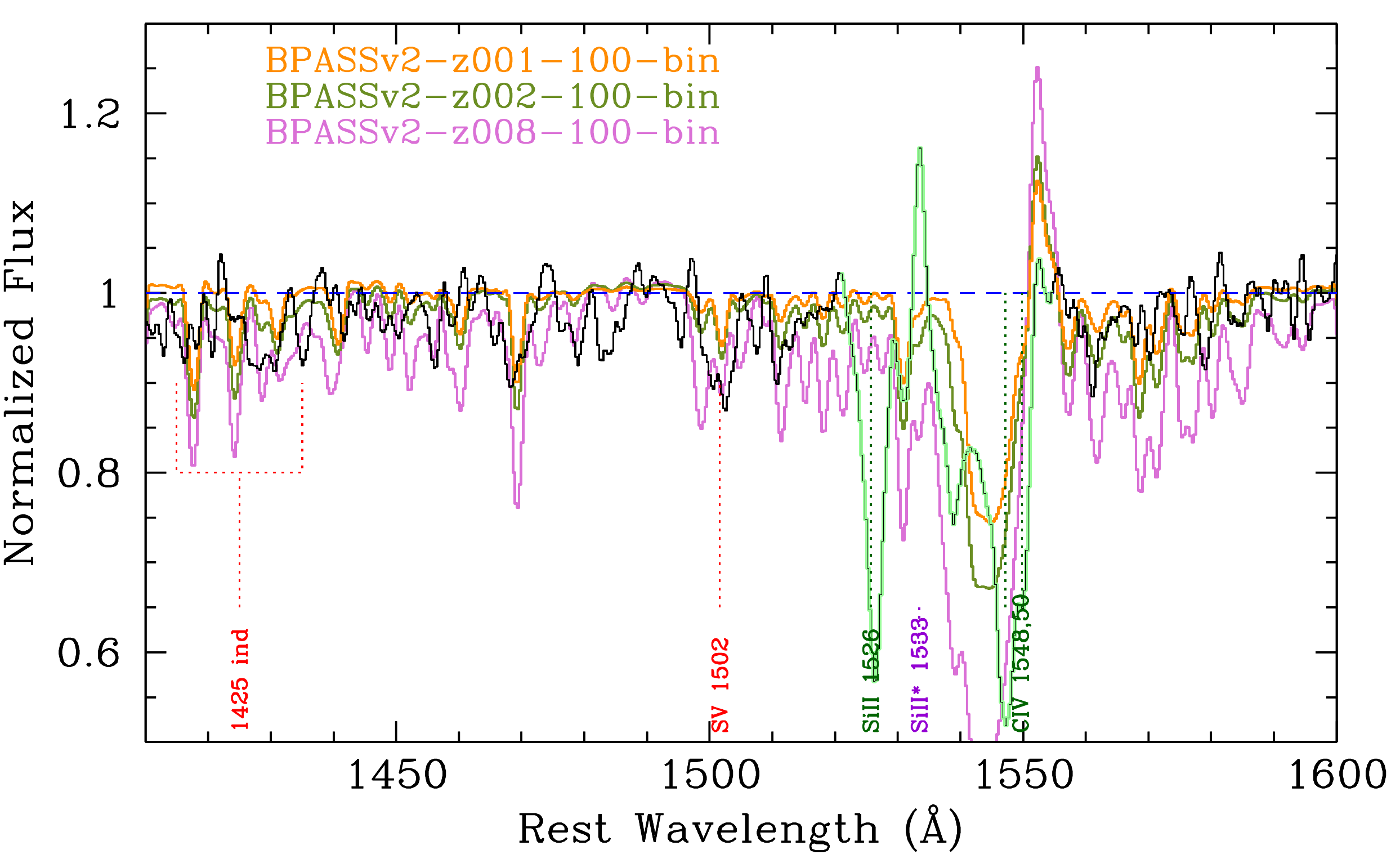}\includegraphics[width=8.5cm]{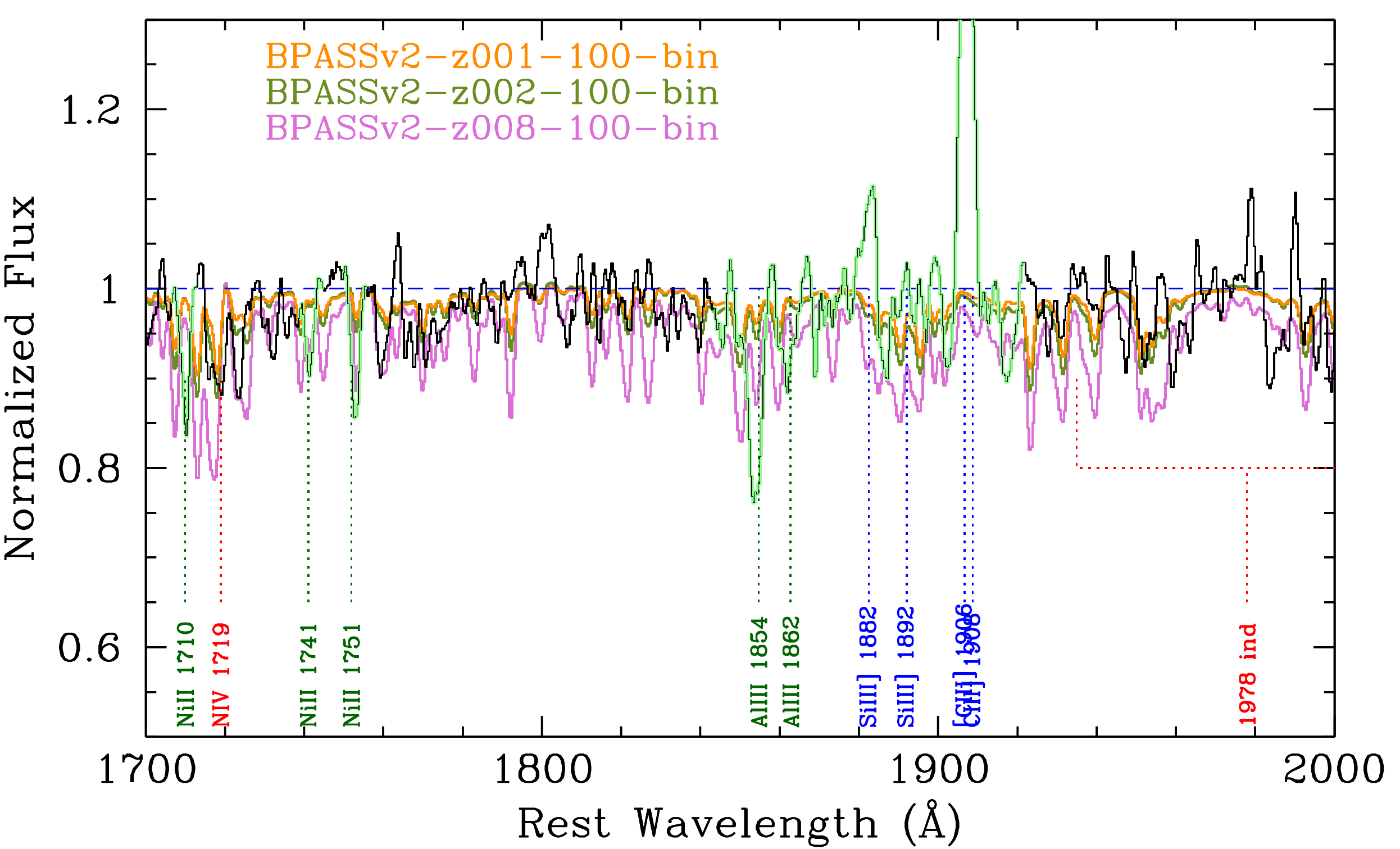}}
\caption{Comparison of the KBSS-LM1 composite spectrum (black histogram) with the predicted stellar spectra
at 3 values of stellar metallicity in the S99 family of models (top panels) and BPASSv2 (bottom panels): 
$Z_{\ast}=0.001$ ($Z_{\ast}/Z_{\odot} = 0.07$), $Z_{\ast}=0.002$ ($Z_{\ast}/Z_{\odot} = 0.14$), and $Z_{\ast}=0.008$ 
($Z_{\ast}/Z_{\odot}=0.56$).
Selected stellar absorption features (including the 1425 and 1978 photospheric absorption 
indices suggested by \citet{rix04} as stellar metallicity indicators) are labeled in red. 
Spectral pixels excluded from the Mask 2 $\chi^2$ minimization are shaded light green, and individual feature
identifications are as in Figure~\ref{fig:all_deep_full}. }  
\label{fig:photos_abs}
\end{figure*}

To facilitate such comparisons, we normalized the observed and model spectra by
the unabsorbed stellar spectrum (i.e., the stellar continuum, in the absence of photospheric absorption
lines). In the case of the S99 models, the WM-Basic high resolution UV spectra include such a normalized
version. For the BPASSv2 models and the observed LM1 composite spectrum, we used the 
continuum shape of the best-fitting S99 model as a first guess for the continuum; the initial continuum
fit was then modified interactively by adding spline points constraining the fit in wavelength regions
deemed relatively free of photospheric absorption (including the regions recommended for this purpose by \citealt{rix04}). 
For the observed spectrum, this procedure was repeated until upward fluctuations relative to the continuum fit
(excluding regions containing nebular, stellar, or excited fine structure emission lines) had a distribution
consistent with those expected from the 1$\sigma$ error spectrum in the same spectral region\footnote{Downward fluctuations 
include both shot noise and the continuum blanketing by stellar absorption features, and so cannot be used for
evaluating the continuum placement.}.  In practice, since the reddened model spectra described in the previous section resulted in
excellent fits to the shape of the observed spectrum (e.g., Figure~\ref{fig:all_deep_full_wmods}), only small adjustments to the theoretical unabsorbed stellar
continuum of the S99 models were needed to fit the continua of the LM1 (observed) and BPASSv2 model spectra.  
Figures~\ref{fig:civ_zoom}~and~\ref{fig:photos_abs} show zoom-in comparisons of continuum-normalized 
data and models in selected spectral regions.  

\subsection{Stellar Wind Features}
\label{sec:stellar_winds}

Both sets of model spectra shown in Figure~\ref{fig:civ_zoom} do reasonably well reproducing the \ion{N}{5}~$\lambda 1240$ wind feature,
though there are systematic uncertainties in the data in this region because the models include no interstellar Lyman-$\alpha$
absorption, and the IGM/CGM correction to the observed spectrum is not expected to fully capture this contribution.\footnote{The IGM/CGM
correction is based on the statistics of \nhi\ for galactocentric distances $d > 50$ kpc (see \citealt{rudie13}).} The \ion{N}{5} 
feature does not appear to be particularly sensitive to stellar metallicity in the BPASSv2 models, as the 3 metallicities shown in Fig.~\ref{fig:civ_zoom}  
are nearly indistinguishable. However, for the S99 models, there is a clear preference for the S99-v00-z002 model, which offers a nearly-perfect
match to the observed \ion{N}{5} profile, in both emission and absorption. The effect of varying the stellar IMF in the context of the S99
models with fixed $Z_{\ast}=0.002$ is shown in the upper-lefthand panel of Figure~\ref{fig:civ_zoom}, where the flatter IMF models
produce \ion{N}{5} profiles that are slightly too strong in both emission and absorption.  

The \ion{C}{4} stellar wind feature has been discussed by many authors as a potential diagnostic because of its
strength and its dependence on stellar metallicity 
(e.g., \citealt{pettini00,leitherer01,shapley03,rix04,steidel04,quider09,quider10,leitherer10}).
The only disadvantages are the substantial contribution of interstellar \ion{C}{4} 
absorption, and the presence of narrow \ion{Si}{2}$^{\ast}\lambda 1533.4$ emission, both of which 
can be difficult to separate from the stellar wind P-Cygni profile at low spectral
resolution and/or S/N (see \citealt{crowther06}.) Nevertheless, the LM1 composite spectrum is of high enough S/N and spectral
resolution that the broad P-Cygni absorption profile 
of the \ion{C}{4} feature is easily distinguishable in spite of the superposition of these narrower spectral features, as
shown in the righthand panels of Figure~\ref{fig:civ_zoom}.  
The top-right panel of Fig.~\ref{fig:civ_zoom} shows a clear preference among the S99 models for metallicity $Z_{\ast}=0.002$, and
a slight preference for a stellar IMF with slope $> -2.3$.  
The BPASSv2 models are less able to match the details of the \ion{C}{4} absorption profile, and only the lowest-metallicity
models (BPASSv2-z001) are consistent with the observed depth of the P-Cygni absorption in the LM1 composite. The same BPASSv2-z001
models appear to over-produce the emission component of the observed spectrum, but it is important to note that 
the apparent emission component of the P-Cygni profile is
dependent on the kinematics, optical depth, and covering fraction of gas giving rise to 
the interstellar \ion{C}{4} {\it absorption} in the vicinity
of the galaxy systemic redshift; such absorption would tend to diminish the apparent strength of the P-Cygni
emission component relative to its intrinsic strength.\footnote{A similar ambiguity is much less important for the \ion{N}{5} feature 
because there is little or no contribution to the observed line profile from interstellar \ion{N}{5} absorption.}      

\begin{figure}[htbp]
\centerline{\includegraphics[width=8.5cm]{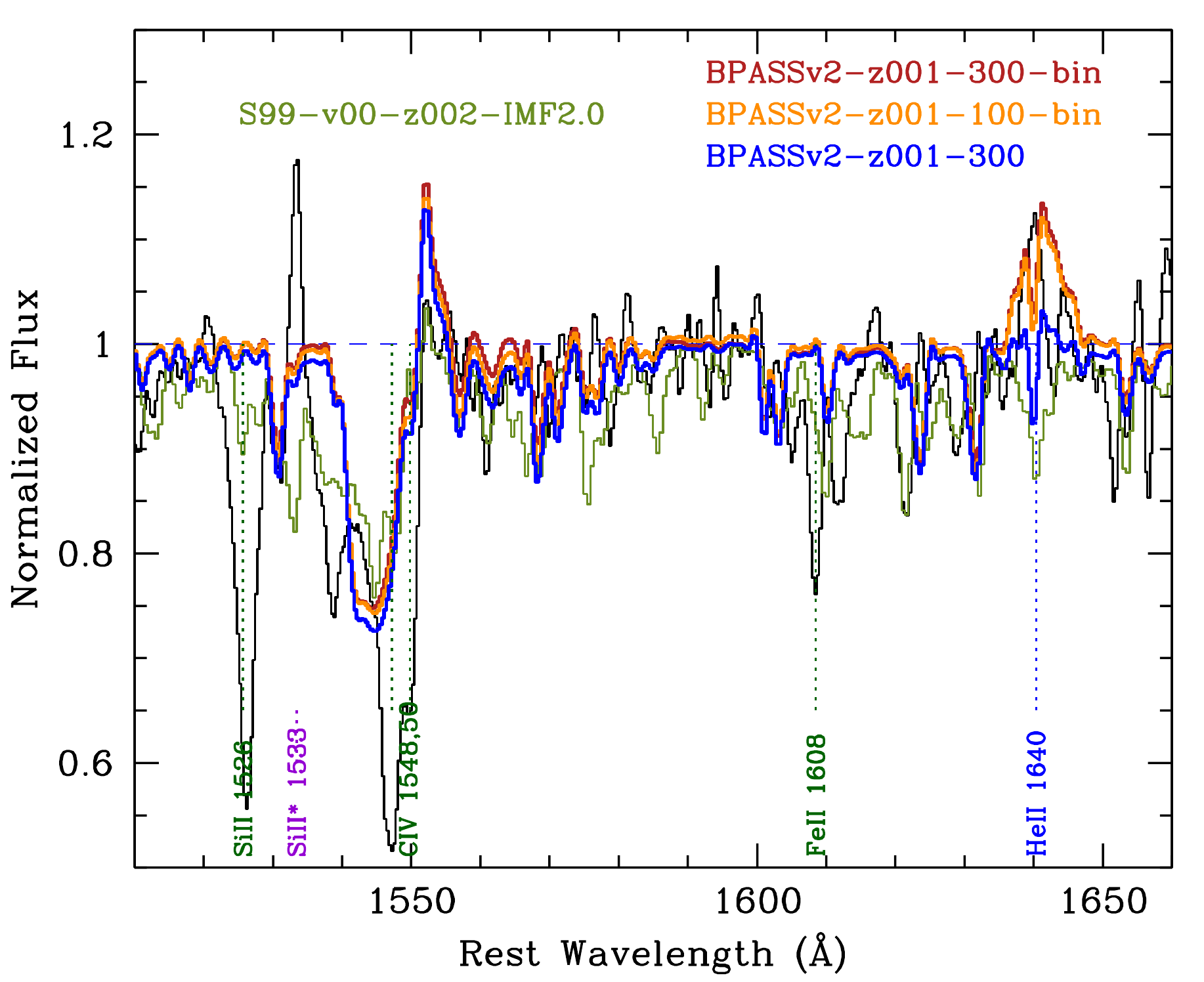}}
\caption{Illustration of the differences in the predictions for stellar \ion{He}{2}$\lambda 1640$ emission for models which otherwise
all produce good fits to the far-UV spectra. Both BPASSv2 models that include binary evolution (orange and dark red) predict nearly identical
stellar \ion{C}{4} and \ion{He}{2} emission; the single star models (S99 with upper mass cutoff of 100 M$_{\odot}$ [green] and BPASSv2
with upper mass cutoff of 300 M$_{\odot}$ [blue]) predict no detectable stellar \ion{He}{2} emission.  }
\label{fig:he2_compare}
\end{figure}

Over the rest wavelength range $\lambda_0 \simeq 1000-2000$ \AA, the most significant difference 
between the S99 and BPASSv2 models including binary evolution is that the latter predict broad \ion{He}{2}$\lambda 1640$
stellar emission that is absent from any of the continuous star formation S99 models (compare the righthand
panels of Figure~\ref{fig:civ_zoom}). Note that this feature is also absent from the single-star BPASSv2-z001-300 model (see Figure~\ref{fig:he2_compare}). 
\ion{He}{2} emission
can be present as either a stellar feature, broadened by processes associated with winds from massive
stars, or as a nebular feature, from recombination in \ion{H}{2} regions; of course, it is possible that both mechanisms
may contribute in any single case (see, e.g., 
\citealt{erb2010}). As illustrated in Figure~\ref{fig:civ_zoom} (and discussed in \S\ref{sec:measurements} below) 
the observed \ion{He}{2} feature in the LM1 LRIS spectrum 
may be attributed almost entirely to stars if the BPASSv2 models are used for the stellar continuum, whereas
adopting the S99 models would imply that the line must be entirely nebular in origin. 
We will return to a discussion of the importance of this distinction in \S\ref{sec:measurements} and \S\ref{sec:HeII} below. 

\subsection{Stellar Photospheric Absorption}
\label{sec:photospheric}

We showed above that global $\chi^2$ minimization of the stellar population synthesis models relative to the LM1 composite 
far-UV (1000-2000 \AA) spectrum favors a stellar metallicity of $Z_{\ast} = 0.001-0.002$ ($Z_{\ast}/Z_{\odot} = 0.07-0.14$) for both
BPASSv2 and S99 models (see Table~\ref{tab:chi2}, Fig.~\ref{fig:all_deep_full_wmods}) using the Mask 1 wavelength intervals
(Table~\ref{tab:masks}), which includes all wavelength pixels that are unaffected by the strongest interstellar absorption and
nebular emission features. Mask 2 includes a subset of Mask 1, confined to a more restricted wavelength interval
1400~\AA~$\le \lambda_0 \le$~2000~\AA\ and which also excludes strong P-Cygni stellar wind features. The smaller wavelength range 
gives less weight to matching the overall continuum shape, in order to increase the weight given to the details of
photospheric line blanketing. We re-fit the same suite of models using Mask 2, again parametrizing the overall shape 
of the SED using E(B-V) in the context of the \citet{calzetti00} attenuation relation and determining the combination
of E(B-V) and overall normalization factor that minimizes $\chi^2$ as defined in equation~\ref{eqn:chi2}. Note that
E(B-V) was used only as a parametrization of spectral shape, and that it was not fixed at the same value that 
minimized $\chi^2$ over the wider wavelength baseline of Mask 1. The results of this exercise are presented in the bottom
half of Table~\ref{tab:chi2}. As expected, the results are similar to those obtained using Mask 1, with the exception that
the best fitting models using Mask 2 have $Z_{\ast}=0.001$ ($Z_{\ast}/Z_{\odot} =0.07$) for both S99 and BPASSv2 models; while
the $Z_{\ast}/Z_{\odot}=0.14$ models produce $\chi^2$ only marginally higher for each, there is a systematic trend among
all of the models that using Mask 2 moves the minimum $\chi^2$ toward slightly lower metallicity models.  

Figure~\ref{fig:photos_abs} compares the KBSS-LM1 spectrum with a subset of the models over wavelength regions included
on Mask 2 to illustrate the results summarized in Table~\ref{tab:chi2}, chosen to highlight regions with the strongest
variations in line blanketing as a function of assumed stellar metallicity $Z_{\ast}$.  
Figures~\ref{fig:civ_zoom}~and~\ref{fig:photos_abs} show that in some
spectral regions the lower-metallicity $Z_{\ast}=0.001$ models provide better matches to individual photospheric absorption
features. An example can be seen within the stellar 
absorption blend at 1415-1435 \AA\ (lefthand panels of Fig.~\ref{fig:photos_abs}), which 
according to \citet{leitherer01} is a blend of \ion{Si}{3}$\lambda 1417$, \ion{C}{3} $\lambda 1427$,
and \ion{Fe}{5} $\lambda 1430$. The deepest absorption in the S99 model spectra (probably associated with the \ion{Fe}{5} feature) 
is over-predicted relative to the observations even for the $Z_{\ast}=0.002$ model, whereas the same model is more consistent with
the observations elsewhere within the same complex. 

A number of authors have used the strength of the broad complex of stellar absorption features spanning 
the wavelength range $1935-2050$~\AA\ (see the righthand panels of Fig.~\ref{fig:photos_abs}) originally proposed as
a metallicity indicator by \citet{rix04}\footnote{The original calibration was subsequently updated by \citet{james14}}. 
The total equivalent width of the blend, which is believed to be dominated by many photospheric lines of \ion{Fe}{3} in the spectra of
early B-stars, was identified as one of the more promising indicators of stellar photospheric metallicity. 
However, as pointed out by \citet{rix04}, measurement of the equivalent width of the feature can be noisy due to 
its sensitivity to continuum placement for such a broad and relatively shallow feature. Figure~\ref{fig:photos_abs} 
shows that the so-called ``1978 index'' is certainly dependent on assumed stellar metallicity (for both the S99 and BPASSv2 
families of models), but there are additional spectral regions 
regions that appear to have similar metallicity sensitivity: e.g.,  $1560-1590$~\AA\ (lefthand
panels of Fig.~\ref{fig:photos_abs}), and $1770-1850$ \AA\ (righthand panels of Fig.~\ref{fig:photos_abs}).    
Given the relatively high resolution of the KBSS-LM1 far-UV spectrum, and the sensitivity to stellar metallicity
evident in many spectral regions shown in Figs.~\ref{fig:civ_zoom}~and~\ref{fig:photos_abs}, we believe that 
a $\chi^2$ minimization similar to that described above makes optimal use of the available spectral information for
estimating photospheric abundances.  

In general, there are clear differences in detail between the photospheric 
absorption predicted by the S99 and BPASSv2 models, but the quantitative results in Table~\ref{tab:chi2} are borne out
by the visual impression of Figure~\ref{fig:photos_abs}: the data are most closely matched by stellar metallicity between $Z_{\ast}=0.001$
and $Z_{\ast}=0.002$ ($0.07 \simlt Z_{\ast}/Z_{\odot} \simlt 0.14$). Stellar metallicities as high as $Z_{\ast}=0.008$ ($Z_{\ast}/Z_{\odot}=0.56$)
are ruled out with a very high significance ($> 10\sigma$) for both families of models. Although stellar metallicity models with
$0.002 < Z_{\ast} < 0.008$ are not yet available within S99 using the most recent stellar tracks, the similarity of the results
for the values of $Z_{\ast}$ in common to the BPASSv2 and S99 models, and the quantitative trend of $\Delta\sigma$ vs. $Z_{\ast}$ 
for BPASSv2 models in the 
$Z_{\ast}=0.003-0.006$ range (Mask 2 in Table~\ref{tab:chi2}), strongly indicate that the most likely stellar metallicity of the
KBSS-LM1 composite is in the range $Z_{\ast}=0.001-0.002$, or $0.07 \simlt Z_{\ast}/Z_{\odot} \simlt 0.14$. 

\subsection{Stellar IMF Dependence} 
\label{sec:imf}

As mentioned in \S\ref{sec:stellar_winds} (see Fig.~\ref{fig:civ_zoom}), for the S99 models at fixed stellar metallicity of $Z_{\ast}=0.002$, the
\ion{N}{5} and \ion{C}{4} stellar wind profiles are modulated by changing the slope of the stellar IMF, with the stronger
dependence exhibited by the \ion{N}{5} feature. The Mask 2 results section of Table~\ref{tab:chi2} shows that varying the stellar
IMF slope over the range between $-2.30$ and $-1.70$ has a barely perceptible effect on the best-fitting stellar photospheric metallicity, though
there is a trend toward slightly increasing the best-matching metallicity as the IMF slope increases (i.e., as it becomes flatter). 
Taken together, in the context of S99 models and the non-ionizing far-UV spectrum, neither the stellar wind lines nor the spectral regions
dominated by photospheric stellar absorption lines offer strong evidence in favor of flatter IMF models over the default slope ($-2.3$) models.  

The upper mass cutoff of the IMF can have a strong effect on the net EUV spectrum of massive stars. The spectra predicted
by models with a high IMF cutoff mass look similar in many ways to those that incorporate massive binaries; the most
dramatic way in which they differ is in the predicted stellar \ion{He}{2}$\lambda 1640$ emission. 
Within the first
few Myr of an instantaneous burst, when the most massive stars are evolving off the main sequence to become Wolf-Rayet
stars,  
the hardness of the EUV spectrum and the strength of stellar \ion{He}{2} reach a maximum relative to the FUV continuum.  
While observation of substantial
stellar \ion{He}{2} emission in the integrated spectrum of massive star clusters {\it can} be explained by the 
presence of evolved stars with $M>100$ M$_{\odot}$ (e.g., \citealt{crowther16}), such a short-lived phenomenon cannot 
explain \ion{He}{2} emission observed in the integrated FUV spectrum of entire galaxies (e.g., \citealt{shapley03,eldridge12}).  
As discussed in \S\ref{sec:intro}, binary evolution of massive stars can produce stars with effective temperatures similar
to those of the most massive Wolf-Rayet stars from much lower initial stellar masses even in low-metallicity environments. 

Indeed, Figure~\ref{fig:he2_compare} shows that the BPASSv2-z001 {\it binary} model with upper mass cutoff 
of 100 M$_{\odot}$
predicts much stronger stellar \ion{He}{2}$\lambda 1640$ emission than the BPASSv2-z001 {\it single star} model with the same IMF slope and
an upper mass cutoff of
300 M$_{\odot}$.\footnote{We also show in \S\ref{sec:cloudy_models} (see especially \S\ref{sec:R23} - \S\ref{sec:euv_shape}) that the EUV continuum spectrum of the M$_{\rm up}=100$ M$_{\odot}$ binary
model is harder than that of the M$_{\rm up}=300$ M$_{\odot}$ single-star model.}  
In fact, we will show in \S\ref{sec:cloudy_models} that, among the models providing a reasonable fit to the observed far-UV spectrum, only
the BPASSv2 {\it binary} models have a sufficiently hard spectral shape in the ionizing EUV to successfully predict all of the 
observed nebular line intensity ratios, {\it and} predict the
observed stellar \ion{He}{2} emission for continuous star formation histories.

\section{Emission Lines in the Composite Spectra}
\label{sec:measurements}

Because of the way in which the rest-optical and rest-UV composite spectra have been constructed (\S\ref{sec:observations}), they are
on the same flux density scale, in the sense that all of the composites have been cross-calibrated so that they can be used
together across the full range of 
observed wavelengths (3400-24000 \AA, or $\sim 1000-7000$ \AA\ in the rest-frame). 
We checked the continuum fluxes (well-detected
for the ensemble in all 3 near-IR bands as well as in the LRIS spectrum) against broad-band optical and near-IR
photometry in hand for the full ensemble to verify that the flux density scale is consistent with that expected from
photometric measurements across all wavelengths. This procedure showed that any residual systematic errors in flux scale between
the LRIS spectra and the MOSFIRE J, H, and K-band composites are $< 2-3$\% (rms).

\begin{deluxetable}{lccc}
\tabletypesize{\scriptsize}
\tablewidth{0pc}
\tablecaption{Nebular Emission Line Measurements\tablenotemark{a}}
\tablehead{
\colhead{Ion} & \colhead{$\lambda_0$ (\AA)\tablenotemark{b}} & \colhead{F($\lambda$)\tablenotemark{c}} & \colhead{I($\lambda$)\tablenotemark{d}}}  
\startdata
\cutinhead{LRIS-B+R} 
Lyman-$\alpha$ & 1215.67 & $2.591\pm0.019$  & $6.064\pm0.044$\\ 
\ion{He}{2} & 1640.42  & $0.096\pm0.009$\tablenotemark{e} & $0.100\pm0.010$\tablenotemark{e} \\
 & 1640.42  & $0.075\pm0.009$\tablenotemark{f} & $0.078\pm0.010$\tablenotemark{f} \\
 & 1640.42  & $0.016\pm0.009$\tablenotemark{g} & $0.017\pm0.010$\tablenotemark{g}\\
${\rm OIII]}$  & 1660.81  & $0.019\pm0.006$ & $0.020\pm0.006$ \\
${\rm OIII]}$  & 1666.15  & $0.054\pm0.006$ & $0.056\pm0.007$ \\
${\rm [SiIII]}$  & 1882.47  & $0.072\pm0.015$ & $0.078\pm0.016$ \\
${\rm [SiIII]}$  & 1892.03  & $0.025\pm0.015$ & $0.028\pm0.016$ \\
${\rm [CIII]}$   & 1906.68  & $0.173\pm0.006$ & $0.190\pm0.007$ \\
${\rm CIII]}$    & 1908.73   & $0.119\pm0.006$ & $0.131\pm0.007$ \\
\cutinhead{MOSFIRE}
${\rm [OII]}$    & 3727.09  & $2.25\pm0.05$ & $1.36\pm0.03$ \\
${\rm [OII]}$    & 3729.88 & $2.38\pm0.06$ & $1.45\pm0.03 $ \\
${\rm [NeIII]}$  & 3869.81  & $0.66\pm0.05$ & $0.43\pm0.03 $ \\
${\rm [OIII]}$\tablenotemark{d}    & 4364.44  & $\le 0.12$   & $\le 0.06 $  \\
H$\beta$     & 4862.69     & $2.15\pm0.04$ & $1.00\pm0.02$  \\
${\rm [OIII]}$     & 4960.30  & $3.14\pm0.03$ & $1.42\pm0.02 $  \\
${\rm [OIII]}$    & 5008.24  & $9.39\pm0.04$ & $4.25\pm0.02 $  \\
H$\alpha$    & 6564.61     & $7.76\pm0.04$ & $2.89\pm0.02 $  \\ 
${\rm [NII]}$    & 6549.84  & $0.26 \pm 0.04$ & $0.10\pm0.02 $\\
${\rm [NII]}$    & 6585.23  & $0.75\pm0.04$ & $0.28\pm0.02 $  \\
${\rm [SII]}$    & 6718.32  & $0.82\pm0.05$ & $0.29\pm0.02 $ \\
${\rm [SII]}$    & 6732.71  & $0.66\pm0.04$ & $0.23\pm0.02 $ \\
\enddata
\tablenotetext{a}{Only emission lines whose observed wavelengths were included in the spectra
of all 30 galaxies have been measured.} 
\tablenotetext{b}{All wavelengths are in vacuum.}
\tablenotetext{c}{All fluxes have units of 10$^{-17}$ ergs s$^{-1}$ cm$^{-2}$.}
\tablenotetext{d}{Extinction corrected line intensities relative to $I(\Hb)$; extinction corrections assume nominal ratio $I(\Ha)/I(\Hb) = 2.89$, 
which implies ${\rm E(B-V)=0.21}$ with the Galactic extinction curve of \citet{cardelli89}.} 
\tablenotetext{e}{For Starburst99 continua; assumes that the feature is entirely nebular.}
\tablenotetext{f}{For BPASSv2-z001-300 single star model, after subtraction of the stellar continuum.}
\tablenotetext{g}{For BPASSv2 binary stellar models; after subtraction of the model, the residual emission is assumed to be the nebular component.}
\label{tab:lines}
\end{deluxetable}

\begin{figure}[htbp]
\centerline{\includegraphics[width=8.5cm]{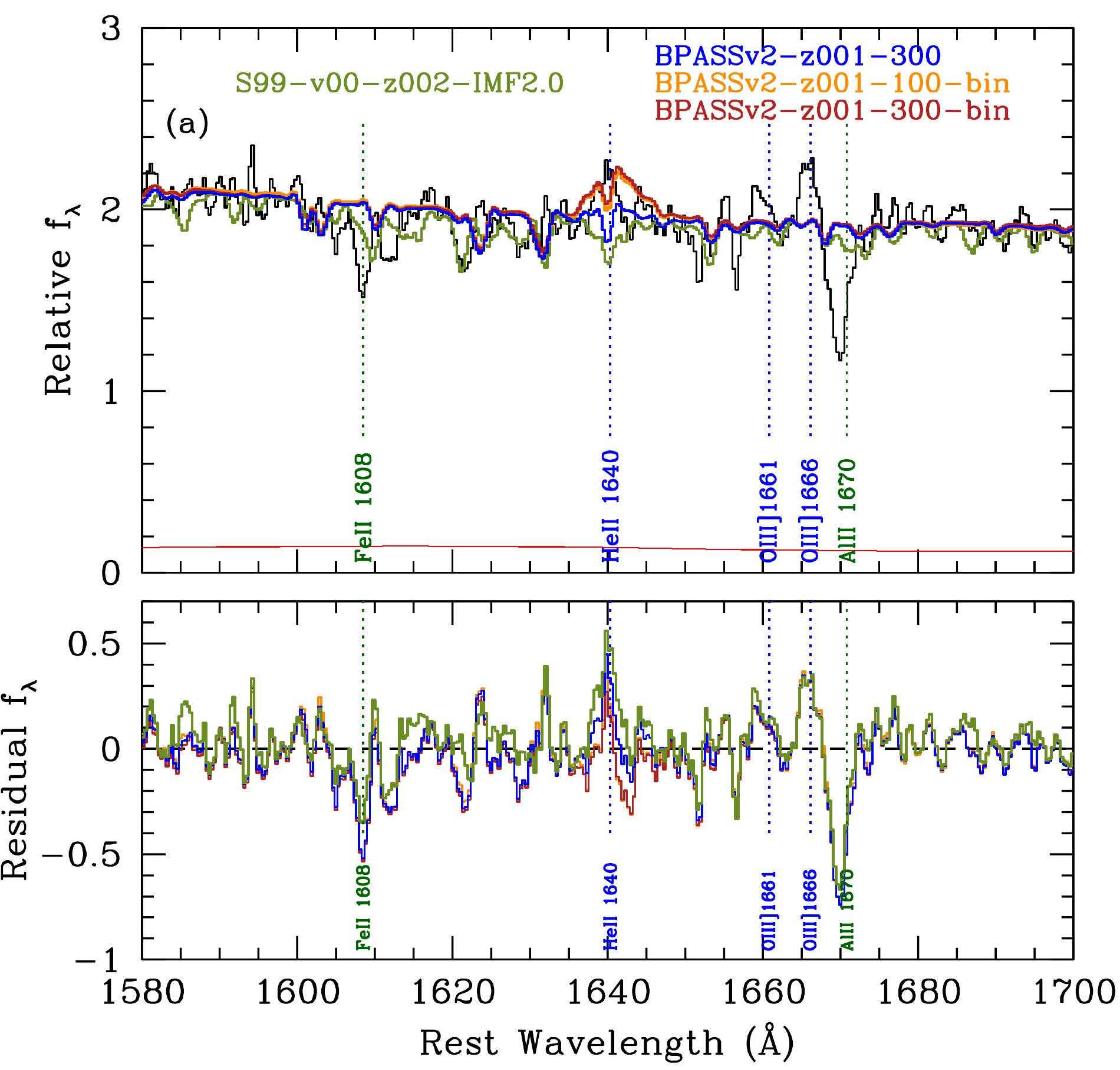}}
\centerline{\includegraphics[width=8.5cm]{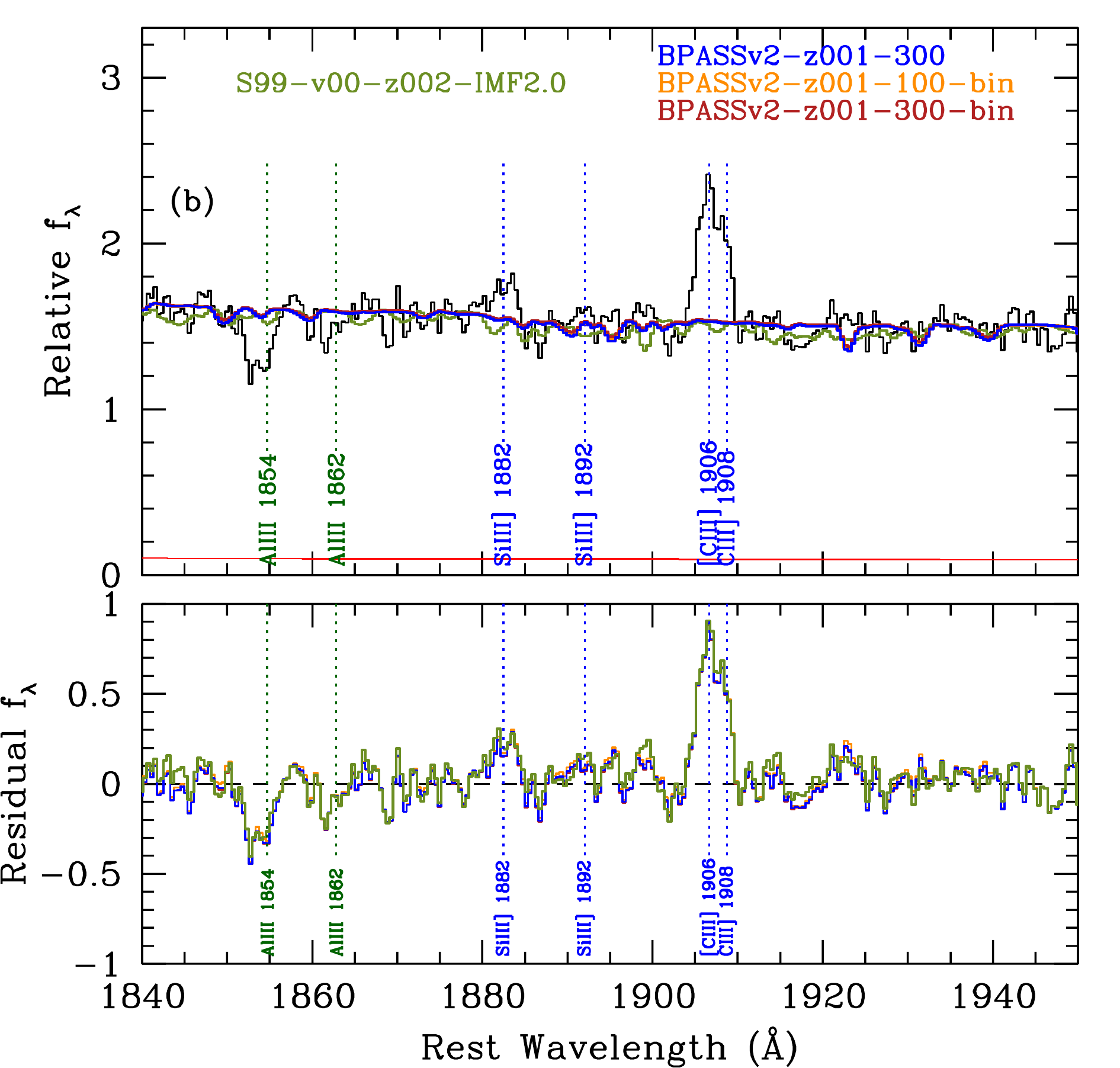}}
\caption{Portions of the LM1 far-UV composite near weak nebular emission lines. The top panel in each case shows
the observed spectrum (black histogram) together various stellar continuum models. The lower panel shows
the residual spectrum after subtraction of the stellar continua; the residuals are plotted in the same color as the models
in the upper panels. 
(a) The region near the \ion{He}{2}$\lambda 1640.42$ 
and \ion{O}{3}] $\lambda\lambda 1660.81$,1666.15 features. Note that for the BPASSv2 binary models most of the \ion{He}{2}
feature is consistent with being stellar, with weak residual emission possibly of nebular origin. For the S99 and BPASSv2 single star models, which
do not exhibit obvious stellar \ion{He}{2} emission, the residuals are much stronger (see Table~\ref{tab:lines}). (b) 
The spectral region near the [\ion{C}{3}]/\ion{C}{3}]$\lambda\lambda 1906.68$, 1908.73 and [\ion{Si}{3}]$\lambda\lambda 1882.47$, 1892.03
features, with color coding as in (a). In this case, all of the stellar models yield similar residual line profiles, with only small variations.  
} 
\label{fig:uv_emlines}
\end{figure}

\subsection{Rest-Frame Optical}

Emission line fluxes were measured directly from composite spectra shifted to the mean redshift of
the sample, $\langle z \rangle = 2.396$. For MOSFIRE spectra, the fluxes were measured using techniques described in detail by \citet{steidel14}, 
with the following exception: because emission lines in the stacked spectra are not necessarily well-described by a single 
Gaussian component (usually adequate to describe the lines in the spectrum of an individual galaxy), we used direct 
integration of the line fluxes after subtraction of a locally-normalized stellar continuum model (so that it includes
appropriate Balmer line absorption- see Figure~\ref{fig:mosfire_spectra}). The measured fluxes are compiled in Table~\ref{tab:lines}, together
with uncertainties estimated from a combination of the statistical errors and possible systematic errors due to ambiguity in the 
normalization of the local continuum; for the weakest lines, the continuum uncertainty accounts for $\sim 50$\% of the
estimated error bars.  

\subsection{Rest-Frame Far-UV}

For the rest-frame UV spectrum, observed emission line fluxes were measured using the {\it splot} task in IRAF, using both
Gaussian fits and direct integration of the line fluxes. Some of the rest-UV emission lines are statistically well-detected 
but are very weak relative to the
continuum, again causing the continuum determination to be one of the largest sources of uncertainty. 
In \S\ref{sec:stellar_pops}, we converged on a set of stellar population synthesis models providing the best fits to the 
observed LM1 far-UV continuum spectrum. These models provide important templates to be used in measuring the 
relatively weak nebular emission lines present in the rest-frame far-UV (Figure~\ref{fig:all_deep_full}).
As discussed in \S\ref{sec:PSMs} above, 
the far-UV stellar continuum contains a large number of stellar photospheric absorption lines, most of which are unresolved and thus produce 
variable line blanketing that depends on the detailed properties of the stellar population.
We used the best-fitting population synthesis models (described in sections~\ref{sec:stellar_pops}~and~\ref{sec:PSMs}) to model and subtract the 
stellar continuum near weak nebular emission lines; 
the uncertainties in Table~\ref{tab:lines} include both statistical uncertainties and
the dispersion in flux measurements after subtraction of the stellar continua produced by the acceptable range of models. 
Figure~\ref{fig:uv_emlines} shows zoomed-in regions near weak nebular emission lines for the KBSS-LM1 rest-UV spectrum, where
the bottom panels show the emission spectrum after subtraction
of the various stellar continuum models; as in previous figures showing portions of the rest-UV spectrum, the notable nebular
emission lines are indicated with blue labels.

The UV intercombination doublet OIII]$\lambda\lambda1661$,1666 is expected to have a fixed intensity ratio
of $\simeq 1:2.5$, but its apparent ratio may be affected by the nearby interstellar \ion{Al}{2}$\lambda 1670.79$ absorption feature; however,
even without fixing the ratio of the two lines, the measured intensities are consistent within the errors with
the expected ratio. As discussed in \S\ref{sec:oh}, this feature is important to the measurement of the nebular electron temperature ($T_{\rm e}$) for
for direct oxygen abundance determination.  

The detection of nebular \ion{He}{2}$~\lambda 1640$ is a potentially important constraint on the ionizing
sources in the galaxies; as discussed in section~\ref{sec:stellar_pops} above, 
this feature can be either nebular or stellar, with possible contribution from both (see, e.g., \citealt{erb2010}). 
It is notoriously difficult to produce measurable nebular \ion{He}{2} emission assuming photoionization by normal
main-sequence O-stars. Table~\ref{tab:lines} includes three measurements of the residual \ion{He}{2}$\lambda 1640$ after
subtraction of three different stellar models with large variations in the predicted strength of {\it stellar} \ion{He}{2}, and
thus very different predictions for the residual \ion{He}{2} emission (Figure~\ref{fig:uv_emlines}a). 
We discuss the implications of the observed \ion{He}{2} residuals in the context of photoionization
models in \S\ref{sec:HeII}. 

In the case of the [\ion{C}{3}]/\ion{C}{3}] forbidden/semi-forbidden line pair, which has a rest-frame separation of only 2.05\AA\ 
and is therefore only marginally resolved at the resolution of the LRIS composite ($R=1500$ corresponds to FWHM$\approx 1.25$ 
\AA\ in the rest frame at 1908 \AA), the lines were measured using a simultaneous fit of 2 Gaussian components constrained
to have a fixed separation and equal width, again after subtraction of several different continuum representations as illustrated in 
Fig.~\ref{fig:uv_emlines}b).

\begin{deluxetable}{lcccc}
\tabletypesize{\scriptsize}
\tablewidth{0pc}
\tablecaption{Excited Fine Structure Emission Lines\tablenotemark{a}}
\tablehead{
\colhead{Ion} & \colhead{$\lambda_0$} & \colhead{$\lambda_{\rm obs}$} & \colhead{$W_{\lambda}$} & \colhead{${\rm \Delta v}$\tablenotemark{a}} \\
\colhead{ }   & \colhead{(\AA)}       & \colhead{(\AA)}               & \colhead{(\AA)} & \colhead{(\kms)}}
\startdata 
SiII*  & 1197.39 &  1197.37 & \nodata &  $-~6.0$  \\
SiII*  & 1265.00 &  1265.06 & 0.23 & $+14.2$\\
SiII*  & 1309.28 &  1309.20 & 0.42 & $-18.3$\\
SiII*  & 1533.43 &  1533.45 & 0.43 & $ +~3.9$ \\ 
       &         &          &      & $\langle \Delta v \rangle = 1.3\pm12.1$ \kms 
\enddata
\tablenotetext{a}{Redshift of observed feature relative to $z_{\rm neb}$ defined by strong rest-optical nebular lines.} 
\label{tab:efs}
\end{deluxetable}

In addition to nebular emission lines from \ion{H}{2} regions, the far-UV also contains numerous emission features from
excited fine structure transitions (purple labels in Figures~\ref{fig:all_deep_full}, \ref{fig:civ_zoom},~and~\ref{fig:photos_abs}).
These features are listed separately in Table~\ref{tab:efs}, 
since they are not ``nebular'', but instead are 
produced by resonant scattering and re-emission of strong ground-state interstellar absorption features (rather than by collisional 
excitation or recombination.) We note that in the LM1 composite, the excited fine structure emission lines agree to within $\sim 12$ \kms\ (rms) with
the nebular redshift  
$z_{\rm neb}$ measured from strong {\it nebular} emission lines in the KBSS-LM1 rest-optical composite spectra. 

\subsection{Nebular Extinction}

\label{sec:nebular_ext}

Since nebular emission lines can be measured accurately from the composite spectra over a relatively 
wide wavelength baseline, full use of the diagnostic power of the line ratios depends on 
accurate estimation of the wavelength-dependent nebular extinction due to dust. 
The standard means for achieving this involves measurement
of the ``Balmer decrement'', i.e., comparing the observed ratio $I(\Ha)/I(\Hb)$ to that expected
in the case of zero nebular reddening.  The expected intensity ratio is usually taken to be
that predicted by ``Case B'' recombination, in the low density limit and assuming electron
temperature $T_e = 10,000$ K, $I(\Ha)/I(\Hb) = 2.86$ (\citealt{osterbrock06}). 
However, for consistency with the analysis
in \S\ref{sec:cloudy_models} below, we adopted the intrinsic Balmer line ratio predicted by 
the photoionization models 
that best-reproduce the observed intensity ratios of lines that have little or no dependence on nebular extinction corrections; 
these models tend to predict slightly higher values of the intrinsic Balmer decrement, 
$I(\Ha)/I(\Hb) \simeq 2.89$. 
Adopting this intrinsic ratio and the \citet{cardelli89} Galactic extinction curve, 
the observed $I(\Ha)/I(\Hb) = 3.61\pm0.07$ maps to $E(B-V)_{\rm neb}  = 0.21\pm0.02$. The 4th 
column of Table~\ref{tab:lines} shows the resulting extinction-corrected line strengths relative to
\Hb. The implications of the observed line intensities and their ratios are discussed in section~\ref{sec:inferences}. 

We showed in \S\ref{sec:stellar_pops} above that the global fits to the FUV LM1 composite yielded 
{\it continuum} reddening $E(B-V)_{\rm cont} = 0.190-0.197$ for the best fitting population synthesis models (cf. Table~\ref{tab:chi2}
and Figure~\ref{fig:all_deep_full_wmods}). 
There is a rapidly developing literature comparing dust attenuation factors estimated
from nebular line ratios to those estimated from continuum reddening (e.g., \citealt{calzetti00,kashino13,price14,reddy15,shivaei16}). 
Since we are primarily concerned with correcting nebular emission line fluxes, this comparison is not essential, but
because of the high quality of the composite spectra in both the rest-UV and rest-optical, it may be of some interest. 
Taken at face value, the continuum reddening inferred from the far-UV spectrum
is slightly smaller than, but consistent within the uncertainties, with the inferred nebular reddening 
$E(B-V)_{\rm neb} = 0.21\pm0.02$ (Table\ref{tab:properties}), bearing in mind that the values refer to
different assumed extinction/attenuation relations (\citealt{calzetti00} for the continuum and \citealt{cardelli89} for the nebulae). 
If we had used purely stellar continuum models (rather than stellar plus nebular continuum), the two values of $E(B-V)$ would have
been indistinguishable.  

The similarity in the inferred $E(B-V)_{\rm cont}$ and $E(B-V)_{\rm neb}$ supports the underlying assumption that the
nebular emission lines are responding to the same massive stars that dominate the far-UV continuum-- i.e., 
that there is not likely to be a significant contribution of very dust-obscured massive stars (with potentially different
stellar metallicity) contributing to nebular line luminosities but remaining obscured in the far-UV. 

\section{Inferences from the Nebular Measurements}

In this section, we describe several relatively model-independent measurements
that provide useful constraints on the range of parameters to be explored in the photoionization models detailed in \S\ref{sec:cloudy_models}. 

\label{sec:inferences}

\begin{deluxetable*}{lll}
\tabletypesize{\scriptsize}
\tablewidth{0pc}
\tablecaption{Measured and Inferred Properties of the KBSS-LM1 Composite Spectra}
\tablehead{
\colhead{Quantity} & \colhead{Value} & \colhead{Units/Notes} }
\startdata
\cutinhead{Extinction and Star Formation Rate (UV and Nebular) }
${\rm F(\Ha)/F(\Hb)}$ & $3.61\pm 0.07$ & observed \\ 
${\rm E(B-V)_{neb}}$ & $0.21\pm0.02$ & ~~a  \\ 
${\rm E(B-V)_{UV}}$ & $0.192\pm0.004$ &  ~~b  \\ 
${\rm SFR_{UV}}$  & $28.6 \pm 0.8$ &  ${\rm M_{\odot}}$ yr$^{-1}$ \\
${\rm SFR_{\Ha}}$  & $26.8 \pm 1.2$  &  ${\rm M_{\odot}}$ yr$^{-1}$ \\
${\rm log~(sSFR)}$  & $0.54 \pm  0.09$& Gyr$^{-1}$  \\
\cutinhead{Electron Density-Sensitive Line Ratios}
${\rm [CIII]~1906/[CIII~1909}$ & $~1.469\pm0.028$ & ${\rm n_e=1470\pm 660~cm^{-3}}$  \\ 
${\rm [OII]~3727/3729}$ & $~0.945\pm0.033$ &  ${\rm n_e=~360\pm 45~cm^{-3}}$\\
${\rm [SII]~6718/6732}$ & $~1.242\pm0.090$ & ${\rm n_e=~160 \pm 90~cm^{-3}}$ \\
\cutinhead{Ionization-Sensitive Line Ratios}
${\rm O3}$    & $~~0.629\pm0.008$ & ${\rm log([OIII]5008/\Hb)}$ \\
${\rm O3_{tot}}$ & $~~0.753\pm0.008$ & ${\rm log([OIII]4960+5008/\Hb)}$ \\
${\rm O2}$    & $~~0.449\pm0.007$ & ${\rm log([OII]3727+3729/\Hb)}$ \\
${\rm O32}$   & $~~0.303\pm0.007$ & ${\rm O3_{tot} - O2}$ \\
${\rm Ne3O2}$ & $-0.812\pm0.027$ & ${\rm log([NeIII]3870/[OII]3727+3729)}$ \\ 
\cutinhead{Abundance-Sensitive Line Ratios}
${\rm O3uv}$  &  $~0.019\pm0.003$ & ${\rm OIII](1661+1666)/[OIII]5008}$ \\
${\rm O3opt}$\tablenotemark{c} &  $~~< 0.014$ & ${\rm OIII]~4364/[OIII]~5008}$ \\
${\rm N2}$    &  $-1.02\pm0.02$ & ${\rm log([NII]6585/\Ha)}$ \\
${\rm S2}$    &  $-0.74\pm0.02$ & ${\rm log([SII](6718+6732)/\Ha)}$ \\
${\rm O3N2}$  &  $~~1.64\pm0.02$  & ${\rm O3 - N2}$ \\
${\rm N2O2}$  &  $-1.00 \pm 0.02$ & ${\rm log~([NII]6585/[OII](3727+3729)}$ \\ 
${\rm N2S2}$  &  $-0.28 \pm 0.03$ & ${\rm log~([NII]6585/[SII](6718+6732))}$ \\
${\rm R23}$   &  $~~0.93 \pm 0.01$ & ${\rm log~([OIII](4960+5008)+[OII](3727+3729)]/\Hb)}$ \\
${\rm C3O3}$  &  $~~0.62 \pm 0.07$   & ${\rm log~(CIII](1907+1909)/OIII](1661+1666))}$ \\  
${\rm Si3O3}$ &  $~~0.00 \pm 0.08$   & ${\rm log~([SiIII]1882/OIII](1661+1666))}$ \\
\cutinhead{Inferred Stellar Metallicity}
 Photospheric Absorption  & $~~Z_{\ast}\simeq 0.001-0.002$  &   $Z_{\ast}/Z_{\odot} \simeq 0.10$ \\
 \ion{C}{4}, \ion{N}{5} Wind   & $~~Z_{\ast}\simeq 0.002$   &   $Z_{\ast}/Z_{\odot} \simeq 0.14$ \\
\cutinhead{Nebular Oxygen Abundances (Direct and Modeled)}
${\rm 12+log(O/H)_{CEL}}$ & $~~8.14\pm0.04$ & $Z_{\rm neb}/Z_{\odot}=0.29\pm0.03$;~~ ${\rm T_e([OIII]) = 12250\pm600~K}\tablenotemark{d}$ \\
${\rm 12+log(O/H)_{REL}}$ & $~~8.38\pm0.05$ & $Z_{\rm neb}/Z_{\odot}=0.49\pm0.05$;~~ Eq.~\ref{eqn:offset}\tablenotemark{e} \\
${\rm 12+log(O/H)_{mod}}$ & $~~8.39\pm0.09$ & $Z_{\rm neb}/Z_{\odot}=0.50\pm0.10$;~~ from photoionization modeling (this work) \\ 
\cutinhead{Nebular Oxygen Abundances (Strong Line Methods)}
${\rm 12+log(O/H)_{R23}}$ & $~~8.20\pm0.03$ & $Z_{\rm neb}/Z_{\odot}=0.32\pm0.03$;~~ lower branch \citet{mcgaugh91} \\
${\rm 12+log(O/H)_{O3N2}}$ & $~~8.23\pm0.02$ & $Z_{\rm neb}/Z_{\odot}=0.35\pm0.02$;~~ Eq.~\ref{eqn:o3n2} \\
${\rm 12+log(O/H)_{N2}}$ & $~~8.29\pm0.02$ & $Z_{\rm neb}/Z_{\odot}=0.40\pm0.02$;~~ Eq.~\ref{eqn:n2} \\
${\rm 12+log(O/H)_{N2O2}}$ & $~~8.23\pm0.02$ & $Z_{\rm neb}/Z_{\odot}=0.35\pm0.02$;~~ Eq.~\ref{eqn:n2o2_met} \\
${\rm 12+log(O/H)_{N/O}}$ & $~~8.13\pm0.02$ & $Z_{\rm neb}/Z_{\odot}=0.28\pm0.02$;~~ ``N/O analogs''; \S\ref{sec:n_o_analogs} \\ 
\cutinhead{Inferred Gas-Phase Abundance Ratios}
${\rm log(N/O)}$ & $-1.24\pm0.04$ & ${\rm ~[N/O]~ = ~-0.38\pm0.04}$;~~Eqs.~1,~2, \S\ref{sec:comparison} \\
${\rm log(C/O)}$        & $-0.60\pm0.09$ & ${\rm ~[C/O]~ = ~-0.34\pm0.09}$;~~Eq.~\ref{eqn:c_over_o} \\
${\rm log(Si/O)}$       & $-1.81\pm0.08$ & ${\rm ~[Si/O]= ~-0.63\pm0.08}$;~~Eq.~\ref{eqn:si_over_o}  
\enddata
\tablenotetext{a}{From the Balmer decrement, assuming the \citet{cardelli89} extinction curve (see Table~\ref{tab:lines}).}
\tablenotetext{b}{From the UV continuum shape, assuming the \citet{calzetti00} attenuation relation. }
\tablenotetext{c}{[OIII]$\lambda 4364$ was observed in only $\sim 50$\% of the galaxies.}
\tablenotetext{d}{Total oxygen abundance calculated assuming that ${\rm T_e([OII]) = 0.7~T_e([OIII)+3000}$ K,
and that ${\rm (O/H) = (O^{++}/H^{+}) + (O^{+}/H^{+})}$.}
\tablenotetext{e}{Total oxygen abundance calculated assuming ${\rm log(O/H)_{REL}-log(O/H)_{CEL}=0.24\pm0.02}$} 
\label{tab:properties}
\end{deluxetable*}

\subsection{Electron Density}

\label{sec:ne}

Three sets of nebular emission line doublets detected in the KBSS-LM1 composites are among those used for estimating electron
density ($n_{\rm e}$) in \ion{H}{2} regions: [\ion{O}{2}]$\lambda\lambda 3727,3729$ and [\ion{S}{2}]$\lambda\lambda 6718,6732$ in the J and K band
MOSFIRE spectra (respectively), and [\ion{C}{3}]$\lambda 1907$/\ion{C}{3}]$\lambda 1909$ in the LRIS UV spectrum.  Conversion of a measured
intensity ratio into a constraint on $n_{\rm e}$ depends weakly on electron temperature $T_{\rm e}$, and strongly on the actual $n_{\rm e}$
compared to the critical densities of the two transitions, which in turn depend on knowledge of the correct rate coefficients 
for collisional excitation/de-excitation.  
We used the measured line ratios (Tables~\ref{tab:lines} and \ref{tab:properties}) to calculate the constraints
on $n_{\rm e}$ provided by each pair of density-sensitive lines. For [\ion{O}{2}] and [\ion{Si}{2}] we used the 
fitting formulae recently calculated by \cite{sanders16}, which incorporates all of the recent updates to the atomic data
for these species, and assumes $T_{\rm e} = 10^4$ K. In the case of the \ion{C}{3} calculation, we used  
a grid of Cloudy models (v13.02; \citealt{ferland13}) with assumed $T_{\rm e} \approx 10000$, with the BPASSv2-z001-100bin model used
as the input ionizing spectrum (see \S~\ref{sec:PSMs}). The results for all 3 measurements are summarized in Table~\ref{tab:properties}. 

Clearly, the values of $n_e$ obtained
from [OII] ($n_{\rm e} = 360\pm45$ cm$^{-3}$) and [SII] ($n_{\rm e} = 160\pm90$ cm$^{-3}$) are marginally consistent with one another, 
with the value of $n_e$ from [OII] being much better constrained 
because of the higher S/N of the line ratio measurement and the larger difference in critical density for the two transitions of
[OII] compared to those of [SII].  

The implied $n_{\rm e} =1470\pm660$ cm$^{-3}$ from the ratio of the \ion{C}{3} lines
is substantially higher than implied by [OII] or [SII]; however, 
the observed ratio  ($I(1907)/I(1909) = 1.469\pm0.028$; see Table~\ref{tab:properties}) is only $\sim 2\sigma$ from the
the theoretical low-density limit ($I(1907)/I(1909) = 1.529$),
and $\sim 1.5\sigma$ from the ratio expected if ${\rm n_{e}(CIII)\simeq 360}$ cm$^{-3}$.  
While it is possible that the
physical conditions in the \ion{H}{2} regions producing the bulk of the [CIII] emission is different from those in which most of the 
[OII] is produced, the high critical density for both [\ion{C}{3}]$\lambda 1907$ 
($\simeq 7.4\times10^4$ cm$^{-3}$) and \ion{C}{3}]$\lambda 1909$ ($\simeq 9.7\times10^8$ cm$^{-3}$) make
their intensity ratio not particularly well-suited for measuring $n_{\rm e}$ of a few hundred
cm$^{-3}$. 


\subsection{N/O Abundance Ratio}
\label{sec:n_over_o}

The ratio of gas-phase nitrogen to oxygen (N/O) is well known to depend sensitively
on (O/H) in \ion{H}{2} regions and star-forming galaxies, but (as discussed by \citealt{steidel14}) 
the precise run of log(N/O) versus log(O/H) is a subject about which there is little 
consensus in the literature. It is straightforward to measure N/O 
because of the similar ionization potential of N and O, so that 
${\rm (N/O) \simeq (N^{+}/O^{+})}$;  differences in the shape or intensity of the
ionizing radiation field thus make only very small differences to the mapping of
the N2O2 line index (see definitions in Table~\ref{tab:properties}) to the N/O abundance ratio. 
The disagreements in the behavior of N/O versus O/H in the literature are dominated by
systematic differences in O/H among the methods used to estimate it.  

We can therefore use a calibration
sample drawn from the local universe for an initial estimate of N/O in the KBSS-LM1
composite.  For reasons discussed in more detail by Strom et al (2016), we use
a sample of 412 extragalactic \ion{H}{2} regions compiled by \citet{pilyugin12}, which 
spans approximately the same range in N2O2 as the KBSS-MOSFIRE sample and includes $T_{\rm e}$-based
abundances of N, O, and S.  
A linear fit
to the \citet{pilyugin12} dataset (see Strom et al 2016) yields
\begin{equation}
{\rm log(N/O) = 0.65*N2O2 - 0.57; ~\sigma = 0.05~dex }
\label{eqn:n2o2}
\end{equation}
where $\sigma = 0.05$ dex refers to the scatter of individual measurements relative to the fit. 

A similar fit using the ratio N2S2 (see Table~\ref{tab:properties}), sensitive to N/S and useful  
as a proxy for N/O (since S and O are believed to be produced by the same nucleosynthetic processes), is given by  
\begin{equation}
{\rm log(N/O) = 0.68*N2S2 - 1.08 ; ~\sigma=0.12~dex.}
\label{eqn:n2s2}
\end{equation}
As expected, the scatter is larger using N2S2 (see also \citealt{pmc09}), but for application to the KBSS-LM1
composite, equation~\ref{eqn:n2s2} has the advantage of 
being independent of cross-band calibrations (both [\ion{N}{2}] and [\ion{S}{2}] are measured in the K band for the KBSS-LM1 
sample) and nearly independent of reddening/extinction.  

Substituting the measured values of N2O2 and N2S2 (see Table~\ref{tab:properties}) into equations~\ref{eqn:n2o2}~and~\ref{eqn:n2s2}
results in values of log(N/O) in very good agreement with one another: (${\rm log(N/O)=-1.22\pm0.02}$ from N2O2 and
${\rm log(N/O)=-1.26\pm0.02}$ from N2S2), suggesting that the values are reasonable
and that our cross-band calibration and reddening corrections are accurate. Hereafter, we adopt the average 
${\rm log(N/O) =-1.24\pm0.03}$ as a fiducial value, but we re-assess the N/O abundance ratio for the KBSS-LM1
composite using a method entirely 
independent of the 
empirical N2O2$\rightarrow$N/O calibration (eqs.~\ref{eqn:n2o2}~and~\ref{eqn:n2s2}) in \S\ref{sec:n_over_o2} below. 

\section{Photoionization Models}

\label{sec:cloudy_models}

Since we are interested in identifying stellar population synthesis models that can simultaneously account
for both the far-UV OB-star continuum and the excitation of nebular emission in the same galaxies, 
we use the same population synthesis models as described in sections~\ref{sec:stellar_pops} and~\ref{sec:PSMs} as the source of the
ionizing radiation field in photoionization models. For this purpose, we used Cloudy (v13.02; \citealt{ferland13}); in
brief, the models assume a constant density $n_{\rm e} = 300$ cm$^{-3}$ (see section~\ref{sec:ne}) plane-parallel
geometry where the radiation field intensity is characterized by the dimensionless ionization parameter $U$,  
\begin{equation}
U \equiv \frac{n_{\gamma}}{n_H} \approx \frac{n_{\gamma}}{n_e} 
\label{eqn:gamma}
\end{equation}
where $n_{\rm H}$ is the number density of hydrogen atoms and $n_{\gamma}$ is the equivalent density of photons capable of ionizing hydrogen 
impinging on the face of the gas layer. 

We varied the nebular abundances from $Z_{\rm neb}/Z_{\odot} = 0.1$ to $Z_{\rm neb}/Z_{\odot} = 1.0$ in
steps of $0.1Z_{\odot}$, using the solar abundance set  
from \citet{asplund09} for all relevant elements. Dust grains were included using the ``Orion'' mix provided by Cloudy, assuming that the grain abundance
(and thus the dust-to-gas ratio) scales linearly with metallicity relative to solar. We allowed ${\rm log~{\it U}}$ to vary over the range ${\rm -3.5 \le log {\it U} \le -1.5}$    
in steps of ${\rm \Delta log {\it U} = 0.1}$
for each grid. 

As mentioned above, we have explicitly decoupled the {\it stellar metallicity} ($Z_{\ast}$) of the population
synthesis models from the ionized gas phase {\it nebular abundances} ($Z_{\rm neb}$), 
so that each population synthesis model was tested over the full range of assumed
gas-phase metallicity.  As output, the final predictions for the nebular spectrum at the edge of the fully-ionized
region were saved, along with the associated nebular continuum emission. As discussed in section~\ref{sec:stellar_pops}, 
the nebular continuum (generated self-consistently along with the line emission) was then added to the high-resolution, purely
stellar models for detailed comparison with the far-UV spectrum. Having generated 
many such grids of photoionization models (210 grid points in $U$ and $Z_{\rm neb}$ space for each of the population synthesis models described
in section~\ref{sec:stellar_pops}) we sought those that could most closely reproduce the observed emission line
intensity ratios. In what follows, we focus primarily on the models that produced the best fits to the global far-UV spectra
in section~\ref{sec:stellar_pops} (Table~\ref{tab:chi2}), and discuss inferences based on combining the observational constraints with the 
photoionization model results.   
As we show below, the stellar models capable of reproducing the observed nebular emission line indices are a subset of those
which provided reasonable matches to the FUV stellar spectra.

\begin{figure}[htbp]
\centerline{\includegraphics[width=8.5cm]{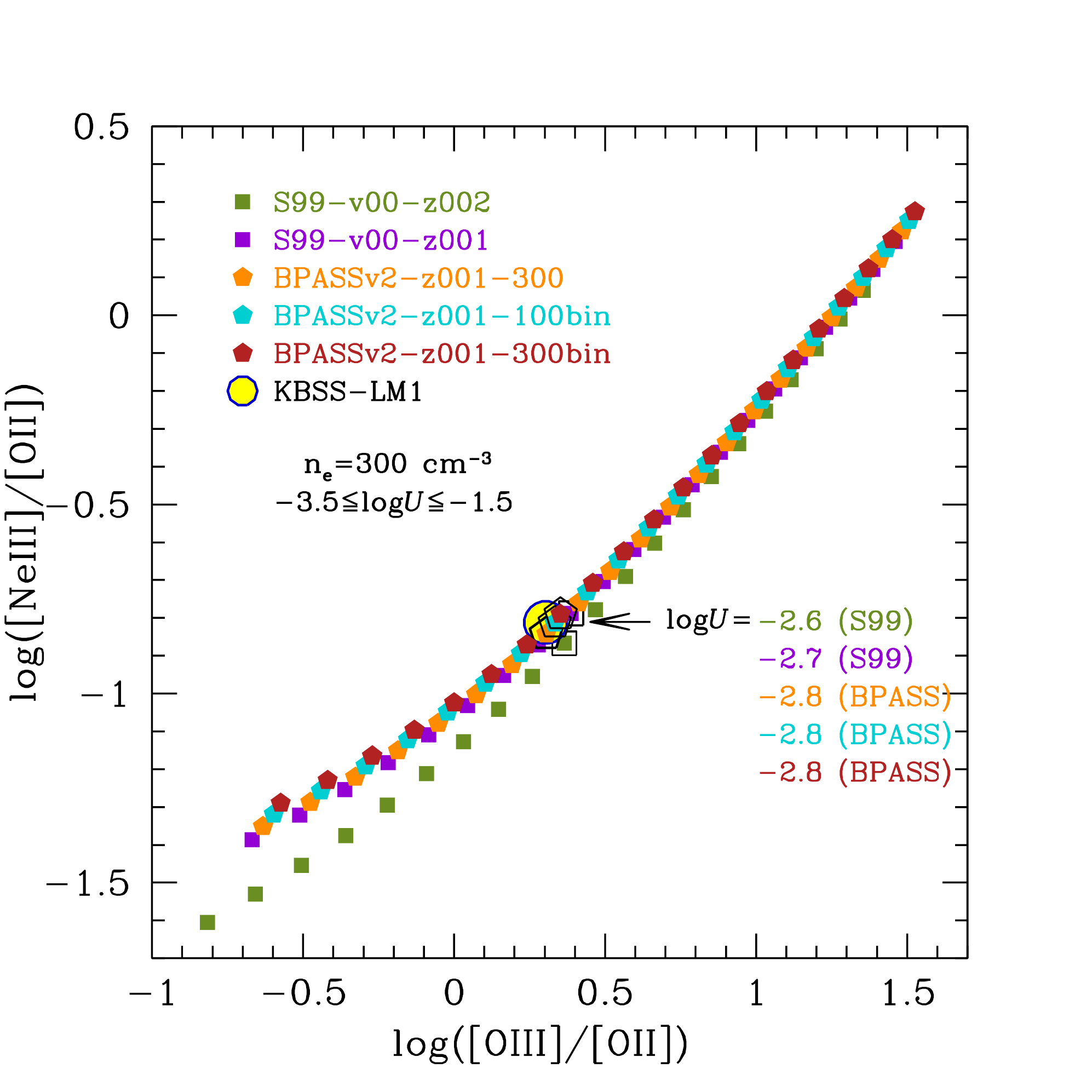}}
\caption{The predictions of photoionization models using the stellar population synthesis models that best-fit the observed
far-UV spectrum, for two nebular line ratios sensitive to ionization parameter $U$. The yellow point is 
measured from the KBSS-LM1 composite (Table~\ref{tab:properties}); its error bars are smaller than the size of the point. The models
shown assume $Z_{\rm neb}/Z_{\odot}=0.5$, but there is only very weak dependence on gas-phase metallicity of the line ratios shown. 
The value of ${\rm log~U}$ that most closely
matches the observed line indices (symbols enclosed by black skeletal lines) is listed for each model shown.  
 }
\label{fig:plot_gamma}
\end{figure}

\subsection{Ionization Parameter}

\label{sec:gamma}

The most useful line ratios for constraining the ionization parameter $U$ are those with different ionization potentials
but fixed elemental abundance ratios- the best such lines observed in the KBSS-LM1 composite spectra are ${\rm log \left( [OIII](4960+5008)/log [OII](3727+3729)\right)
\equiv O32}$ and ${\rm log\left([NeIII]3870/[OII](3727+3729)\right)\equiv Ne3O2}$ (see Table~\ref{tab:properties}). The advantage of the former is
that both lines derive from the same element; the advantage of
the latter is the proximity of the two features in wavelength -- both appearing in the J band over the redshift range of the LM1 sample--
meaning that 
uncertainties in relative slit losses are absent, and relative extinction corrections
are small. The abundance ratio (Ne/O) is also not expected to vary significantly with (O/H).  
A successful photoionization model 
should correctly predict the observed values of both O32 and Ne3O2 with a single value of ${\rm log {\it U}}$.  

Figure~\ref{fig:plot_gamma} compares the predictions of the photoionization models using the population synthesis models  
that provide the closest match to the observed far-UV spectrum as described in section~\ref{sec:stellar_pops}.
The observed point (labeled KBSS-LM1) can be reproduced by both S99 and BPASSv2 models, but the value of
${\rm log U}$ at which the models produce their closest match is model-dependent: 
higher values of ${\rm log {\it U}}$ (by $\simeq 0.1-0.2$ dex) are needed for the S99 models, as indicated in Fig.~\ref{fig:plot_gamma}.  
We find that the predicted locus of 
O32 versus Ne3O2 for a given population synthesis model is insensitive to assumed gas-phase
metallicity; Figure~\ref{fig:plot_gamma} shows grid points for 
$Z_{\rm neb}/Z_{\odot}= 0.5$ (12+log(O/H)$=8.39$) for illustration purposes and for consistency with the discussion in \S\ref{sec:bpt} 
and \S\ref{sec:R23}. 

\begin{figure*}[htbp]
\centerline{\includegraphics[width=8.5cm]{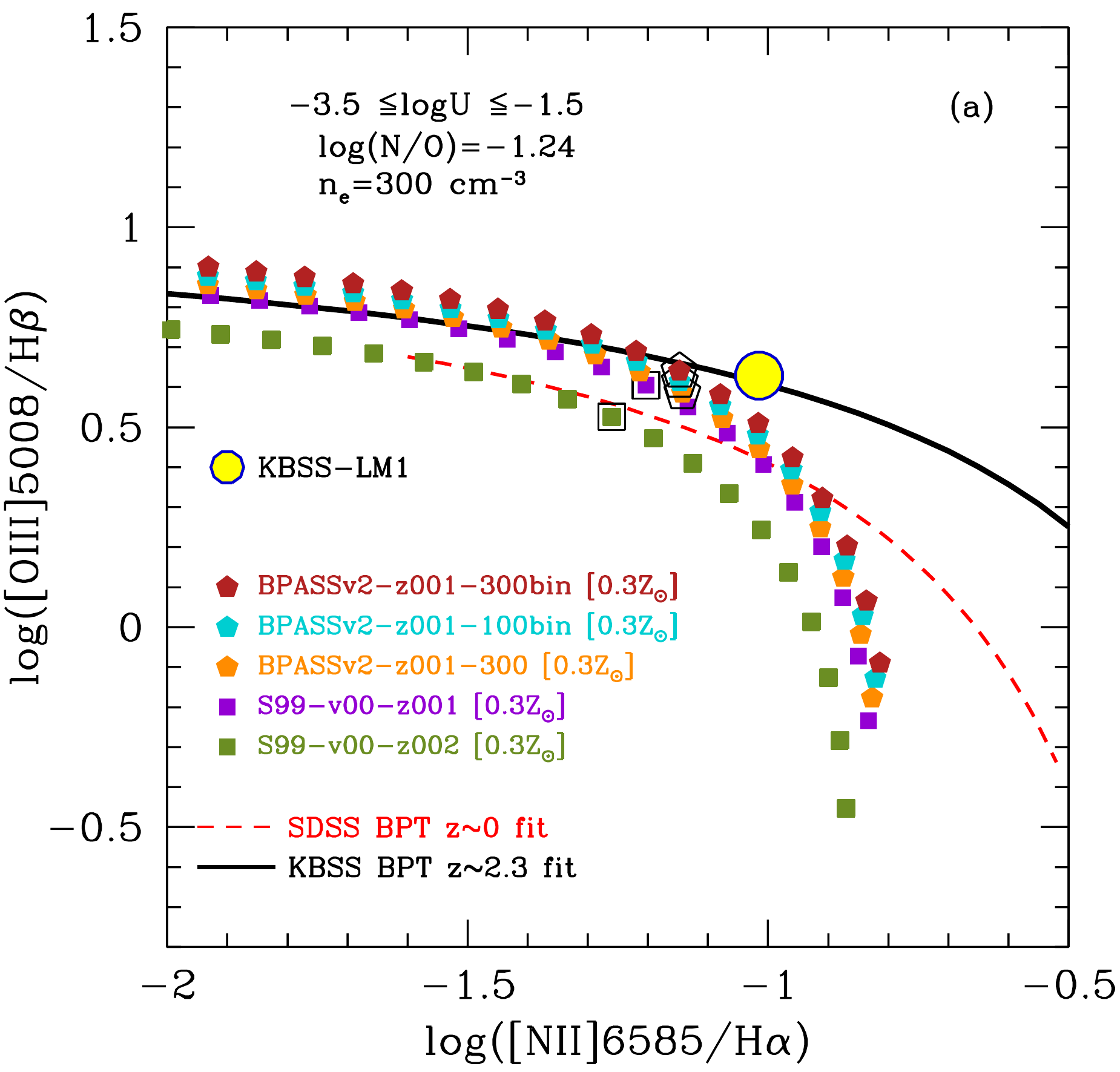}\includegraphics[width=8.5cm]{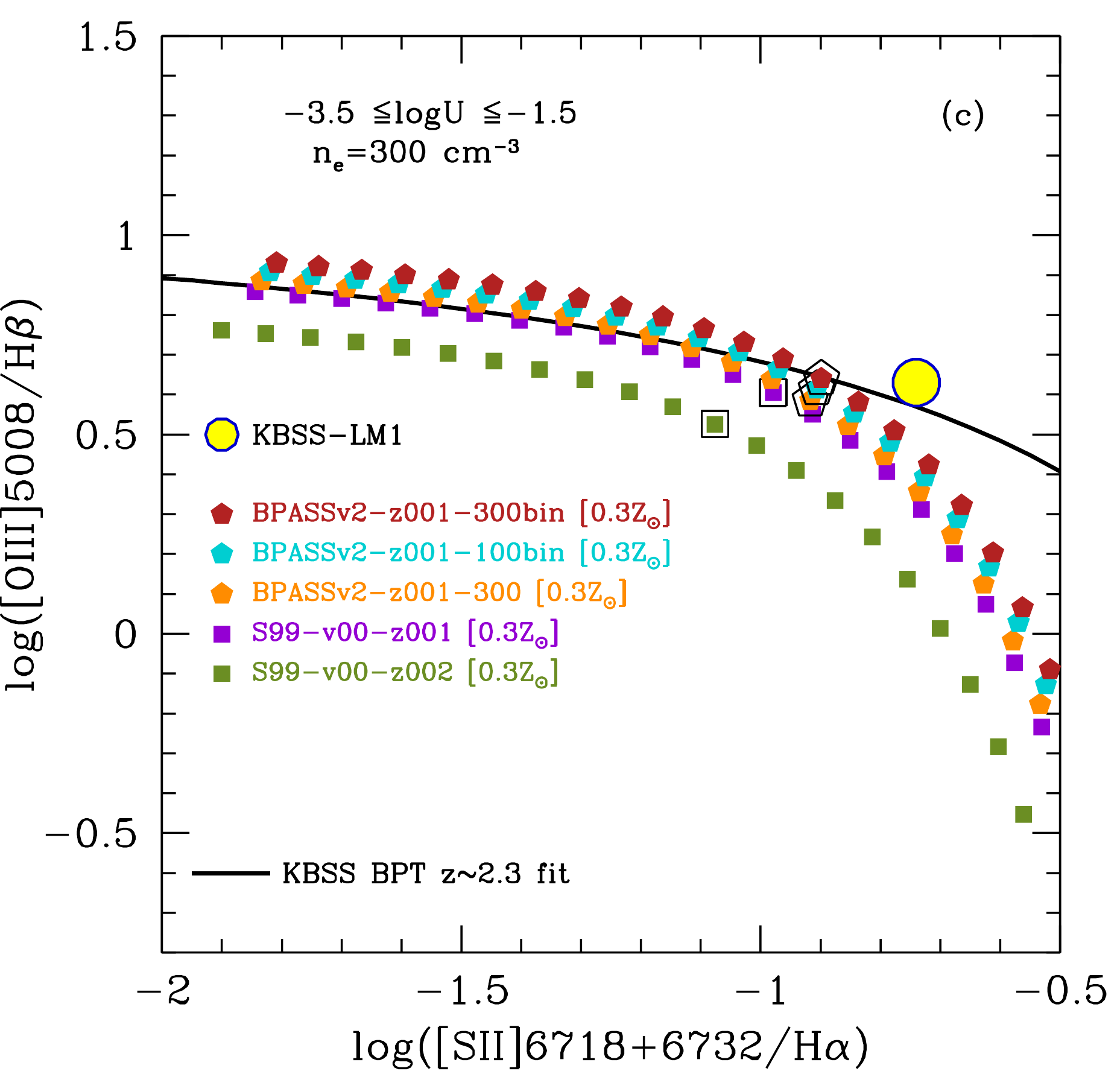}}
\centerline{\includegraphics[width=8.5cm]{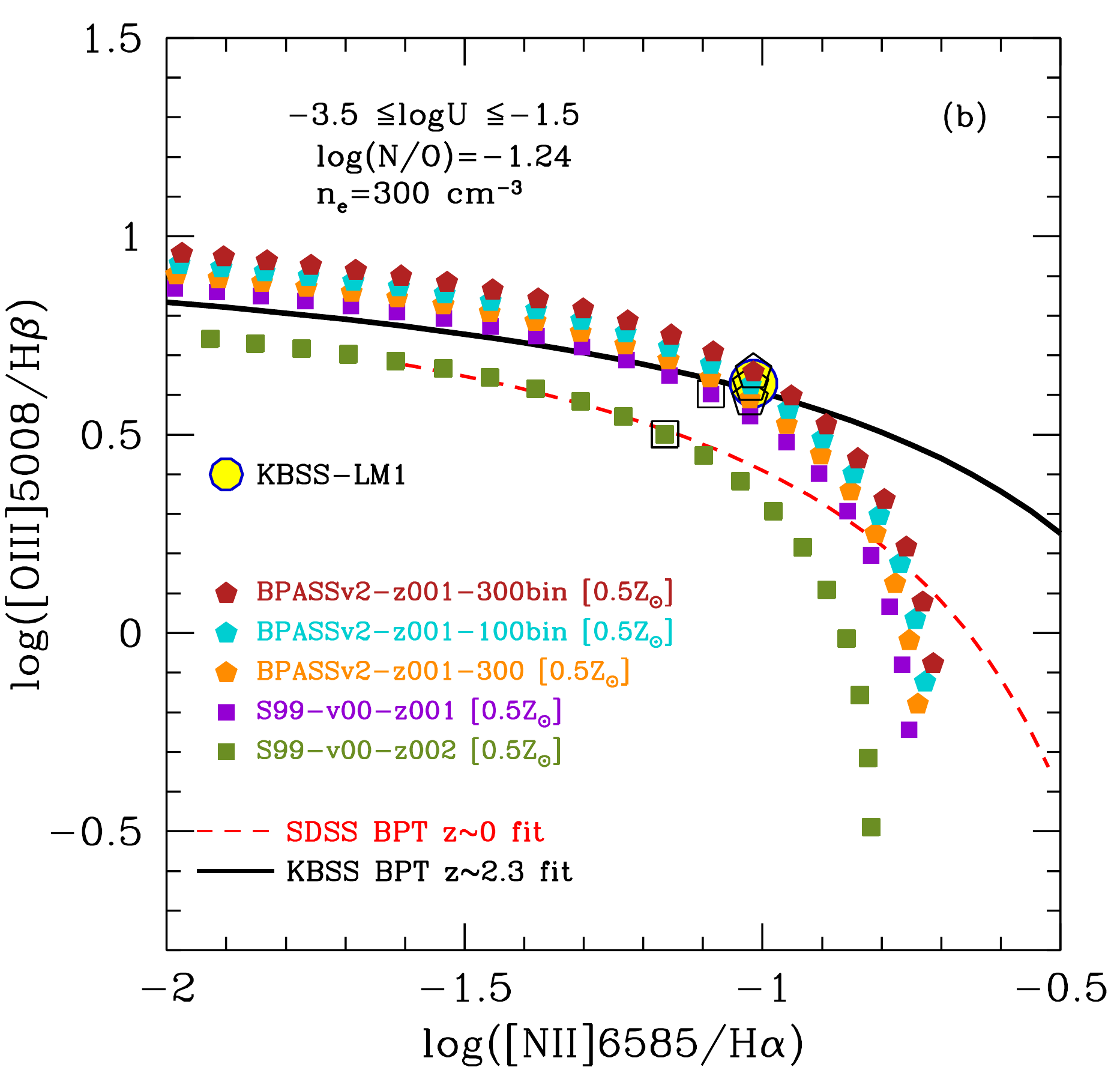}\includegraphics[width=8.5cm]{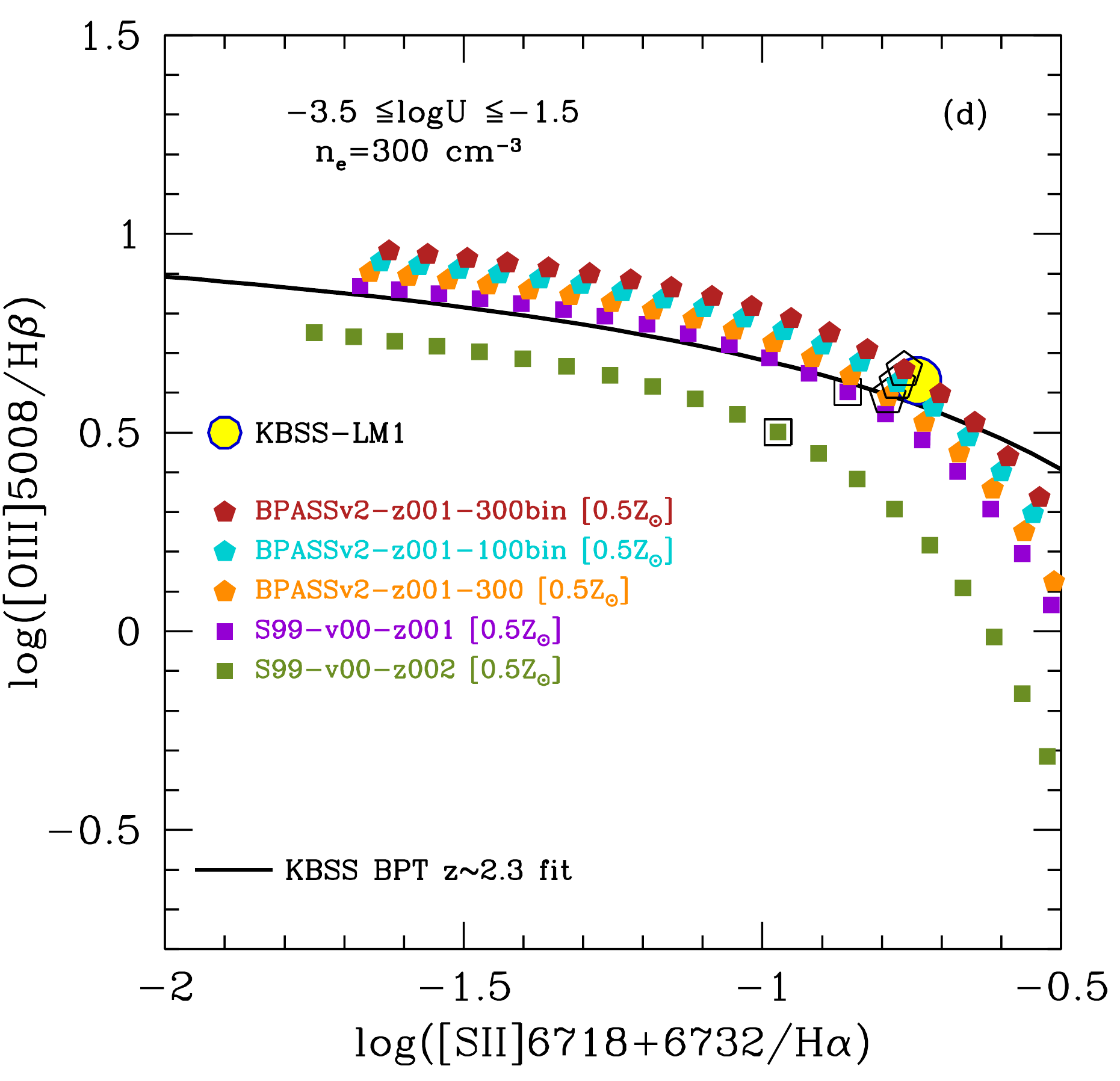}}
\caption{Predictions of the same
set of models as in Figure~\ref{fig:plot_gamma} in the N2 (lefthand panels) and S2 (righthand panels) BPT plane, compared to the KBSS-LM1 measurement. The same ${\rm log {\it U}}$ grid points that most
closely matched the observed line ratios in Figure~\ref{fig:plot_gamma} are indicated with skeletal black symbols. As in Figure~\ref{fig:plot_gamma}, the error
bars for the LM1 measurements are smaller than the size of the point. The upper panels show the predictions assuming $Z_{\rm neb}/Z_{\odot} = 0.3$, while
the lower panels assume $Z_{\rm neb}/Z_{\odot}=0.5$; note that the higher gas-phase metallicity is strongly preferred.
} 
\label{fig:plot_bpt}
\end{figure*}

\subsection{N2 and S2 BPT Planes}
\label{sec:bpt}
Figure~\ref{fig:plot_bpt} shows the predictions of the Cloudy models 
in the N2 and S2 BPT planes. In the top panel, the model grid assumes ${\rm log(N/O) = -1.24}$ as 
obtained in \S\ref{sec:n_over_o}
using the N2O2 calibration in eqn.~\ref{eqn:n2o2}. 
Note first that the KBSS-LM1 measurement falls directly on the fit to the BPT
locus of the full $z \sim 2.3$ KBSS-MOSFIRE sample (\citealt{steidel14}; Strom et al 2016); i.e., the composite is representative
of the full KBSS sample. The grid point corresponding
to the best value of $U$ for each model determined from Ne3O2 and O32 (Figure~\ref{fig:plot_gamma}) 
is indicated with a black skeletal symbol in each panel. The BPASSv2-z001-100bin and BPASSv2-z001-300bin population synthesis models with
${\rm log}~U=-2.8$ both predict values in excellent agreement with the observations  
for $Z_{\rm neb}/Z_{\odot}=0.5$; the predictions for the BPASSv2-z001-300 single star model come close to
the observations in both BPT planes for $Z_{\rm neb}/Z_{\odot} = 0.5$. In contrast, 
the S99 models cannot simultaneously match the O3 and N2 or O3 and S2 constraints at any gas-phase metallicity
or value of log$U$;  
the S99 models, when ${\rm log} U$ is fixed by the O32 vs. Ne3O2 constraint (Figure~\ref{fig:plot_gamma}), 
underpredict both N2 and O3 ($\simeq 0.15$ dex in each). The S99 models are capable of producing O3 as high
as observed, but only if log U were $\simeq 0.5$ dex higher than the value constrained by Ne3O2 and O32; such high
values of log$U$ would predict much {\it lower} N2 (or S2) than observed. 

\begin{figure}[htbp]
\centerline{\includegraphics[width=8.5cm]{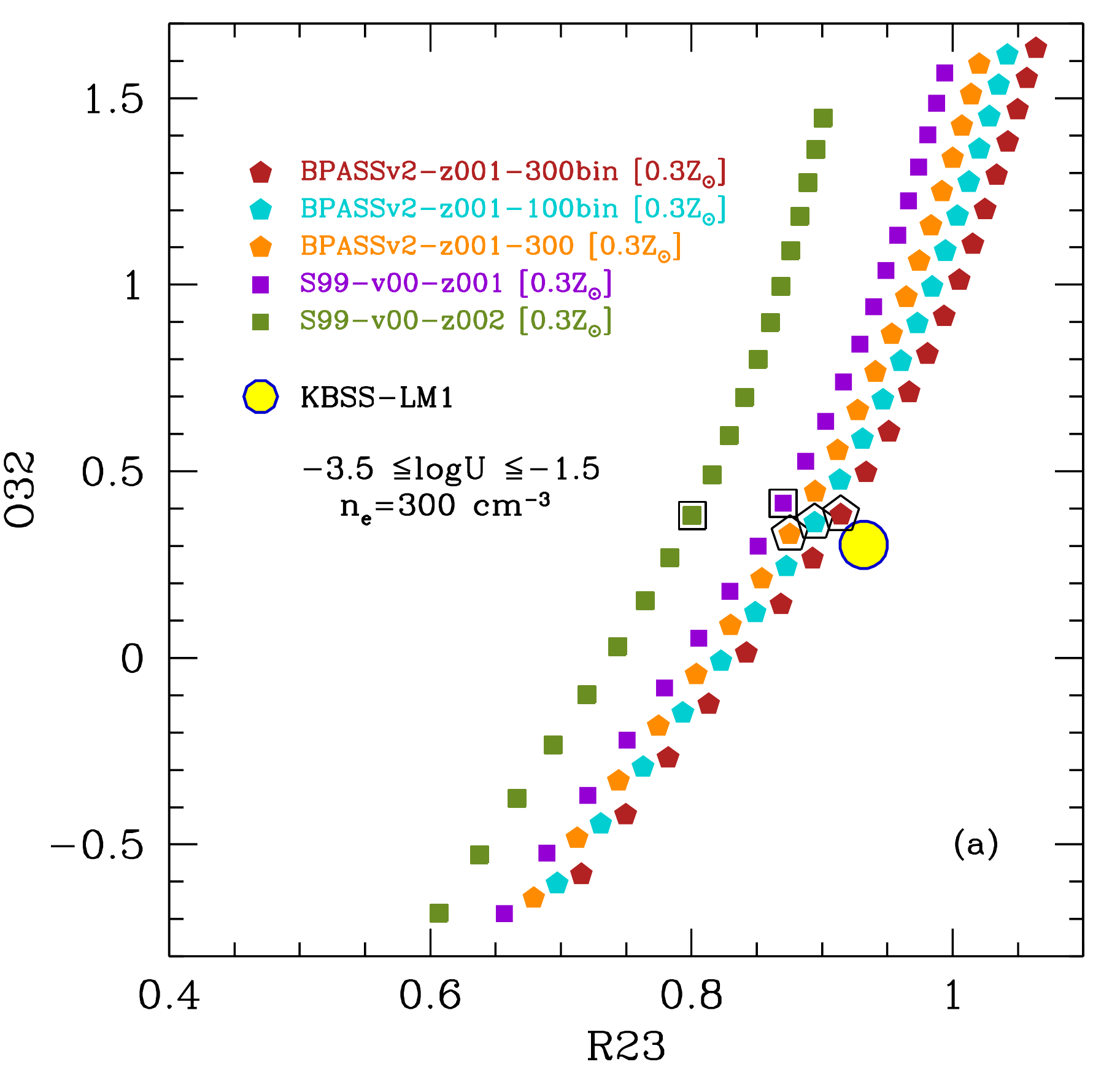}}
\centerline{\includegraphics[width=8.5cm]{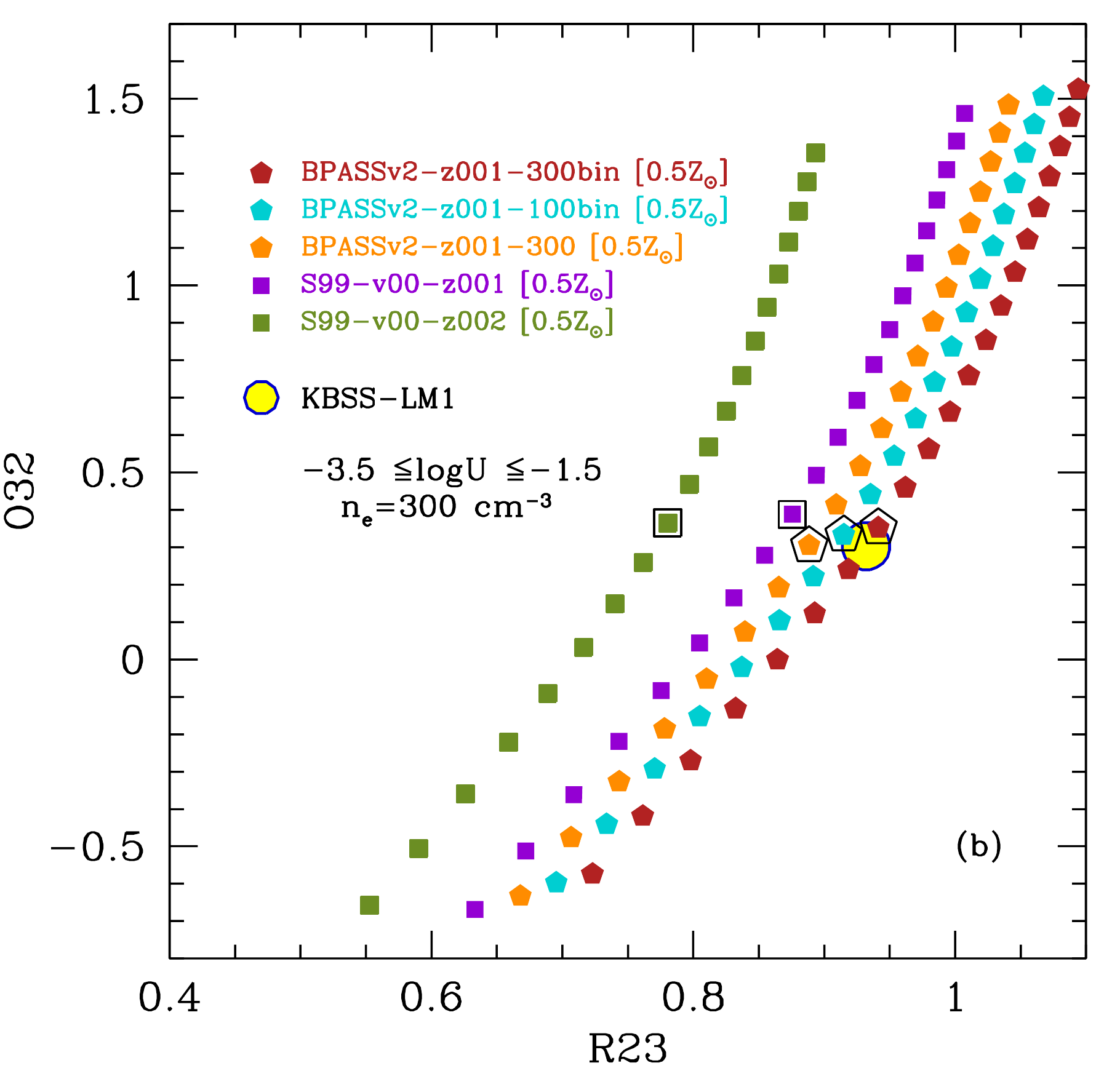}}
\caption{(a) As in Figures~\ref{fig:plot_gamma} and \ref{fig:plot_bpt} (the LM1 error bars are again smaller than the size of the point),
a comparison of the Cloudy model predictions for various assumed
input stellar population models in the O32 vs. R23 plane. 
All of the models shown can reproduce the observed value of O32 with ${\rm -2.8 \le log {\it U} \le -2.6}$
(as in Figure~\ref{fig:plot_gamma}, where the point corresponding to the value of $U$ that most closely reproduces the
observed Ne3O2 and O32 ratios is indicated with a black open box/pentagon), but R23 is matched only by 
the BPASSv2 binary models with $Z_{\rm neb}/Z_{\odot} = 0.5$. 
Note that the O32 vs. R23 plane provides the greatest sensitivity to the hardness of the stellar EUV continuum, 
and that (unlike the BPT diagrams in Figure~\ref{fig:plot_bpt}) both BPASSv2 binary models are clearly favored
over the BPASSv2-z001-300 single star model.
}
\label{fig:plot_R23}
\end{figure}

\subsection{R23 vs. O32}
\label{sec:R23}
Figure~\ref{fig:plot_R23} examines the efficacy of the photoionization models in
the O32 versus R23 plane. A combination of R23 and O32 has been proposed as an indicator of gas-phase oxygen
abundance by a number of authors (e.g., \citealt{mcgaugh91,kobulnicky99,kewley02}), where O32 is used
to set the ionization parameter whereupon R23 is sensitive to overall O/H. 
The same models illustrated in Figures~\ref{fig:plot_gamma}
~and~\ref{fig:plot_bpt} 
are shown in Figure~\ref{fig:plot_R23}; 
the S99 models cannot produce
values of R23 as large as those observed at fixed O32, for {\it any} assumed gas-phase (O/H).
Note that the BPASSv2 models are more sensitive to gas-phase metallicity for values of log~U
fixed by O32 and Ne3O2 ratios-- 
the emission in the [\ion{O}{2}] and [\ion{O}{3}] nebular lines continues to increase until reaching a peak at   
$Z_{\rm neb}/Z_{\odot} \simeq 0.5$ (${\rm 12+log(O/H) = 8.39}$), while the S99 models produce their maximum R23 closer to $Z_{\rm neb}/Z_{\odot} \simeq 0.3$.
Comparison of Figure~\ref{fig:plot_R23} with Fig.~\ref{fig:plot_bpt} shows that the O32 vs. R23 plane is most
sensitive to differences in spectral hardness among the models shown: note that  
at fixed O32, R23 increases monotonically with increasing spectral hardness -- and that only the BPASSv2-z001 {\it binary} models
produce predicted line indices consistent with the observations, and then only when $Z_{\rm neb}/Z_{\odot} =0.5$. 

\begin{figure}
\centerline{\includegraphics[width=8.5cm]{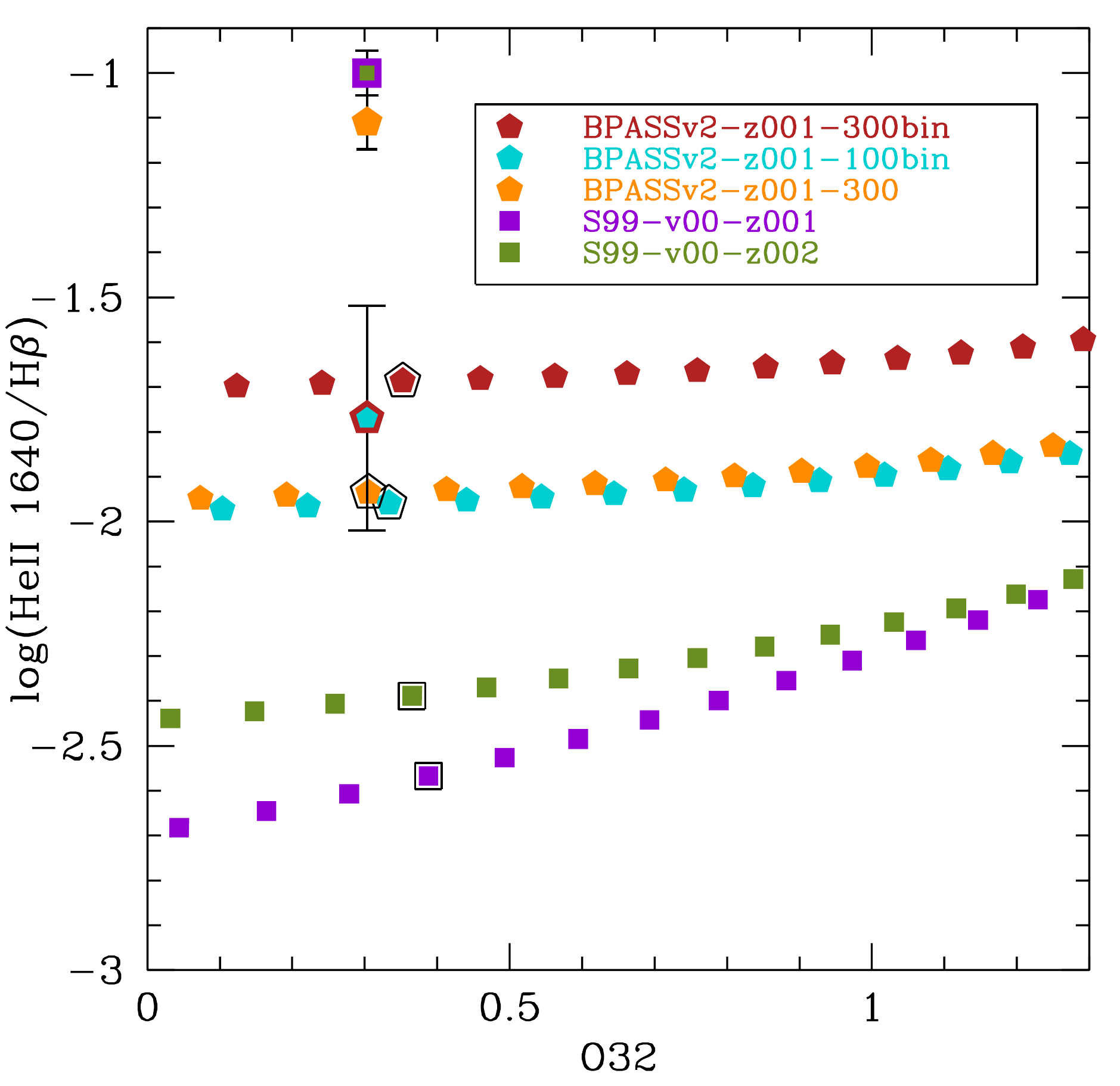}}
\caption{Illustration of the substantially different predictions for {\it nebular} \ion{He}{2} $\lambda 1640$ emission relative
to \Hb.  The colored points with error bars show the residual \ion{He}{2}$\lambda 1640$/\Hb\ measurements after
subtraction of the corresponding stellar model. Note that the S99 models and the BPASSv2-z001-300 single star
model have much larger \ion{He}{2} residual emission than the BPASSv2 binary models because of much weaker
predicted {\it stellar} \ion{He}{2}. Only the BPASSv2-z001-100bin and BPASSv2-z001-300bin models predict
internally consistent stellar and nebular contributions to the observed \ion{He}{2} $\lambda 1640$ spectral feature.  
The colored symbols with skeletal black points surrounding them have the same meaning as in previous figures. 
}
\label{fig:he2}
\end{figure}

\subsection{\ion{He}{2} Emission}
\label{sec:HeII}
Figure~\ref{fig:he2} shows that the largest differences between BPASSv2 binary and S99 population synthesis model predictions
are for the ratio of \ion{He}{2}~$\lambda1640/ \Hb$.  As discussed in \S\ref{sec:imf} above, the BPASSv2 binary models with 
continuous star formation
predict a stellar \ion{He}{2} feature that is absent from the single-star (S99 or BPASSv2) models-- this can be understood qualitatively
as a direct result of the evolution of massive stars in binaries, which produce an extended evolutionary
phase during which stellar effective temperatures can be as high as 10$^5$ K over a much longer period than 
classical Wolf-Rayet stars (see, e.g., \citealt{eldridge12}). Massive star binaries naturally produce a harder
ionizing-UV spectrum with a higher duty cycle than single star models, which can produce \ion{He}{2}-ionizing
photons only during brief intervals when very high mass Wolf-Rayet stars are present (see, e.g., \citealt{shirazi12,crowther16}). 

In any case, Figure~\ref{fig:he2} compares the predictions of the various models for the ratio of
{\it nebular} \ion{He}{2} $\lambda 1640$/\Hb\ with the observed residuals after subtraction of the
corresponding predicted {\it stellar} \ion{He}{2} emission (Table~\ref{tab:lines}, Fig.~\ref{fig:uv_emlines}). 
The points with error bars are the measured residuals from Table~\ref{tab:lines}, color-coded by the stellar 
model subtracted from the LM1 FUV spectrum.  A disagreement between the predicted ratio \ion{He}{2}$\lambda 1640$/\Hb\ 
(locus of color-coded points without error bars) and the residual \ion{He}{2}/\Hb\ indicates an inconsistency
between the model and the observations. 
Only the BPASSv2-z001 binary models (dark red and turquoise pentagons) suggest consistency between
the predicted stellar \ion{He}{2} and the strength of the residual emission when interpreted as the
nebular component of \ion{He}{2}$\lambda 1640$. 
When the BPASSv2-z001 binary population synthesis models are subtracted from the observed
KBSS-LM1 spectrum (Fig.~\ref{fig:uv_emlines}), the {\it residual} \ion{He}{2} feature is narrow and is reduced in intensity by a factor of
$\sim 6$ (turquoise and dark red pentagon with error bar); though its significance is marginal, the residual \ion{He}{2} emission is consistent with that predicted by
the photoionization models using the same BPASSv2 spectra. The BPASSv2-z001-300 single star model predicts nebular \ion{He}{2}/\Hb\ 
comparable to that of the BPASSv2-z001-100bin model shown, but, because it predicts negligible stellar \ion{He}{2}, the residual emission
(orange pentagon with error bar) is much higher than predicted for nebular \ion{He}{2} by the same model. 

One issue that becomes relevant in the current context is that photospheric absorption lines in the stellar spectrum 
are present at wavelengths coincident with the broad stellar \ion{He}{2} feature (see Fig.~\ref{fig:uv_emlines}a)-- this
``blanketing'' might conceivably mask the presence of a broad \ion{He}{2} line in spectra with low S/N and/or low
spectral resolution, and it also introduces some degree of model dependence to the measurement of \ion{He}{2} emission
in galaxy spectra (since the blanketing depends on the details of the photospheric absorption lines). 

\begin{figure}[htb]
\centerline{\includegraphics[width=8.5cm]{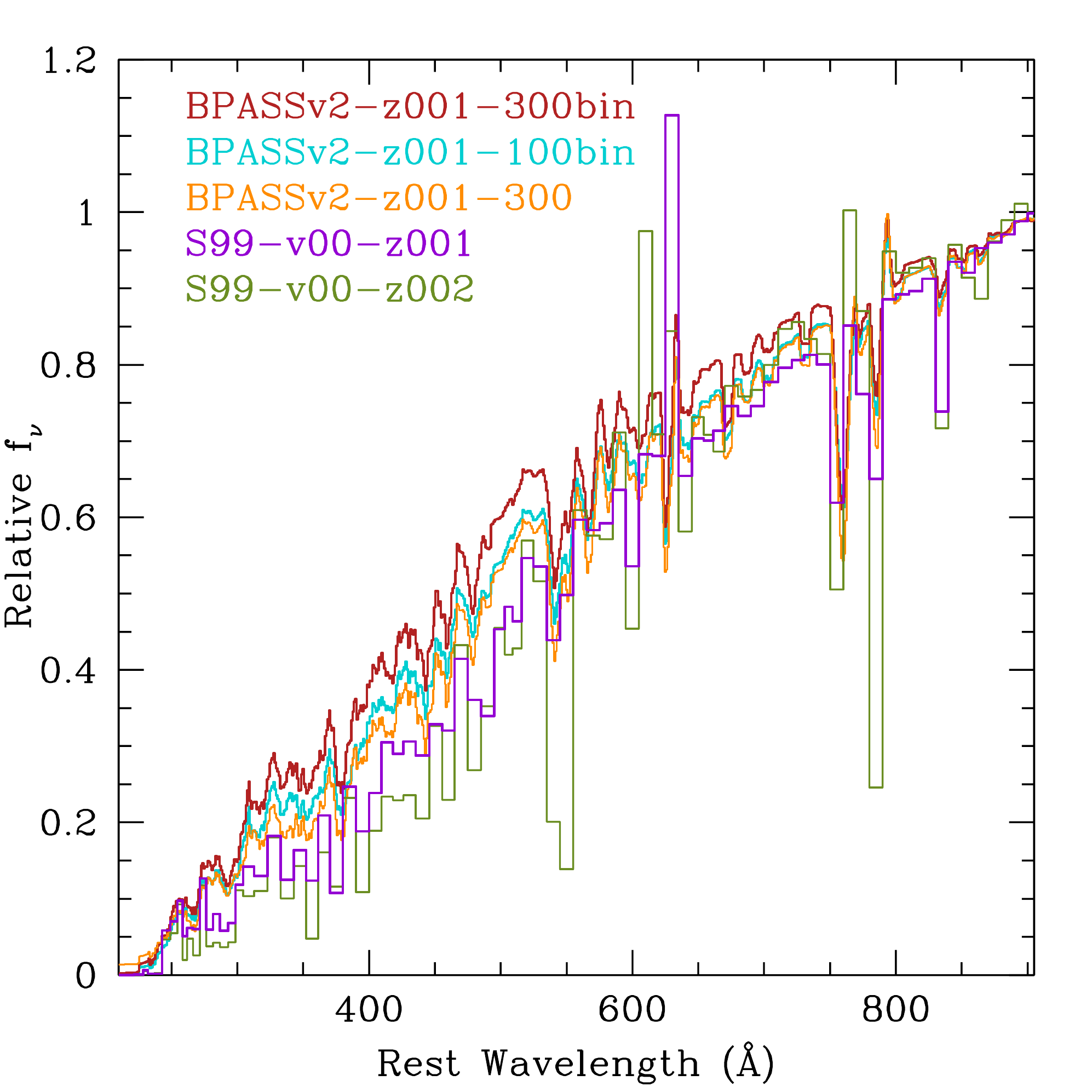}}
\caption{Comparison of the EUV 1-4 Rydberg spectra for the same models shown in Figs.~\ref{fig:plot_gamma} through \ref{fig:he2}.
All of the models assume constant SFR over $10^8$ yrs, and are normalized so that
$f_{\nu}=1.0$ at $\lambda_0 = 910$ \AA. The models that come
closest to reproducing the observed nebular emission line ratios are BPASSv2-z001-300bin and BPASSv2-z001-100bin, the models
with the hardest EUV spectra. The other models do not produce
enough photons at $\simeq 2-4$ Ryd (225-450 \AA) relative to 1 Ryd (912 \AA) to match the observed
nebular excitation.  
}
\label{fig:euv_spectra}
\end{figure}

\subsection{Spectral Shape in the FUV and EUV}

\label{sec:euv_shape} 

The preceding analysis of the nebular emission line ratios essentially amounts to constraining the spectral
shape of the ionizing EUV stellar continuum.  Using O32 and Ne3O2 to measure $U$ fixes the intensity
of the ionizing radiation field in the 1-3 Ryd range, so that if the spectral shape of a model is off, the predictions
for the relative intensities of other lines (e.g., O3, N2, S2, \ion{He}{2}/\Hb) will not match the observations.  
To summarize, of the population synthesis models that provide the best matches to both the global far-UV spectrum
and to the details of the stellar wind and photospheric absorption lines, 
the BPASSv2 {\it binary} models appear to be capable of reproducing simultaneously all of the line ratios observed, 
at a fixed value of $U$. 
Evidently, the stellar ionizing spectrum of this particular sample of (typical) star-forming galaxies
at $z \sim 2.4$ needs to satisfy the following:

$\bullet$ It must be modestly reddened by dust, with $E(B-V)_{\rm cont}\simeq 0.19$, or $A_{\rm V} \simeq 0.76$ under the assumption
of the \citet{calzetti00} starburst attenuation relation. 

$\bullet$ It must have low photospheric abundances ($Z_{\ast}/Z_{\odot} \sim 0.1$), to reproduce the limited blanketing of the stellar far-UV continuum.

$\bullet$ The strongest stellar wind lines (most notably, that of \ion{C}{4}) suggest $Z_{\ast}/Z_{\odot} \simeq 0.14$. 

$\bullet$ It must be capable of producing nebular emission lines characterized by R23$\sim 0.93$ when O32$\simeq 0.3$.

$\bullet$ It must have a sufficiently hard ionizing UV spectrum to produce substantial stellar \ion{He}{2}$\lambda 1640$ emission and
(probably) nebular ${\rm HeII~\lambda 1640/\Hb \simeq 0.02}$ (section~\ref{sec:HeII}). 

The principal reason that the BPASSv2 binary models are more successful overall is the harder ionizing spectrum
under conditions of continuous star formation (i.e., star formation timescales of $t_{\rm d} \simgt 20-30$ Myr; see \S\ref{sec:intro}). 
Figure~\ref{fig:euv_spectra} shows a direct
comparison of the EUV ionizing spectra for the models discussed in this section. 
As discussed in \S\ref{sec:R23}, the sequence of increasing EUV hardness is reflected most directly by the R23
parameter at fixed O32 (Figure~\ref{fig:plot_R23}b); in both Figures~\ref{fig:plot_R23}b~and~\ref{fig:euv_spectra}, the EUV spectral
hardness decreases from top to bottom in the figure legends. 


\begin{figure}[htbp]
\centerline{\includegraphics[width=8.5cm]{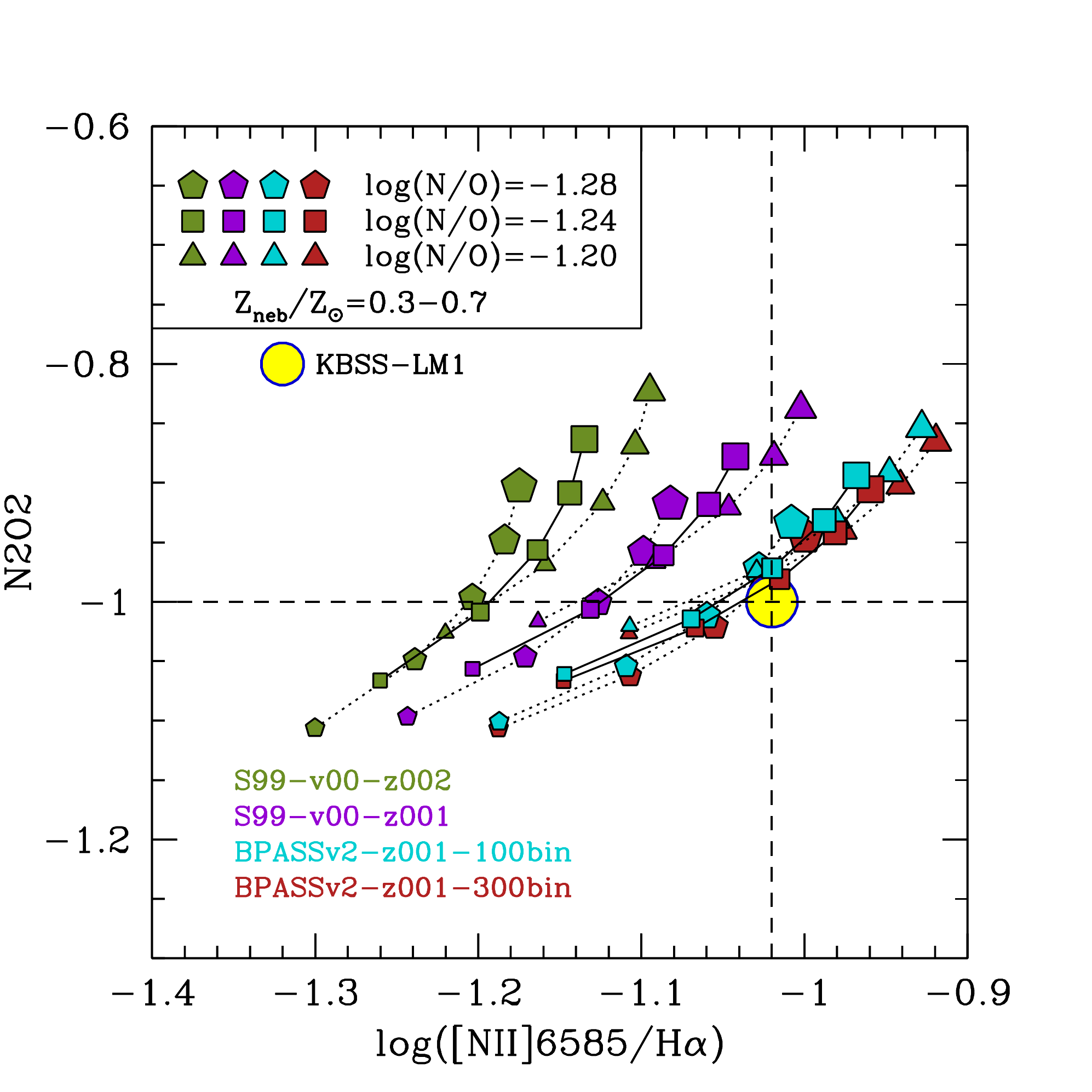}}
\caption{The predictions of a subset of the population synthesis models (each assuming the value of ${\rm log U}$ constrained by
the observed O32 and Ne3O2 values shown in Figure~\ref{fig:plot_gamma}) for the N2O2 index with respect to N2${\rm \equiv log([NII]\lambda 6585/\Ha)}$. 
The large yellow point is the LM1 measurement, where the radius of the circle represents the uncertainties on the
measurements. 
Each model is plotted for a range of assumed $Z_{\rm neb}/Z_{\odot} = 0.3-0.7$, in linear steps of 0.1,
where the symbol size scales with the value. Outside of the range ${\rm log(N/O) = -1.24 \pm 0.04}$ with $Z_{\rm neb}/Z_{\odot} = 0.5\pm 0.1$, 
no combination of N/O and O/H ($Z_{\rm neb}/Z_{\odot}$) can reproduce the observed point. 
The BPASSv2-z001 binary models   
match the observations, with  ${\rm log(N/O) = -1.24}$ when $Z_{\rm neb}/Z_{\odot} = 0.5$; the marginal 
values of ${\rm log(N/O) = (-1.28, -1.20)}$ match when $Z_{\rm neb}/Z_{\odot}=(0.6, 0.4)$. The allowed range, ${\rm log(N/O)=-1.24\pm0.04}$, 
is in excellent agreement with that obtained from the local calibration of N2O2$\rightarrow$N/O (\S\ref{sec:n_over_o}).   
}
\label{fig:plot_n2o2}
\end{figure} 

\subsection{N/O Revisited}
\label{sec:n_over_o2}

The constraints on the spectral shape of the EUV ionizing radiation field provided by 
the combination of population synthesis models and ionization parameter as described in \S\ref{sec:gamma} 
allow us to remove the last remaining dependency on calibrations established at low-redshift: the
mapping of the N2O2 index (which is observed) to N/O.  We showed in section \S\ref{sec:bpt} that the
model predictions for the N2, S2, and O3 indices are very well-matched under the assumption that
${\rm log(N/O) = -1.24}$ as implied by eqs.~\ref{eqn:n2o2} and \ref{eqn:n2s2}. We now examine  
the constraints on log(N/O) based only on measured quantities and the population synthesis+photoionization 
models. 

To accomplish this, we allowed N/O to vary independently of the overall scaling of the nebular abundances $Z_{\rm neb}/Z_{\odot}$
in the models, fixed the ionization parameter as described in \S\ref{sec:gamma}, 
and required that the predicted N2O2 and N2 line indices both agree with the observed point within the errors
(see Table~\ref{tab:properties}); the results are shown in Figure~\ref{fig:plot_n2o2}.
We find that only the BPASSv2-z001 binary models
can reproduce the observed line indices simultaneously, and then {\it only} if ${\rm log(N/O)=-1.24\pm0.04}$
{\it and} $Z_{\rm neb}/Z_{\odot} = 0.5\pm0.1$. 
Figure~\ref{fig:plot_n2o2} shows the model predictions for the N2O2 index versus the N2 index (see
Table~\ref{tab:properties} for assumed gas-phase oxygen abundances in the range $Z_{\rm neb}/Z_{\odot}=0.3-0.7$.

In Fig.~\ref{fig:plot_n2o2}, the points connected by the solid curve correspond to the ${\rm log(N/O) = -1.24}$ for
each color-coded model, for which the $Z_{\rm neb}/Z_{\odot}=0.5$ metallicity point matches the KBSS-LM1 observation.
Also shown are curves assuming the $\pm 1 \sigma$ marginal values of ${\rm log(N/O)}$ that are still consistent
with the observation, i.e.  
${\rm log(N/O) = (-1.20, -1.28)}$; assuming the marginal values yields best-matching oxygen abundance  
$Z_{\rm neb}/Z_{\odot} = (0.4, 0.6)$, respectively. 
Thus, the most successful population synthesis+photoionization models predict that $Z_{\rm neb}/Z_{\odot} = 0.5\pm0.1$, or 
${\rm 12+log(O/H) = 8.39^{+0.08}_{-0.10}}$; we return to a detailed discussion of the nebular oxygen abundance in \S\ref{sec:oh}. 

The abundance ratio N/O, constrained only by the $z\sim 2.4$ KBSS-LM1 observations and 
the population synthesis+photoionization models that successfully reproduce all other strong
line ratios {\it is identical to N/O obtained from local \ion{H}{2} region calibrations.}
Since both methods yield consistency using either the N2O2 or N2S2 indices, it also implies that 
nebular S/O${\rm = (S/O)_{\odot}}$ to within $\le 0.1$ dex for the KBSS-LM1 ensemble. 
The N/O abundance ratio for the full $z \sim 2.3$ KBSS-MOSFIRE sample is investigated in a companion paper (Strom et al 2016). 

\section{Ionized Gas Phase Abundances}
\label{sec:metallicity}

We showed in \S\ref{sec:cloudy_models} that the strong line ratios observed in the nebular spectra of the KBSS-LM1 ensemble 
are reproduced by a model assuming nebular abundances for O (and S) of $Z_{\rm neb}/Z_{\odot}=0.5\pm0.1$ and 
${\rm log(N/O) = -1.24\pm0.04}$ (${\rm [N/O]=-0.38\pm0.04}$, or ${\rm \simeq 0.42 (N/O)_{\odot}}$). In this section, we compare the
model results for O/H with various other commonly-used methods, and present additional abundance ratio determinations 
made possible by the combination of the KBSS-LM1 spectra and the successful population synthesis+photoionization models. 

\subsection{Nebular Oxygen Abundance} 

\label{sec:oh}

Table~\ref{tab:properties} summarizes the oxygen abundance measurements based on the nebular emission lines in the
KBSS-LM1 composite spectra. 
In this section, we report a direct $T_{\rm e}$ measurement of O/H,  
and compare it to the results of the stellar population
synthesis+photoionization modeling presented in \S\ref{sec:cloudy_models} above. Comparisons of the direct
and modeled oxygen abundances to several estimates based on strong line indices are presented in
\S\ref{sec:strong}.   

\subsubsection{Direct Method}

\label{sec:direct}
The KBSS-LM1 nebular spectra are of high enough quality to
yield a $T_{\rm e}$ measurement from the observed ratio ${\rm O3uv \equiv OIII](1661+1666)/[OIII]\lambda 5008}$ 
(see Table~\ref{tab:properties}), which gives $T_{\rm e}({\rm OIII}) = 12250 \pm 600$ K. Assuming that 
${\rm T_e(OII) = 0.7~T_e(OIII)+3000}$ (e.g., \citealt{campbell86,garnett92,izotov06,pilyugin09}; 
hereinafter referred to as the ``$T_2-T_3$'' relation) and that ${\rm (O/H) = (O^{++}/H^{+}) + (O^{+}/H^{+})}$,
the corresponding oxygen abundance is 
\begin{align}
\begin{split}
{\rm 12+log(O/H)_{dir}} & =8.14\pm0.04~; \\ 
Z_{\rm neb}/Z_{\odot} & = 0.29\pm0.03 .
\end{split}
\label{eqn:direct}
\end{align}  
The error bars account both
for the uncertainty in the line fluxes (dominated by the uncertainty in the UV \ion{O}{3}] doublet) and
in the extinction correction from the Balmer decrement measurement. However, the quoted error may
underestimate the true uncertainty given that we have assumed an extinction curve that may not
be the correct one for nebular extinction in the high redshift objects\footnote{The inferred $T_{\rm e}$-based oxygen
abundances assuming alternative extinction curves (for values of E(B-V) that reproduce the same assumed intrinsic Balmer line ratio) 
are 12+log(O/H)=[7.98, 8.12, 8.11] for SMC (\citealt{gordon03}), \citet{calzetti00}, and \citet{reddy15}, respectively.}.
A smaller extrapolation would be required 
to use the ${\rm [OIII]\lambda 4364}$ auroral line instead of the UV \ion{O}{3}] intercombination doublet, but the former was covered in the
observed wavelength range for only $\sim 50$\% of the KBSS-LM1 galaxies; we have therefore 
treated a tentative detection as a 3$\sigma$ upper limit 
on the electron temperature, $T_{\rm e} < 13370$ K (see Table~\ref{tab:properties}), consistent with
the $T_{\rm e}$ measurement based on the UV \ion{O}{3}] feature.  

Another potential source of systematic error is that
we have not {\it measured} the electron temperature associated with the [OII]-emitting gas; the same problem
has affected essentially all other high-redshift estimates, and most estimates at low redshift, in the literature. 
In the absence of a measurement of $T_{\rm e}$ for a missing ionization stage, it is common to use
the $T_2-T_3$ relation mentioned above, which was originally established using photoionization models but has been generally supported 
by observations of local galaxy \ion{H}{2} regions (e.g., \citealt{brown14,berg15}). Nevertheless,
the applicability 
of this relation is a source of some controversy; for example, 
\citet{andrews+martini13} (AM13) have shown that for composite SDSS galaxy spectra in bins of $M_{\ast}$ and SFR, the standard
$T_2-T_3$ relation systematically over-predicts ${\rm T_{\rm e}(OII)}$ at a given ${\rm T_{e}(OIII)}$ for the vast majority
of the bins in ${\rm M_{\ast}}$ and ${\rm M_{\ast}-SFR}$. The exceptions are for objects with ${\rm log SFR > 1.0}$ (rare in the SDSS
sample), which are generally 
in good agreement with the $T_2-T_3$ formulation. The implication for the KBSS-LM1 sample is unclear, since 
most of the $z \sim 2.4$ galaxies have ${\rm log(SFR) > 1.0}$.

However, it is interesting to ask how the possible offset of $T_2$ as found 
by AM13 would
affect the direct method oxygen abundance for the KBSS-LM1 composite. For what AM13 called their
``${\rm M_{\ast}-SFR}$'' stacks, they found an average offset of $\Delta T_2=-1300$ K relative to the $T_2-T_3$ relation.
Applying this offset to the KBSS-LM1 composite would change the direct method oxygen abundance ${\rm 12+log(O/H)_{dir}}$ from 8.14 to 8.25 (i.e., from $Z_{\rm neb}/Z_{\odot}=0.28$ to
$Z_{\rm neb}/Z_{\odot}=0.36$). 


\subsubsection{Comparison to Population Synthesis+Photoionization Model Results} 

\label{sec:comparison}

There is a well-known  
tendency for measurements of $T_{\rm e}$ based on collisionally-excited lines (CELs) to systematically under-estimate 
nebular O/H (see e.g. \citealt{peimbert02} and references therein); this 
effect is generally ascribed to the fact that, for nebulae having zones of differing temperature, the
high temperature regions will be over-represented due to the temperature sensitivity of the emissivity of CELs, leading to
an over-estimate of $T_{\rm e}$. Nebular recombination lines (RELs), on the other hand, are much less temperature sensitive,
and when both CELs and RELs can be measured for the same systems, the inferred oxygen abundances based
on RELs are systematically higher, and generally in better agreement with the {\it stellar} O/H measured from nearby regions
in the same galaxy-- including the well-studied Orion and 30 Doradus regions in the Galaxy and LMC, respectively.  
Recently, 
\citet{esteban14} used high-quality echelle spectra of star-forming ``knots'' in nearby galaxies to measure nebular abundances
from both CELs and RELs, finding 
a systematic offset 
\begin{equation}
{\rm log(O/H)_{REL} - log(O/H)_{CEL}= 0.24\pm0.02~ dex}  
\label{eqn:offset}
\end{equation}
(median and inter-quartile range),
consistent with other similar measurements for individual \ion{H}{2} regions in the literature (e.g., \citealt{blanc15}.)  

For some purposes, it is sufficient simply to be consistent in the method one uses to estimate nebular O/H, as the
{\it relative} values of O/H will be preserved within a sample. Because the metallic recombination lines are extremely weak, 
it is far more difficult to measure O/H from RELs -- such measurements may remain
out of reach for high-redshift galaxies for the foreseeable future. It therefore makes sense to use CEL-based direct measurements (or
their strong-line proxies) for wholesale measurements of gas-phase oxygen abundances.
However, for sensible comparisons between stars and ionized gas in the same galaxies (of paramount interest
in this paper), it is important
to acknowledge that CEL-based $T_{\rm e}$ {\it measurements} are likely to yield numerical values of log(O/H) 
lower than the {\it actual} gas-phase O/H (or the stars
in the same regions) by $\simeq 0.24$ dex.
In this context, applying the offset from eq.~\ref{eqn:offset} to the KBSS-LM1 (CEL-based) $T_{\rm e}$ 
measurement (eqn.~\ref{eqn:direct}) gives 
\begin{align}
\begin{split}
{\rm 12+log(O/H)_{REL}} & =8.38\pm0.04  \\ 
Z_{\rm neb}/Z_{\odot} & = 0.49\pm0.05. 
\label{eqn:true}
\end{split}
\end{align}
This is the same  
value for the nebular oxygen abundance favored by the the population synthesis+photoionization
modeling described in \S\ref{sec:cloudy_models}; thus, we adopt $Z_{\rm neb}/Z_{\odot} = 0.5$ as the most 
probable value of the ``true'' nebular oxygen abundance of the KBSS-LM1 ensemble.     

Recall that the most likely {\it stellar} metallicity obtained from matching the observed spectrum of KBSS-LM1 to
population synthesis models 
(\S\ref{sec:PSMs}) is $Z_{\ast}/Z_{\odot} = 0.07-0.14$, lower by a factor of $\simeq 4-5$. 
This apparent discrepancy between the metallicity of the {\it stars} and that of the {\it ionized gas} 
required to simultaneously match the FUV stellar spectra and the FUV/optical nebular emission in the same galaxy ensemble is discussed in section~\ref{sec:discussion}
below.   

\subsubsection{Strong Line Methods}

\label{sec:strong}

For completeness, we have evaluated the nebular oxygen abundance based on various strong-line indices commonly used for that purpose,
summarized in \S\ref{tab:properties}.
By construction, the strong-line methods are intended to reproduce the direct $T_{\rm e}$ metallicity scale, for use 
when spectra are of insufficient quality to allow an actual measurement of $T_{\rm e}$-sensitive line ratios.

Among the strong-line abundance measurements listed in Table~\ref{tab:properties}, the
only one that is not calibrated using the 
\citet{pilyugin12} \ion{H}{2} region data set is
that of R23, which instead applies  
the low-metallicity branch R23 calibration from \citet{mcgaugh91}.  This calibration 
used a combination of photoionization models and $T_{\rm e}$ measurements, and accounts
for the ionization parameter through the measured value of O32,
\begin{multline}
{\rm 12+log(O/H)_{R23} = 12-4.944+0.767*R23+0.602*(R23)^2} \\
     {\rm -O32*[0.290+0.332*R23-0.331*(R23)^2]} 
\label{eqn:R23met}
\end{multline}
where R23 and O32 are as defined in Table~\ref{tab:properties}. 
From this, we obtain for the KBSS-LM1 composite:
\begin{align}
\begin{split}
{\rm 12+log(O/H)_{R23}} & = 8.20\pm0.03~; \\
Z_{\rm neb}/Z_{\odot}  & =0.32\pm0.03,  
\end{split}
\label{eqn:R23met_val}
\end{align} 
where the quoted error includes uncertainties in the 
R23 and O32 line indices and their nebular extinction corrections. Systematic errors were estimated by \citet{mcgaugh91} to be $\sim\pm0.1$ dex. 

For the other strong-line estimates of oxygen abundance, we use calibrations 
based on the same \citet{pilyugin12} \ion{H}{2} region sample 
discussed in \S\ref{sec:n_over_o}. Strom et al (2016) 
present linear fits of the strong-line O3N2 and N2 indices to $T_{\rm e}$-based oxygen abundances:
\footnote{Only the \ion{H}{2} regions with ${\rm 12+log(O/H)_{\rm dir} \ge 8.0}$ were included in the fit; objects
with lower metallicity would not be compatible with the observed locations of the KBSS-MOSFIRE sample on the BPT diagram.}
\begin{align}
\begin{split}
{\rm 12+log(O/H)_{O3N2}} & = 8.56-0.20*{\rm O3N2} \\
   \sigma & =0.08~{\rm dex}
\end{split}
\label{eqn:o3n2}
\end{align}
and
\begin{align}
\begin{split}
{\rm 12+log(O/H)_{N2}} & = 8.68+0.38*{\rm N2} \\ 
           \sigma & = 0.09~{\rm dex},
\end{split}
\label{eqn:n2}
\end{align}
where the uncertainties are the (global) rms scatter between the fit and individual measurements in the calibration set. 
One can also use the \citet{pilyugin12} data set to calibrate the N2O2 index (used in \S\ref{sec:n_over_o} for an initial
estimate of N/O) directly
onto O/H, for 12+log(O/H)$\ge 8.0$:
\begin{align}
\begin{split}
{\rm 12+log(O/H)_{N2O2}} & = 8.63 + 0.40*{\rm N2O2} \\
~\sigma & = 0.09~{\rm dex}.
\end{split}
\label{eqn:n2o2_met}
\end{align}

The calibrations given in equations~\ref{eqn:o3n2} and \ref{eqn:n2} 
use somewhat different parameters compared to the commonly-used calibrations of \citet{pettini04} (PP04); however,
for the line indices measured
in the KBSS-LM1 composite the results are very similar, as we show below.
Error bars assigned to all strong-line metallicity estimates below (see also Table~\ref{tab:properties}) reflect 
uncertainties in the measurement of the line indices (including extinction corrections, where relevant) as well as the
rms scatter in the calibration itself, accounting for the fact that the KBSS-LM1 measurements are based
on an ensemble, rather than a single galaxy. They do not attempt to account
for possible systematic errors caused by differences between the
calibration sample and the KBSS-LM1 ensemble, and all are based on the simplest linear relationship
between line index and direct $T_{\rm e}$ (CEL) oxygen abundance in the calibration set.   

For O3N2, we find
\begin{align}
\begin{split}
 {\rm 12+log(O/H)_{O3N2}}& =8.23\pm0.02 \\ 
Z_{\rm neb}/Z_{\odot} & = 0.35\pm0.02.
\end{split}
\label{eqn:o3n2_met}
\end{align}
An identical value would be obtained using the PP04 O3N2 metallicity calibration.

For N2, we obtain
\begin{align}
\begin{split}
{\rm 12+log(O/H)_{N2}} & = 8.29\pm0.02 \\ 
Z_{\rm neb}/Z_{\odot} & =0.40\pm0.02~.
\end{split}
\label{eqn:n2met}
\end{align}
Applying the (linear) PP04 N2 calibration instead  
would yield an oxygen abundance {\it higher} by 0.03 dex ($\sim 10$\%).  

The results for O3N2 and N2 differ in the same sense for both the PP04 calibration and those
in equations~\ref{eqn:o3n2} and \ref{eqn:n2}-- with N2 yielding higher log(O/H) than O3N2 -- but 
the amplitude of the difference is reduced: ${\rm log(O/H)_{N2}-log(O/H)_{O3N2} = 0.06}$ dex for 
the new calibration, compared to ${\rm log(O/H)_{N2}-log(O/H)_{O3N2} = 0.12}$ dex using PP04 calibrations. 
As discussed by \citet{steidel14}, the smaller difference is due in large part to a closer match
between the range of line indices represented in the calibration sample and those in the sample being calibrated. 

Using the measured N2O2 index for the KBSS-LM1 spectrum, 
equation~\ref{eqn:n2o2_met} yields
\begin{align}
\begin{split}
{\rm 12+log(O/H)_{N2O2}} & = 8.23\pm0.03 \\ 
Z_{\rm neb}/Z_{\odot} & =0.35\pm0.03~,
\end{split}
\label{eqn:n2o2_met_val}
\end{align}
identical to the estimate based on the O3N2 index. 

\subsubsection{N/O ``Analogs''}
\label{sec:n_o_analogs}

Alternatively, within the \citet{pilyugin12} calibration data set
there are 42 \ion{H}{2} regions ($\simeq 10$\% of the full sample) with ${\rm -1.28 \le log(N/O) \le -1.20}$, thus matching
the range of N/O inferred for the KBSS-LM1 composite discussed in \S\ref{sec:n_over_o2}; hereafter, we refer to this subsample
as ``N/O Analogs'' (cf. \S\ref{sec:n_over_o}) for KBSS-LM1.  
We have already pointed out that using the N2S2 and N2O2 indices with the local calibration sample predicts the
same value of N/O obtained using our photoionization modeling; here we can test whether the modeled N/O successfully
predicts the direct $T_{\rm e}$ abundances in the calibration sample. 
The median direct $T_{\rm e}$ (CEL) oxygen abundance of the N/O analog sample is 
\begin{align}
\begin{split}
{\rm 12+log(O/H)_{N/O}} &= 8.13 \pm 0.02 \\
     Z_{\rm neb}/Z_{\odot} & = 0.28\pm0.02 
\end{split}
\label{eqn:analogs}
\end{align}
where the errors represent the uncertainty in the median of the calibration sample of
42.

In other words, matching the {\it modeled} N/O for the KBSS-LM1 composite to analogs in
the \ion{H}{2} region calibration sample predicts a $T_{\rm e}$ (CEL) oxygen abundance within
0.01 dex of the value {\it measured} in KBSS-LM1 (equation~\ref{eqn:direct}). We have already shown that the oxygen
abundance predicted by the modeling, without reference to the local calibration sample, is 
the same value obtained from the direct (CEL) measurement after converting to the REL
abundance scale using eqn.~\ref{eqn:offset}.  
By extension, there appears to be full consistency in the mapping between N/O and O/H of the calibration sample
and the measurements (both direct and modeled) of the KBSS-LM1 composite. 
{\it The level of consistency implies that N/O at fixed O/H (or O/H at fixed N/O), 
is nearly identical in $z \simeq 0$ extragalactic \ion{H}{2} regions and $z \sim 2.4$ star-forming galaxies.}

Since the measurement of N/O via N2O2 is not strongly dependent on ionization
parameter and/or the shape of the EUV ionizing radiation field (unlike O32, R23, N2, and O3N2), and {\it is} strongly
dependent on O/H, the most reliable strong-line measurement of O/H (when information is limited) may be obtained through N/O. 
 
\subsubsection{Implications for Strong-Line Measurements}

To summarize, estimates of the oxygen abundances based on the measurement
of various strong line ratios-- anchored by the same \ion{H}{2} region calibration data set-- span
the range ${\rm 8.20 \le 12+log(O/H) \le 8.29}$ depending on the method used.   
All of the strong-line methods 
appear to 
over-estimate, by varying degrees, the $T_{\rm e}$ (CEL) abundance measured in the high redshift LM1 composite. 
Use of the N2 index results in the largest over-estimate ($\simeq 0.15$ dex). Nevertheless, we have shown that
a higher N2 index {\it at fixed N/O and O/H} is expected when 
the EUV spectrum of ionizing stars is systematically harder compared to the calibration sample at the same gas-phase O/H. 
Once this has been accounted for, there is no evidence for redshift-dependent behavior of N/O vs. O/H. 

The comparatively poor performance of ${\rm N2 \equiv [NII]\lambda 6585/\Ha}$ for predicting
the $T_{\rm e}$ (CEL) oxygen abundance of the high redshift composite can be understood
by referring to Figure~\ref{fig:plot_n2o2}: for fixed N/O, increasing the hardness of the radiation
field (e.g. between the S99-v00-z002 model and the BPASSv2-z001 binary models shown) increases the N2 index by $\sim 0.2$ dex while leaving
N2O2 nearly unchanged at fixed O/H. In contrast, the similarity (in both slope and normalization) of eqns.~\ref{eqn:n2} and \ref{eqn:n2o2_met} 
suggests that the
N2 and N2O2 indices are nearly indistinguishable in the way they map to direct oxygen abundances {\it within the calibration sample.} 
Harder ionizing spectra at fixed O/H and N/O at high redshift cause asymmetric shifts in the N2 and N2O2 line indices 
relative
to the calibration sample, causing a systematic shift in the mapping between strong line index and abundance. 
The O3N2 index comes closer than N2 to predicting the $T_{\rm e}$ oxygen abundance because both
O3 and N2 {\it increase} with increasing spectral hardness (at fixed O/H and N/O)-- so that the amplitude of the
systematic shift is reduced simply because the relevant line index is the difference, ${\rm O3 -N2}$.  

It is important to note that the amplitude of the systematic errors (relative to the $T_{\rm e}$/CEL abundances) introduced by the strong-line
methods will depend on the details of the differences in excitation at fixed O/H between the
calibration sample and the sample to which the calibration is applied. 
For reasons discussed in detail in \S\ref{sec:discussion}, the amplitude of systematic errors 
caused by such ``calibration sample mis-match'' is likely to increase with $Z_{\rm neb}/Z_{\odot}$.  


\subsection{C/O}

\label{sec:co}
The use of the rest-UV line ratio ${\rm C3O3}$ (see Table~\ref{tab:properties}) 
as an indicator of the C/O abundance ratio in low metallicity \ion{H}{2} regions was originally proposed by \citet{garnett95a},
and has been employed in a number of studies of high redshift galaxies (e.g., \citealt{shapley03,erb2010}) thanks to the relative
ease with which far-UV spectra of such objects can be obtained.  
In the rest-frame far-UV, the \ion{C}{3}] and \ion{O}{3}] features are close enough in wavelength that they are relatively
immune to uncertainties in extinction, and at the same time C$^{++}$ and O$^{++}$ are expected to be by far the dominant
ions for both C and O, so that ionization corrections needed to infer C/O from ${\rm C^{++}/O^{++}}$
are small and easily modeled.  

Rather than using the approximations given 
by \citet{garnett95a}, we make use of the photoionization grid point that combines the BPASSv2-z001 binary population synthesis 
model with ${\rm log {\it U} = -2.8}$ (Fig.~\ref{fig:plot_gamma}) and oxygen abundance in the range discussed
in the previous section ($Z_{\rm neb}/Z_{\odot} = 0.5\pm0.1$, or ${\rm 12+log(O/H) \simeq 8.4\pm0.1}$); 
this allows us to apply internally
consistent ionization corrections and to include the expected dependence of C3O3 on O/H in estimating the uncertainties. 
We re-ran the photoionization models, varying C/O while holding O/H fixed within the range $Z_{\rm neb}/Z_{\odot} = 0.5\pm0.1$ as
discussed in \S\ref{sec:comparison} and ${\rm log~U}$ fixed at the same value shown in Figure~\ref{fig:plot_gamma}. 
The observed ratio (Table ~\ref{tab:properties})
is ${\rm C3O3_{obs} = 0.62\pm0.07}$, which we find is reproduced by the BPASSv2-z001 binary models when ${\rm log(C/O) = -0.60\pm 0.09}$,
where the error bars include the uncertainties in inferred $Z_{\rm neb}/Z_{\odot}$ as well as the observational uncertainties in the C3O3 index.  
Comparing the inferred C/O with the solar ratio ${\rm log(C/O)_{\odot} = -0.26}$ (\citealt{asplund09}), 
\begin{equation}
{\rm [C/O]  = C3O3_{obs} - C303_{\odot} = -0.34\pm0.09}
\label{eqn:c_over_o}
\end{equation} 

\begin{figure}[tbhp]
\centerline{\includegraphics[width=8.5cm]{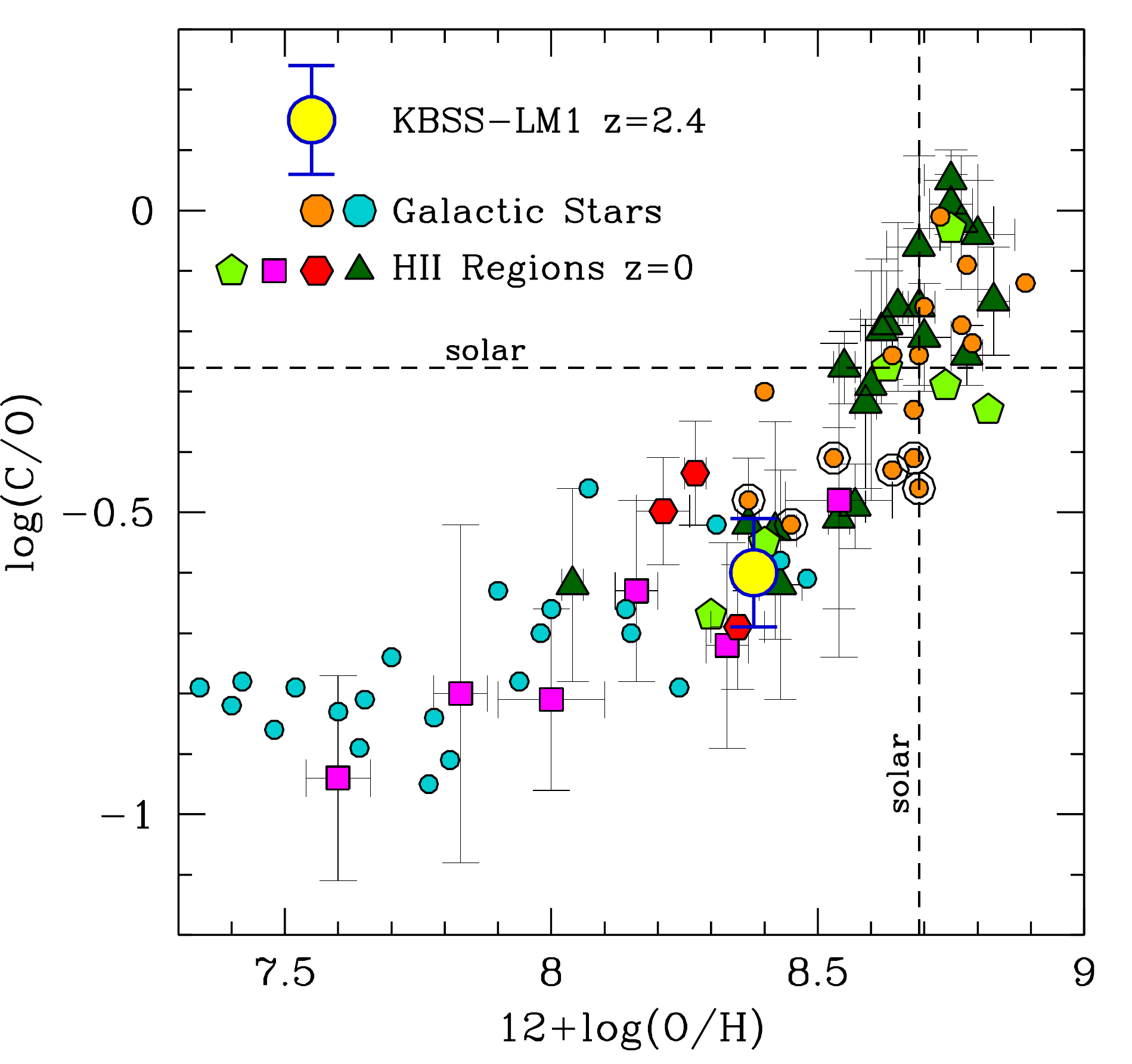}}
\caption{Comparison of log(C/O) vs. 12+log(O/H) for Galactic stars, low metallicity dwarf galaxies, and extragalactic \ion{H}{2} regions
with the measurement from the $z\sim 2.4$ KBSS-LM1 composite (large yellow point).  The low redshift data include metal-poor halo
(\citealt{akerman04}; turquoise points) and ``thick disk'' (\citealt{bensby06}; orange points) stars, dwarf galaxies (\citealt{garnett95a, kobulnicky98}; magenta squares 
and red hexagons, respectively),
local spiral galaxy \ion{H}{2} regions (\citealt{garnett99}; light green pentagons), and measurements of bright emission
line knots in local star-forming galaxies (\citealt{esteban14}; dark green triangles). All nebular values of O/H are based on $T_{\rm e}$ measurements
from collisionally-excited emission lines, and have been adjusted to the REL oxygen abundance scale (presumed to match the {\it stellar} abundance scale) by adding 
${\rm \Delta log(O/H) = 0.24}$ dex (equation~\ref{eqn:offset}; see text for discussion).
Stars with ${\rm [O/Fe] > +0.4}$ from the \citet{bensby06} sample are surrounded by an additional black circle.  
The solar values of log(O/H) and log(C/O) are indicated with dashed lines. }
\label{fig:plot_co}
\end{figure}

Figure~\ref{fig:plot_co} shows that this value is very much in line with the trend of C/O vs. O/H observed in Galactic stars
and \ion{H}{2} regions in nearby dwarf and spiral galaxies. 
Note that the value of log(O/H) for the KBSS-LM1 point has systematic 
uncertainty for the reasons discussed in section~\ref{sec:oh}; the same is true for local \ion{H}{2} region measurements.
For internal consistency, all of the nebular values in 
Figure~\ref{fig:plot_co} are based on $T_{\rm e}$ (CEL) measurements from collisionally excited emission lines, but
have been corrected to the REL abundance scale according to equation~\ref{eqn:offset} by adding 0.24 dex (see the discussion
in \S~\ref{sec:comparison})\footnote{The
values of C/O should require no systematic correction, since nebular measurements of the ratio have only a weak dependence on the assumed oxygen abundance--
see also the discussion in \citet{esteban14}.}.
As discussed by \citet{esteban14}, a similar offset to that given in equation~\ref{eqn:offset} is likely to be present 
between CEL-based direct determinations
and the {\it stellar} abundance scale; thus, we include samples of Galactic halo and thick-disk stars with accurate photospheric abundance
measurements of C/O and O/H in Figure~\ref{fig:plot_co} as well.
Clearly, correcting the nebular measurements to the REL oxygen abundance scale brings them into excellent agreement with
the locus occupied by Galactic stars over the full range in common; in addition, the locus of both stars and nebulae crosses 
${\rm (C/O)_{\odot}}$  at ${\rm (O/H)_{\odot}}$.  

\begin{figure}[htbp]
\centerline{\includegraphics[width=8.5cm]{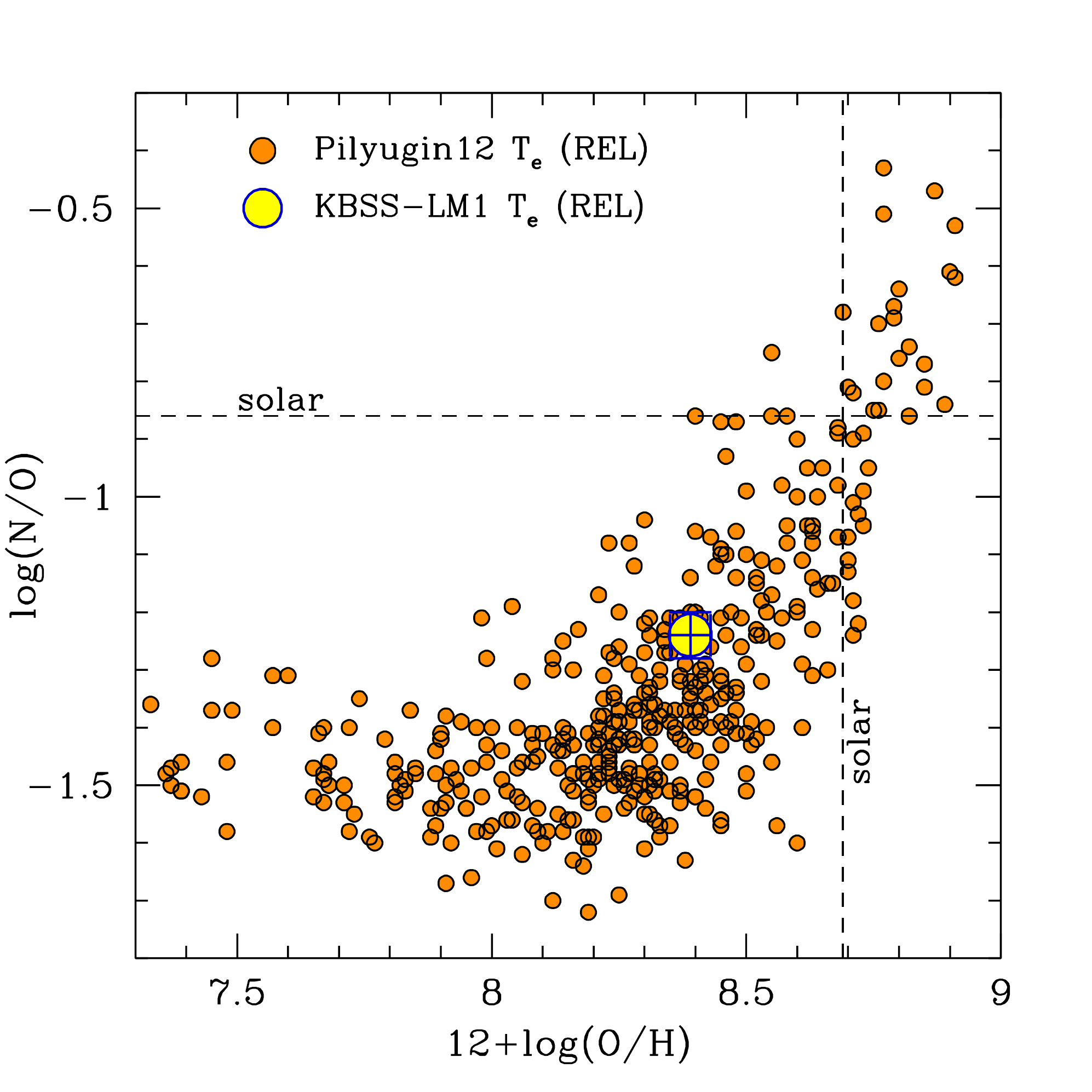}}
\caption{Similar to Figure~\ref{fig:plot_co}, but for N/O vs. 12+log(O/H). As in Figure~\ref{fig:plot_co}, both the local measurements
(in this case, the \citealt{pilyugin12} compilation of extragalactic \ion{H}{2} regions) and the KBSS-LM1 $T_{\rm e}$ measurements
have been shifted by $+0.24$ dex in log(O/H) as in eqn.~\ref{eqn:offset}. 
Recall that the oxygen abundance favored by
the best-fitting photoionization model (\S\ref{sec:n_over_o2}) is identical to the $T_{\rm e}$ oxygen abundance  
after applying this shift from the CEL to REL scale.
The KBSS-LM1 value of N/O is as determined in \S\ref{sec:n_over_o3}. }
\label{fig:plot_pil_rel}
\end{figure}

There have been several previous estimates of C/O at high redshifts: the most directly comparable to the KBSS-LM1
composite is that of \citet{shapley03}, who measured C3O3 from composite
spectra of $z\simeq 3$ LBGs, finding that ${\rm log(C/O)=-0.68\pm0.13}$ for the full galaxy sample and 
${\rm log(C/O) = -0.74\pm0.14}$ for the subset with the strongest \lya\ emission lines. Both values assumed that 
${\rm C/O = C^{++}/O^{++}\times ICF}$ with assumed ${\rm ICF=1.0}$\footnote{The ionization correction
factor implied by the best-fitting model for KBSS-LM1 is ${\rm ICF=0.78}$, which if applied to the \citet{shapley03} results
would {\it lower} the inferred value of ${\rm log(C/O)}$ by $\sim 0.11$ dex.}. \citet{erb2010} derived a value
of ${\rm log(C/O)=-0.62\pm0.09}$ with ${\rm ICF}=1.7\pm0.2$ for a single galaxy, Q2343-BX418 (see also \citealt{steidel14}),
for which a much higher ionization parameter was assumed (log$ U = -1.0\pm0.15$)\footnote{The emission line ratios of Q2343-BX418 
from more recent KBSS-MOSFIRE observations can be reproduced with ${\rm log {\it U} \approx -2.3}$ using the same BPASSv2 population synthesis model
as in the present work, which predicts ${\rm ICF \simeq 0.97}$; a new LRIS observation gives ${\rm C3O3 =0.48\pm0.03}$, and
inferred ${\rm log(C/O) = -0.62\pm0.05}$-- in very good agreement with that of the LM1 composite (Table~\ref{tab:properties}).}.
These previous measurements are all consistent with KBSS-LM1 within their observational and
modeling uncertainties; the new KBSS-LM1 measurement has considerably smaller systematic uncertainties, however, because it is derived based
on a stellar+nebular model that simultaneously reproduces the far-UV and optical line intensity ratios as well as
the far-UV stellar spectrum.

\subsection{N/O}
\label{sec:n_over_o3}

We have already discussed the determination of N/O for the KBSS-LM1 composite in \S\ref{sec:n_over_o} and \S\ref{sec:n_over_o2}, and
have shown that, in spite of the fact that the N2 index over-estimates the $T_{\rm e}$ oxygen abundance by $\sim 0.15$ dex, 
the measured N/O and the $T_e$ (CEL) measurement of O/H are in excellent agreement with the locus occupied by local \ion{H}{2} regions in
the calibration sample (Fig.~\ref{fig:plot_pil_rel} and \S\ref{sec:direct}). 

In view of the results for C/O given in \S\ref{sec:co} above,
it is instructive to cast the N/O measurements in similar terms.   
Figure~\ref{fig:plot_pil_rel} shows the the \citet{pilyugin12} calibration sample and the KBSS-LM1 point after applying the
offset given in eqn.~\ref{eqn:offset} to the measured $T_{\rm e}$ (CEL) oxygen abundances, as was done for C/O and local comparison
samples in Figure~\ref{fig:plot_co}. As was the case for C/O vs. O/H, the shift applied to log(O/H) from eqn.~\ref{eqn:offset}
causes the locus of \ion{H}{2} regions to cross ${\rm (O/H)_{\odot}}$ when ${\rm (N/O)\simeq (N/O)_{\odot}}$.     

The axis scales in Figure~\ref{fig:plot_pil_rel} have been deliberately adjusted to show the same range in 
N/O as for C/O in Figure~\ref{fig:plot_co}, to emphasize the similarity of the behavior of C/O and N/O vs.
O/H. Note that the KBSS-LM1 composite exhibits nearly identical underabundance of C and N with respect to O:
${\rm [C/O]=-0.34\pm0.09}$ compared to ${\rm [N/O]=-0.37\pm0.04}$
(see Table~\ref{tab:properties}). The overall patterns of C/O and N/O are remarkably similar for the
whole range of O/H in common; while the nucleosynthetic origins of C and N are beyond the scope of this
work, comparison of Figs.~\ref{fig:plot_co}~and~\ref{fig:plot_pil_rel} does suggest a common origin
of N and C in which both are closely tied to O/H.   

\subsection{Si/O}

\label{sec:SiO}

As for the C/O abundance determination, we make use of the observed ratio ${\rm Si3O3}$ (Table~\ref{tab:properties})
in conjunction with the most successful photoionization model. The solar ratio of silicon to oxygen according to \citet{asplund09} 
is ${\rm log(Si/O)_{\odot} = -1.18}$, for which the BPASSv2-z002 model (log$U=-2.8$) predicts 
${\rm 0.61 \le Si3O3_{\odot} \le 0.65}$ for $0.4 \le Z_{\rm neb}/Z_{\odot} \le 0.6$. 
The measurement of ${\rm Si3O3_{obs} = 0.00\pm0.08}$ (Table~\ref{tab:properties}) 
thus implies that 
\begin{equation}
{ \rm log(Si/O) = Si3O3_{obs}-Si3O3_{\odot} = -1.81\pm0.10}
\label{eqn:si_over_o} 
\end{equation}
or ${\rm [Si/O] = -0.63\pm0.10}$ in the ionized gas phase of the $z\simeq 2.4$ galaxies.  
Since Si and O are generally believed to be produced by the same processes, the observed [Si/O] under-abundance relative to solar
is likely dominated by depletion of Si onto dust grains. Adopting the definition
\begin{equation}
{\rm [X/Y]_{gas} \equiv log(X/Y)_{neb} - log(X/Y)_{tot}}
\label{eqn:depletion}
\end{equation}
where X and Y are elements, log(X/Y)$_{\rm neb}$ is the nebular (gas-phase) abundance ratio, 
and log(X/Y)$_{\rm tot}$ is the element ISM ratio (gas+dust), we find  
${\rm [Si/H]_{gas} \simeq -0.63}$. This value  
represents an intermediate depletion level relative to the full range seen in Galactic sightlines (${\rm -1.36 \le [Si/H]_{gas} \le -0.22}$; \citealt{jenkins09};
see \S\ref{sec:depletion}).   

Nebular measurements of Si/O at low redshift are few, owing to the need for far-UV spectroscopic observations.  
In an HST/UV spectroscopic study of \ion{H}{2} regions in dwarf galaxies, \citet{garnett95b} found a weighted average 
value ${\rm \langle log(Si/O) \rangle = -1.59\pm0.07}$, and saw no obvious trend with (O/H) over the range 
probed. At the time of that work, the solar ratio ${\rm log(Si/O)_{\odot}}$ was believed to be $\sim 0.2$ dex {\it lower}
than the best current values, which led the authors to speculate that depletion of Si onto dust grains was significantly
lower than expected by then-current grain formation models;  however, with the currently-accepted value of ${\rm (Si/O)_{\odot}}$, 
the dwarf galaxy measurements (${\rm [Si/H]_{gas} \simeq -0.4}$) fall well within the range inferred from 
interstellar absorption line measurements, and have slightly
lower depletion than inferred for the KBSS-LM1 composite. 

\subsection{Effects of Depletion in the Ionized Gas}

\label{sec:depletion}
The inferred depletion of Si onto dust grains (section~\ref{sec:SiO}) can be used to estimate the 
degree to which other elements-- in this case, we are interested primarily in C, N, O, and S-- are likely
to be depleted as well.  Assuming gas-phase depletion ${\rm [Si/H]_{gas}= -0.63}$, the depletion expected
for the other elements can be calculated from the fits given by \citet{jenkins09} with the parameter
$F_{\ast} =0.36\pm0.09$ (see his equation 10 and Table 4) determined from Si:
${\rm [O/H]_{gas} \simeq -0.09}$, ${\rm [C/H]_{gas} \simeq -0.15}$, ${\rm [N/H]_{gas} \simeq -0.11}$, and
${\rm [S/H]_{gas} \simeq -0.15}$, with uncertainties of $\pm0.05$ dex dominated by the scatter in
the calibration.  If these values are approximately correct, then the total oxygen abundance would
be systematically higher by $\sim 0.09$ dex ($\sim 20$\%), but the inferred nebular abundance {\it ratios} C/O, N/O, S/O 
would be essentially unaffected by depletion ($\le 0.05$ dex).   

Note that increasing the nebular oxygen abundance by 0.1 dex would {\it increase} the apparent difference between stellar and nebular
metallicity for the KBSS-LM1 composite spectra. 

\subsection{Abundances: Summary}

$\bullet$ The stellar abundances are traced directly by photospheric and wind lines in the far-UV spectrum, and indirectly by the 
inferred EUV spectrum constrained  
by the nebular line ratios. Both suggest that $Z_{\ast}/Z_{\odot} \simeq 0.10 \pm 0.03$. 

$\bullet$ The nebular oxygen abundance lies in the range $0.3 \le Z_{\rm neb}/Z_{\odot} \le 0.6$ according to the best-fitting photoionization
models, a direct $T_{\rm e}$ measurement of \ion{O}{3}, and several locally-calibrated strong-line estimates. 
However,
an ionized-gas-phase metallicity of $Z_{\rm neb}/Z_{\odot} =0.5$
is favored by the photoionization modeling, which does not rely on the applicability of local calibrations and has been subjected to the strongest constraints
imposed by the requirement of internal consistency. $ Z_{\rm neb}/Z_{\odot}=0.5$ is also inferred from the direct $T_{\rm e}$ measurement
assuming that collisionally-excited emission lines (CELs) under-estimate the true gas-phase oxygen abundance by ${\rm \Delta log (O/H)=0.24}$ dex as
in the local universe. 

$\bullet$ Using far-UV emission lines of \ion{C}{3}] and \ion{O}{3}], whose ratios are insensitive
to extinction corrections and only mildly dependent on O/H, we have measured ${\rm log(C/O) = -0.60\pm0.09}$.   
The measurement is consistent with the observed trend of C/O vs. O/H in Galactic stars and the \ion{H}{2} regions of
nearby galaxies, and provides external support for the inferred O/H. Together, the C/O and O/H measurements
suggest that ${\rm [C/H] \simeq -0.7}$, i.e., ${\rm (C/H) \simeq 0.2 (C/H)_{\odot}}$. 

$\bullet$ The nebular abundance ratio ${\rm [Si/O]=-0.63}$ is presumably due to depletion onto dust grains, suggesting that ${\rm [Si/H]_{gas} \simeq -0.63}$
if ${\rm [Si/O]_{tot} \simeq 0.0}$. The level of Si depletion implies gas-phase depletions of C, N, O, and S of ${\rm [X/H]_{gas} \simeq -0.1}$.
Thus, inferred metallic abundance ratios N/O, C/O, and S/O should be largely unaffected by dust grains. 

\section{Summary and Discussion}
\label{sec:discussion}

In this paper, we have demonstrated that spectroscopic measurements of the rest-frame far-UV stellar continuum, 
{\it and} nebular emission lines in the far-UV ($1000 \simlt \lambda_0 \simlt 2000$ \AA)
and optical ($3700 \simlt \lambda_0 \simlt 7000$ \AA) can be used to extract detailed physical and chemical
properties of stars and ionized gas in high-redshift star-forming galaxies. Perhaps the most significant development
is that the observations can be understood in the context of a self-consistent model of the massive star population 
and ionized gas-phase physical conditions.  

The results for the KBSS-LM1 composite spectra offer a global cross-check on the physical interpretation of 
the most readily-accessible observational signatures of high redshift galaxies, and are almost entirely
independent of calibrations anchored at low redshift.  
Below, we briefly summarize and discuss the most significant results and their implications. 

\subsection{Which Metallicity?} 

\label{sec:which}

We saw above that the most successful model in simultaneously predicting the observed stellar FUV and nebular far-UV+optical
spectrum of the $z=2.4$ KBSS-LM1 composite has $Z_{\ast}/Z_{\odot} \simeq 0.1$ and $Z_{\rm neb}/Z_{\odot} \simeq 0.4-0.5$;
that is, a factor of $\simeq 4-5$ difference between stellar and nebular ``metallicity''! 

However, 
such behavior is actually {\it expected} for galaxies whose ISM has been enriched primarily by Type II (core collapse) supernovae,
by the following argument: 
first, the FUV and EUV spectra of massive stars are most sensitive to the elements dominating the opacity in stellar 
interiors, atmospheres, and winds-- Fe being especially important -- whereas elements acting as important coolants in ionized
gas with $T \simeq10^4$ K, and thus producing strong emission lines-- notably O -- play only a minor role 
within stars. Similarly, in spite of its importance in stars, Fe is depleted onto dust grains in the ISM
and therefore relatively unimportant as a coolant in most \ion{H}{2} regions. 

It is interesting 
to consider the limiting case that stellar opacity and mass loss rates depend only on stellar Fe/H, and nebular 
emission lines only on gas-phase O/H. Assuming constant SFRs, the typical inferred ages of high redshift
star-forming galaxies is a few hundred Myr; since the rate of Fe enrichment of the ISM by Type Ia supernovae
depends on stars that formed some $\sim 1$ Gyr earlier than the epoch at which the massive stars are observed, 
the metals deposited in the ISM over the star-forming lifetime of the galaxy (and thus, the abundances
of massive stars at the epoch of observation) would be dominated by
the nucleosynthetic products of core-collapse supernovae. 
Moreover, it has been argued (e.g., \citealt{papovich11,reddy12})
that typical high-redshift galaxies have SFRs that {\it increase} with time at $z \simgt 2$ -- rising star formation
histories could easily extend the timescale over which core-collapse supernovae dominate the gas-phase enrichment to well
beyond a Gyr, since once again the production
rate of Fe by Type Ia supernovae depends on the galaxy's SFR {\it as it was $\sim 1$ Gyr in the past}.   

According to \citet{nomoto06}, the Salpeter IMF-averaged yields of O and Fe 
from core-collapse supernovae are ${\rm [O/Fe] = (0.74, 0.67, 0.60)}$ dex for initial metallicity
$Z_{\ast} = (0.001, 0.004, 0.020)$. That is, the predicted enhancement in abundance of O relative to Fe is 
${\rm (O/Fe)=(5.5, 4.7, 4.0)\times (O/Fe)_{\odot}}$-- remarkably close to the estimated factor of $\simeq 4-5$ 
difference between $Z_{\rm neb}$ and $Z_{\ast}$ inferred from the observations of KBSS-LM1.      

\begin{figure}[thbp]
\centerline{\includegraphics[width=8.5cm]{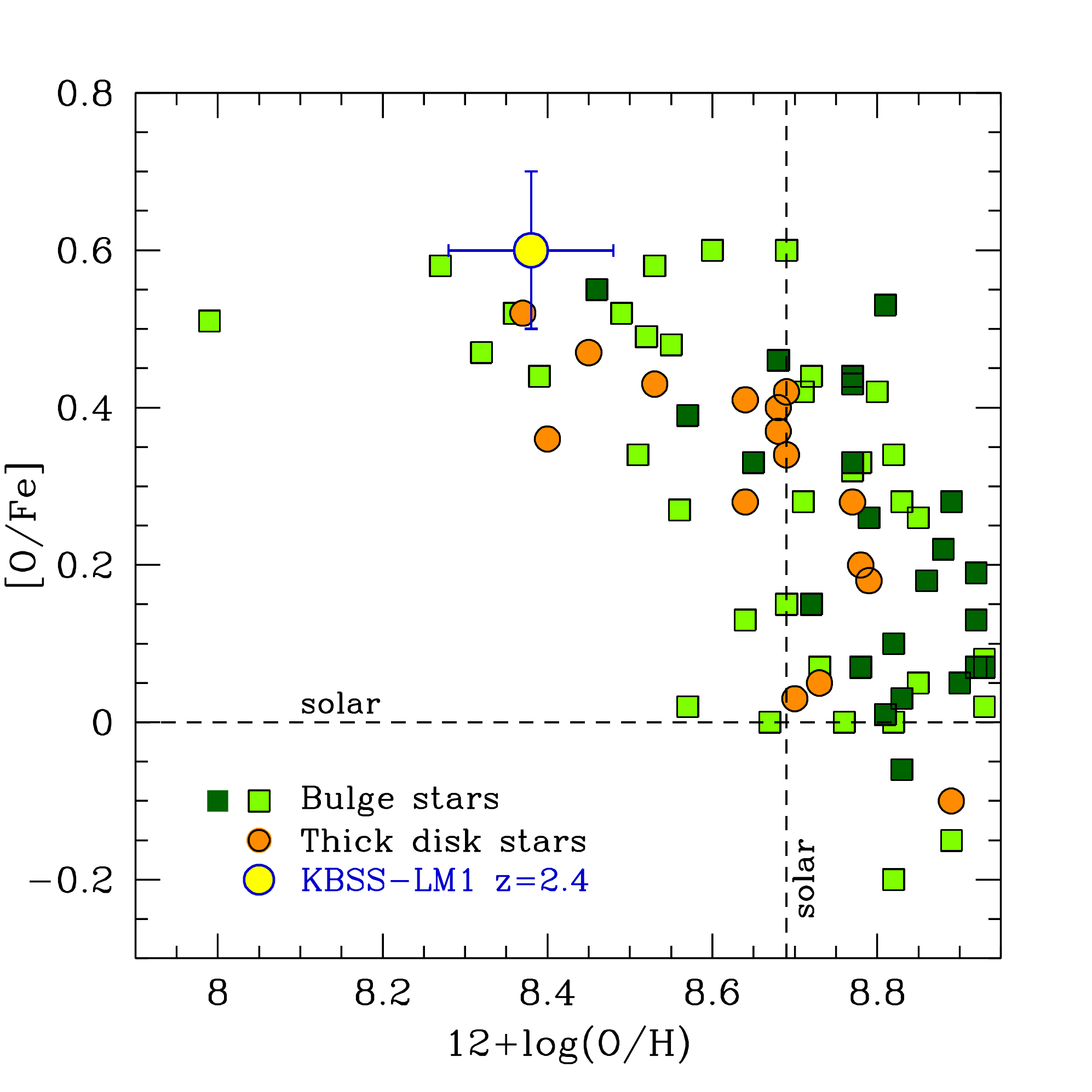}}
\caption{The enhancement in O/Fe relative to solar as a function of {\it stellar} O/H for the
same Galactic thick disk stars as in Fig.~\ref{fig:plot_co} (orange points; \citealt{bensby06}) and for two
samples of Galactic bulge stars:  
micro-lensed dwarfs (light green squares; \citealt{bensby13}) and K-giants (dark green squares; \citealt{lecureur07}). The point
representing KBSS-LM1 (large yellow point) is shown for comparison, assuming ${\rm (O/Fe)\approx(4\pm1)\times (O/Fe)_{\odot}}$ 
and the nebular oxygen abundance (\S\ref{sec:comparison}) from direct $T_{\rm e}$ measurements and photoionization modeling, 
12+log(O/H)$=8.38\pm0.10$ ($Z_{\rm neb}/Z_{\odot} \approx 0.5\pm0.1$). 
}
\label{fig:plot_ofe_oh}
\end{figure}

Figure~\ref{fig:plot_ofe_oh} shows three samples of Galactic stars with measurements of stellar O and Fe abundance, 
where we show the O/Fe abundance ratio relative to solar, [O/Fe], as a function of log(O/H), rather than
the more commonly plotted [X/Fe] versus [Fe/H].  The sample represented by orange points is the same
``thick disk'' sample of \citet{bensby06} as in Fig.~\ref{fig:plot_co}; the squares
are Galactic bulge stars: micro-lensed dwarfs from \citet{bensby13} (light green), and K-giants from \citet{lecureur07} (dark green).  
Also shown is our estimate
for KBSS-LM1 assuming that the inferred $Z_{\ast}$ tracks Fe/H
and that $Z_{\rm neb}$ tracks O/H. The oxygen abundance values for all points in Fig.~\ref{fig:plot_ofe_oh} 
are given using the nebular convention 12+log(O/H)
to emphasize the following point: { \it essentially all of the Galactic stars in the thick disk and bulge that formed out of material
with similar O/H as inferred for the high redshift galaxies are also significantly enhanced in O/Fe.} 
In Fig.~\ref{fig:plot_co}, the thick disk stars with C/O similar to those of KBSS-LM1 
are those with the highest values of O/Fe. Also noteworthy in Fig.~\ref{fig:plot_ofe_oh} is that Galactic thick disk and bulge stars remain 
relatively Fe-poor
even as O/H approaches ${\rm (O/H)_{\odot}}$, suggesting that massive stars forming out of the same gas -- which of course disappeared 
long ago -- would have produced \ion{H}{2} regions of significantly higher excitation than those in 
present-day disks with the same nebular oxygen abundance (see \S\ref{sec:why}).

One important implication, practically speaking, is that in attempting to reproduce the stellar and nebular spectra of high redshift galaxies 
using massive star population synthesis and photoionization models, one should {\it not} assume that
the ``stellar abundance'' and the `'gas-phase abundance'' are the same! 
While it is likely that O/H in the gas phase is very close to O/H in the nearby massive stars, the most 
easily-observed properties of the massive stars -- the far-UV stellar spectra and excitation of the nebular emission --
will appear more like those with much lower overall metallicity. 
Until models of massive stars with {\it non-solar initial abundance ratios} become
available in the future, 
successful models of high redshift systems may require decoupling the nebular and stellar metallicity. 
{\it To a first approximation, the stellar population synthesis model should match the expected abundance of Fe, while the
gas phase abundances should reflect the abundance of O. }

\subsection{Abundance Ratios} 

We have shown that the KBSS-LM1 composite spectra suggest modest under-abundances of N and C relative to O in the ionized gas-- by
$\simeq 0.38$ dex  and $\simeq 0.34$ dex compare to solar, respectively (see Table~\ref{tab:properties}). Both values are
consistent with those of \ion{H}{2} regions in the local
universe with similar values of O/H (Figures~\ref{fig:plot_co},~and~\ref{fig:plot_pil_rel}).  
The C abundance inferred from nebular C/O (\S\ref{sec:co}) and O/H (\S\ref{sec:oh}) 
is in broad agreement with that inferred from the strength of the \ion{C}{4} P-Cygni stellar wind feature, and together
they imply ${\rm C/H \simeq 0.20\pm0.08~(C/H)_{\odot}}$. 
This relatively low abundance of C, which is apparently typical of high-redshift galaxies with bolometric luminosities
${\rm 10^{11}-10^{12}~L_{\odot}}$ from which the KBSS-LM1 sample is drawn, may have significant implications for their detectability in atomic and molecular 
lines by sub-mm/mm observations. {\it In particular, if they were to mimic local galaxies with similar gas-phase C/H, the 
appropriate value of the L(CO)-to-M$_{\rm mol}$ conversion factor $\alpha_{\rm CO}$ could be 1-2 orders of magnitude
higher (i.e., lower L(CO) for a given M$_{\rm mol}$) than applies to more chemically evolved galaxies (see, e.g., \citealt{bolatto13,hunt15}). }

The behavior of N/O among the high redshift galaxy samples has been somewhat controversial 
in the recent literature. Anomalously high N/O at a given O/H has been advocated as the primary driver of the 
N2 BPT ``offset'' (\citealt{masters14,shapley15,sanders15,jones15}). Part of the argument in favor of this interpretation
is the apparent
absence of a similar offset in the S2 BPT diagram. However, we have shown that the location of the LM1 composite
in both diagrams is as expected for normal N/O and N/S (given O/H): the same (normal) N/O is 
obtained by applying the best local calibration (based on direct $T_{\rm e}$ measurements) of 
N2O2$\rightarrow$(N/O) (\S\ref{sec:n_over_o}) and by the photoionization model that reproduces the observed O32 and R23, which do not depend on N/O (\S\ref{sec:R23}) 
. A more thorough treatment of the behavior of N/O versus O/H for the full KBSS-MOSFIRE sample is presented in a companion paper (Strom et al 2016).  

\subsection{Population Synthesis Models}
\label{sec:PSMs_v2}

Population synthesis models without treatment of binary evolution cannot currently produce
nebulae that resemble a large fraction of those observed at high redshift. Single star models cannot produce
steady-state ionizing radiation fields sufficiently hard to produce high [OIII]/\Hb\ and R23 at
fixed O32 (\S\ref{sec:R23}), nor can they produce the observed stellar \ion{He}{2}$\lambda 1640$ emission, unless
the galaxy is an extremely young starburst caught at a particular time post-burst when Wolf-Rayet stars briefly
shine. This post-burst timing argument may be adequate to explain rare dwarf galaxies in the local universe (see, e.g., \citealt{shirazi12}),
but the high redshift examples have $\sim 10-100$ times larger stellar masses and thus much longer dynamical
times that make the required burst timescale uncomfortably short. Perhaps most convincingly, the high redshift galaxies
we have discussed are {\it typical}, not rare, indicating long duty cycles for the periods of high nebular excitation. 
Moreover, we {\it know} that binary evolution of massive stars is
the rule rather than the exception, thus models that do not include it cannot be correct, unless by accident. 

Although the BPASSv2 models highlighted above work very well to explain the observed nebular emission and
reasonably well for the stellar far-UV continuum, it does not necessarily follow that the details of the model -- many of which
remain uncertain-- are correct. {\it However, we can say that the ionizing spectra of the stars that
dominate the nebular excitation in typical star forming galaxies at $z \sim 2.4$ must have EUV spectra 
similar to those of the best-matching BPASSv2 models (see Figure~\ref{fig:euv_spectra}.)}.   


\subsection{What (if anything) is different at high redshift? }
\label{sec:why}

Although analogs of galaxies that comprise the KBSS-LM1 composite 
can be found at all redshifts (e.g., \citealt{shirazi14,brown14,steidel14,jones15,bian16}),
galaxies with very high specific star formation rates and high excitation
are common at
high-redshift 
but rare in the relatively local universe. 
Rarer still are nearby galaxies exhibiting these properties with  
${\rm M_{\ast}}$ (${\rm  \sim 10^{10}~M_{\odot}}$) and SFR (${\rm \sim 30~M_{\odot}yr^{-1}}$)
comparable to the galaxies comprising the LM1 sample (Table~\ref{tab:sample}). 
From the standpoint of nebular astrophysics, it is likely that the low-redshift analogs 
can be explained in a manner similar to those we have advocated for the high-redshift
systems: hard EUV ionizing spectra and moderately sub-solar oxygen abundances. But the
presence of large numbers of relatively high mass galaxies with high excitation nebulae
requires more explanation, since one cannot rely on the same duty cycle arguments that might account
for a rare population of objects as at low redshift. 

We have argued above that galaxies whose ISM metallicity
is dominated by metals produced within the past $\sim$Gyr by core-collapse supernovae 
can attain near-solar O/H while still manifesting massive star population
properties 
normally associated with much lower metallicities. The time period over which ISM enrichment
remains dominated by Type II SNe yields can easily extend over longer periods 
for a given galaxy if its SFR continues to increase as a function of time as expected
for galaxies during their most rapid period of growth.
During this phase, 
newly-formed OB stars  
will continue to appear ``chemically young'', since the production of O always outpaces that of Fe. 
The low inferred Fe/H in the  
active O-star population results in weak stellar winds, lower mass loss rates, and less photospheric line blanketing,
while massive stars in binaries are much more likely to undergo quasi-homogeneous evolution (QHE) for similar reasons (see \S\ref{sec:intro})-- 
all of which favor harder EUV ionizing spectra.   

As mentioned in \S\ref{sec:which} above, Figure~\ref{fig:plot_ofe_oh} suggests that these effects might well have been important at the 
time of formation of the Galactic bulge and thick disk-- enhanced [O/Fe] 
is observed in all bulge and thick disk stars with ${\rm (O/H) \simlt (O/H)_{\odot}}$. Contemporaneous massive
stars forming out of the same ISM material with 
${\rm (O/H) \simlt (O/H)_{\odot}}$ would have had ${\rm (Fe/H)\simlt 0.25 (Fe/H)_{\odot}}$; in terms of the parametrization of
stellar metallicity in the population synthesis models discussed above, this is equivalent to $Z_{\ast} \simlt 0.0035$-- low
enough to allow for the most dramatic effects of massive binary evolution (including QHE).  
For this reason, even at solar gas-phase oxygen abundance, the \ion{H}{2} regions excited by such massive stars 
would be observed to have much higher excitation than a typical \ion{H}{2} region in a present-day
Galactic disk, where ${\rm [O/Fe]} \simeq 0$. 
Similarly, galaxies
with declining star formation histories and/or low sSFRs (i.e., where
the current star formation is a small perturbation on the past integral) are the rule at low redshift; these 
would be much less likely to have super-solar O/Fe, and, at a given O/H,  would have softer EUV spectra
due to increased photospheric line blanketing. Because of much higher mass loss rates, 
the probability of quasi-homogeneous evolution (QHE) in massive binaries would be dramatically reduced (e.g., \citealt{eldridge11,brott11}). 

It seems likely that a similar qualitative argument might explain the otherwise puzzling observation that, at $z\sim 2.3$, 
the correlation between strong line indices sensitive to nebular excitation (e.g., O3, O3N2, R32) and ${\rm M_{\ast}}$ is {\it tighter} than 
could reasonably be expected even from a ``stellar mass-metallicity'' relation (MZR) with zero intrinsic scatter, given the uncertainties
in mapping the line indices to oxygen abundance (e.g. equation~\ref{eqn:o3n2})-- 
see the discussion in \S8 of \citet{steidel14}.  
Excitation is controlled primarily by the spectrum of the ionizing stars, and we have argued that the EUV is modulated
by the abundance of Fe, and not O.  The enrichment pattern of O/Fe in the ISM of star forming galaxies depends very strongly on
the star formation history.  Galaxies with increasingly high ${\rm M_{\ast}}$ have lower sSFR and longer star formation histories where an
increasing fraction of the stellar mass was accumulated in the distant past. O-stars forming in an increasingly Fe-rich ISM will 
be less capable of producing high excitation nebular emission lines; thus, the oldest (and most massive) galaxies
have the lowest-excitation nebulae almost independently of the nebular oxygen abundance. An obvious corollary is that 
strong-line indices measured in the local universe may be systematically different from those measured 
at high redshift {\it for the same gas-phase O/H}; conversely, the same excitation may be observed at systematically different O/H 
if the star-formation histories (or, equivalently, the O/Fe abundance pattern) of any two samples are significantly different.   

\subsection{Suggestions for Future Work} 

In this paper we have shown that a consistent interpretation of the most easily-observed physical properties
of high redshift galaxies can be obtained by simultaneously modeling the EUV-FUV spectra of the stars and
the nebular emission lines from the same galaxies. Among the areas where future work would lead to measurable improvements
in our understanding of stellar populations, chemistry, ISM physics, and other physical properties of high redshift star-forming galaxies  
include:

$\bullet$ Construction of stellar population synthesis models that include all of the following a) binary evolution of massive stars b) a range of non-solar
stellar abundance ratios (particularly varying Fe/O, C/O, and N/O) and c) detailed, moderate to high resolution FUV spectra for direct
comparison with observations.  

$\bullet$ Creation of a streamlined method for simultaneous ``fitting'' of FUV stellar spectra and FUV-optical nebular spectra of the same 
objects or samples. This would involve a more comprehensive grid of stellar and nebular abundances and abundance patterns than currently exists,
and will eventually enable direct constraints on the physics and chemistry of the ISM, forming massive stars, and ionized nebulae at high redshift--independent
of any assumptions imposed by our understanding at low-redshift. 

$\bullet$ Assembly of very deep rest-frame far-UV spectra for large samples of galaxies which also have excellent nebular measurements from the new
generation of near-IR multi-object spectrographs. The results of our pilot program have been a first step in that direction. 

It is worth emphasizing that progress made in these areas will be immediately relevant to the planning and interpretation of future observations of
distant galaxies by the {\it James Webb Space Telescope}, which will be observing the same far-UV continuum and far-UV/optical nebular
spectra of galaxies beyond $z \sim 4$ and into the reionization era.

\acknowledgments
We thank Jarle Brinchmann, J.J. Eldridge, Elizabeth Stanway, Eric Pellegrini, Selma de Mink, Evan Kirby, Jim Fuller, Phil Hopkins, 
and Maryam Shirazi for very informative conversations. 
We benefited significantly from talks, discussions, and conversations during the First Carnegie Symposium in Honor of Leonard Searle, ``Understanding
Nebular Emission in High-Redshift Galaxies'', held at the Carnegie Observatories in Pasadena, 2015 July 13-17.  
T. Bensby, C. Esteban, and D. Erb kindly provided machine-readable data from their published work. 
This work has been supported in part by the US National Science Foundation through grants
AST-0908805 and AST-1313472 (CCS, ALS, RFT), as well as by an NSF Graduate Student Research Fellowship (ALS). 
Finally, we wish to extend thanks to those of Hawaiian ancestry on whose sacred mountain we are privileged
to be guests.

\bibliographystyle{apj}
\bibliography{refs_comb}

\begin{thebibliography}{}
\expandafter\ifx\csname natexlab\endcsname\relax\def\natexlab#1{#1}\fi

\bibitem[{{Akerman} {et~al.}(2004){Akerman}, {Carigi}, {Nissen}, {Pettini}, \&
  {Asplund}}]{akerman04}
{Akerman}, C.~J., {Carigi}, L., {Nissen}, P.~E., {Pettini}, M., \& {Asplund},
  M. 2004, \aap, 414, 931

\bibitem[{{Andrews} \& {Martini}(2013)}]{andrews+martini13}
{Andrews}, B.~H., \& {Martini}, P. 2013, \apj, 765, 140

\bibitem[{{Asplund} {et~al.}(2009){Asplund}, {Grevesse}, {Sauval}, \&
  {Scott}}]{asplund09}
{Asplund}, M., {Grevesse}, N., {Sauval}, A.~J., \& {Scott}, P. 2009, \araa, 47,
  481

\bibitem[{{Bensby} \& {Feltzing}(2006)}]{bensby06}
{Bensby}, T., \& {Feltzing}, S. 2006, \mnras, 367, 1181

\bibitem[{{Bensby} {et~al.}(2013){Bensby}, {Yee}, {Feltzing}, {Johnson},
  {Gould}, {Cohen}, {Asplund}, {Mel{\'e}ndez}, {Lucatello}, {Han}, {Thompson},
  {Gal-Yam}, {Udalski}, {Bennett}, {Bond}, {Kohei}, {Sumi}, {Suzuki}, {Suzuki},
  {Takino}, {Tristram}, {Yamai}, \& {Yonehara}}]{bensby13}
{Bensby}, T., {Yee}, J.~C., {Feltzing}, S., {et~al.} 2013, \aap, 549, A147

\bibitem[{{Berg} {et~al.}(2015){Berg}, {Skillman}, {Croxall}, {Pogge},
  {Moustakas}, \& {Johnson-Groh}}]{berg15}
{Berg}, D.~A., {Skillman}, E.~D., {Croxall}, K.~V., {et~al.} 2015, \apj, 806,
  16

\bibitem[{{Bian} {et~al.}(2016){Bian}, {Kewley}, {Dopita}, \&
  {Juneau}}]{bian16}
{Bian}, F., {Kewley}, L., {Dopita}, M., \& {Juneau}, S. 2016, ArXiv e-prints,
  arXiv:1603.05275

\bibitem[{{Blanc} {et~al.}(2015){Blanc}, {Kewley}, {Vogt}, \&
  {Dopita}}]{blanc15}
{Blanc}, G.~A., {Kewley}, L., {Vogt}, F.~P.~A., \& {Dopita}, M.~A. 2015, \apj,
  798, 99

\bibitem[{{Bolatto} {et~al.}(2013){Bolatto}, {Wolfire}, \& {Leroy}}]{bolatto13}
{Bolatto}, A.~D., {Wolfire}, M., \& {Leroy}, A.~K. 2013, \araa, 51, 207

\bibitem[{{Brott} {et~al.}(2011){Brott}, {de Mink}, {Cantiello}, {Langer}, {de
  Koter}, {Evans}, {Hunter}, {Trundle}, \& {Vink}}]{brott11}
{Brott}, I., {de Mink}, S.~E., {Cantiello}, M., {et~al.} 2011, \aap, 530, A115

\bibitem[{{Brown} {et~al.}(2014){Brown}, {Croxall}, \& {Pogge}}]{brown14}
{Brown}, J.~S., {Croxall}, K.~V., \& {Pogge}, R.~W. 2014, \apj, 792, 140

\bibitem[{{Bruzual} \& {Charlot}(2003)}]{bc03}
{Bruzual}, G., \& {Charlot}, S. 2003, \mnras, 344, 1000

\bibitem[{{Calzetti} {et~al.}(2000){Calzetti}, {Armus}, {Bohlin}, {Kinney},
  {Koornneef}, \& {Storchi-Bergmann}}]{calzetti00}
{Calzetti}, D., {Armus}, L., {Bohlin}, R.~C., {et~al.} 2000, \apj, 533, 682

\bibitem[{{Campbell} {et~al.}(1986){Campbell}, {Terlevich}, \&
  {Melnick}}]{campbell86}
{Campbell}, A., {Terlevich}, R., \& {Melnick}, J. 1986, \mnras, 223, 811

\bibitem[{{Cantiello} {et~al.}(2007){Cantiello}, {Yoon}, {Langer}, \&
  {Livio}}]{cantiello07}
{Cantiello}, M., {Yoon}, S.-C., {Langer}, N., \& {Livio}, M. 2007, \aap, 465,
  L29

\bibitem[{{Cardelli} {et~al.}(1989){Cardelli}, {Clayton}, \&
  {Mathis}}]{cardelli89}
{Cardelli}, J.~A., {Clayton}, G.~C., \& {Mathis}, J.~S. 1989, \apj, 345, 245

\bibitem[{{Chabrier}(2003)}]{chabrier03}
{Chabrier}, G. 2003, \pasp, 115, 763

\bibitem[{{Crowther}(2007)}]{crowther07}
{Crowther}, P.~A. 2007, \araa, 45, 177

\bibitem[{{Crowther} {et~al.}(2006){Crowther}, {Prinja}, {Pettini}, \&
  {Steidel}}]{crowther06}
{Crowther}, P.~A., {Prinja}, R.~K., {Pettini}, M., \& {Steidel}, C.~C. 2006,
  \mnras, 368, 895

\bibitem[{{Crowther} {et~al.}(2016){Crowther}, {Caballero-Nieves}, {Bostroem},
  {Ma{\'{\i}}z Apell{\'a}niz}, {Schneider}, {Walborn}, {Angus}, {Brott},
  {Bonanos}, {de Koter}, {de Mink}, {Evans}, {Gr{\"a}fener}, {Herrero},
  {Howarth}, {Langer}, {Lennon}, {Puls}, {Sana}, \& {Vink}}]{crowther16}
{Crowther}, P.~A., {Caballero-Nieves}, S.~M., {Bostroem}, K.~A., {et~al.} 2016,
  \mnras, 458, 624

\bibitem[{{Eldridge} {et~al.}(2011){Eldridge}, {Langer}, \&
  {Tout}}]{eldridge11}
{Eldridge}, J.~J., {Langer}, N., \& {Tout}, C.~A. 2011, \mnras, 414, 3501

\bibitem[{{Eldridge} \& {Stanway}(2012)}]{eldridge12}
{Eldridge}, J.~J., \& {Stanway}, E.~R. 2012, \mnras, 419, 479

\bibitem[{{Eldridge} \& {Stanway}(2016)}]{eldridge16}
---. 2016, ArXiv e-prints, arXiv:1602.03790

\bibitem[{{Erb} {et~al.}(2010){Erb}, {Pettini}, {Shapley}, {Steidel}, {Law}, \&
  {Reddy}}]{erb2010}
{Erb}, D.~K., {Pettini}, M., {Shapley}, A.~E., {et~al.} 2010, \apj, 719, 1168

\bibitem[{{Erb} {et~al.}(2006){Erb}, {Steidel}, {Shapley}, {Pettini}, {Reddy},
  \& {Adelberger}}]{erb+06b}
{Erb}, D.~K., {Steidel}, C.~C., {Shapley}, A.~E., {et~al.} 2006, \apj, 646, 107

\bibitem[{{Esteban} {et~al.}(2014){Esteban}, {Garc{\'{\i}}a-Rojas}, {Carigi},
  {Peimbert}, {Bresolin}, {L{\'o}pez-S{\'a}nchez}, \&
  {Mesa-Delgado}}]{esteban14}
{Esteban}, C., {Garc{\'{\i}}a-Rojas}, J., {Carigi}, L., {et~al.} 2014, \mnras,
  443, 624

\bibitem[{{Faucher-Gigu{\`e}re} {et~al.}(2013){Faucher-Gigu{\`e}re},
  {Quataert}, \& {Hopkins}}]{fauch13}
{Faucher-Gigu{\`e}re}, C.-A., {Quataert}, E., \& {Hopkins}, P.~F. 2013, \mnras,
  433, 1970

\bibitem[{{Ferland} {et~al.}(2013){Ferland}, {Porter}, {van Hoof}, {Williams},
  {Abel}, {Lykins}, {Shaw}, {Henney}, \& {Stancil}}]{ferland13}
{Ferland}, G.~J., {Porter}, R.~L., {van Hoof}, P.~A.~M., {et~al.} 2013, RMxAA,
  49, 137

\bibitem[{{Garnett}(1992)}]{garnett92}
{Garnett}, D.~R. 1992, \aj, 103, 1330

\bibitem[{{Garnett} {et~al.}(1995{\natexlab{a}}){Garnett}, {Dufour},
  {Peimbert}, {Torres-Peimbert}, {Shields}, {Skillman}, {Terlevich}, \&
  {Terlevich}}]{garnett95b}
{Garnett}, D.~R., {Dufour}, R.~J., {Peimbert}, M., {et~al.} 1995{\natexlab{a}},
  \apjl, 449, L77

\bibitem[{{Garnett} {et~al.}(1999){Garnett}, {Shields}, {Peimbert},
  {Torres-Peimbert}, {Skillman}, {Dufour}, {Terlevich}, \&
  {Terlevich}}]{garnett99}
{Garnett}, D.~R., {Shields}, G.~A., {Peimbert}, M., {et~al.} 1999, \apj, 513,
  168

\bibitem[{{Garnett} {et~al.}(1995{\natexlab{b}}){Garnett}, {Skillman},
  {Dufour}, {Peimbert}, {Torres-Peimbert}, {Terlevich}, {Terlevich}, \&
  {Shields}}]{garnett95a}
{Garnett}, D.~R., {Skillman}, E.~D., {Dufour}, R.~J., {et~al.}
  1995{\natexlab{b}}, \apj, 443, 64

\bibitem[{{Gonz{\'a}lez Delgado} {et~al.}(2002){Gonz{\'a}lez Delgado},
  {Leitherer}, {Stasi{\'n}ska}, \& {Heckman}}]{gonzalez02}
{Gonz{\'a}lez Delgado}, R.~M., {Leitherer}, C., {Stasi{\'n}ska}, G., \&
  {Heckman}, T.~M. 2002, \apj, 580, 824

\bibitem[{{Gonz{\'a}lez Delgado} \& {P{\'e}rez}(2000)}]{gonzalez2000}
{Gonz{\'a}lez Delgado}, R.~M., \& {P{\'e}rez}, E. 2000, \mnras, 317, 64

\bibitem[{{Gordon} {et~al.}(2003){Gordon}, {Clayton}, {Misselt}, {Landolt}, \&
  {Wolff}}]{gordon03}
{Gordon}, K.~D., {Clayton}, G.~C., {Misselt}, K.~A., {Landolt}, A.~U., \&
  {Wolff}, M.~J. 2003, \apj, 594, 279

\bibitem[{{Hopkins} {et~al.}(2014){Hopkins}, {Kere{\v s}}, {O{\~n}orbe},
  {Faucher-Gigu{\`e}re}, {Quataert}, {Murray}, \& {Bullock}}]{hopkins14}
{Hopkins}, P.~F., {Kere{\v s}}, D., {O{\~n}orbe}, J., {et~al.} 2014, \mnras,
  445, 581

\bibitem[{{Hunt} {et~al.}(2015){Hunt}, {Garc{\'{\i}}a-Burillo}, {Casasola},
  {Caselli}, {Combes}, {Henkel}, {Lundgren}, {Maiolino}, {Menten}, {Testi}, \&
  {Weiss}}]{hunt15}
{Hunt}, L.~K., {Garc{\'{\i}}a-Burillo}, S., {Casasola}, V., {et~al.} 2015,
  \aap, 583, A114

\bibitem[{{Izotov} {et~al.}(2006){Izotov}, {Stasi{\'n}ska}, {Meynet}, {Guseva},
  \& {Thuan}}]{izotov06}
{Izotov}, Y.~I., {Stasi{\'n}ska}, G., {Meynet}, G., {Guseva}, N.~G., \&
  {Thuan}, T.~X. 2006, \aap, 448, 955

\bibitem[{{James} {et~al.}(2014){James}, {Pettini}, {Christensen}, {Auger},
  {Becker}, {King}, {Quider}, {Shapley}, \& {Steidel}}]{james14}
{James}, B.~L., {Pettini}, M., {Christensen}, L., {et~al.} 2014, \mnras, 440,
  1794

\bibitem[{{Jenkins}(2009)}]{jenkins09}
{Jenkins}, E.~B. 2009, \apj, 700, 1299

\bibitem[{{Jones} {et~al.}(2015){Jones}, {Martin}, \& {Cooper}}]{jones15}
{Jones}, T., {Martin}, C., \& {Cooper}, M.~C. 2015, \apj, 813, 126

\bibitem[{{Kashino} {et~al.}(2013){Kashino}, {Silverman}, {Rodighiero},
  {Renzini}, {Arimoto}, {Daddi}, {Lilly}, {Sanders}, {Kartaltepe}, {Zahid},
  {Nagao}, {Sugiyama}, {Capak}, {Carollo}, {Chu}, {Hasinger}, {Ilbert},
  {Kajisawa}, {Kewley}, {Koekemoer}, {Kova{\v c}}, {Le F{\`e}vre}, {Masters},
  {McCracken}, {Onodera}, {Scoville}, {Strazzullo}, {Symeonidis}, \&
  {Taniguchi}}]{kashino13}
{Kashino}, D., {Silverman}, J.~D., {Rodighiero}, G., {et~al.} 2013, \apjl, 777,
  L8

\bibitem[{{Kewley} \& {Dopita}(2002)}]{kewley02}
{Kewley}, L.~J., \& {Dopita}, M.~A. 2002, \apjs, 142, 35

\bibitem[{{Kobulnicky} {et~al.}(1999){Kobulnicky}, {Kennicutt}, \&
  {Pizagno}}]{kobulnicky99}
{Kobulnicky}, H.~A., {Kennicutt}, Jr., R.~C., \& {Pizagno}, J.~L. 1999, \apj,
  514, 544

\bibitem[{{Kobulnicky} \& {Skillman}(1998)}]{kobulnicky98}
{Kobulnicky}, H.~A., \& {Skillman}, E.~D. 1998, \apj, 497, 601

\bibitem[{{Kroupa}(2001)}]{kroupa01}
{Kroupa}, P. 2001, \mnras, 322, 231

\bibitem[{{Kudritzki} \& {Puls}(2000)}]{kudritzki+puls2000}
{Kudritzki}, R.-P., \& {Puls}, J. 2000, \araa, 38, 613

\bibitem[{{Langer}(2012)}]{langer12}
{Langer}, N. 2012, \araa, 50, 107

\bibitem[{{Lecureur} {et~al.}(2007){Lecureur}, {Hill}, {Zoccali}, {Barbuy},
  {G{\'o}mez}, {Minniti}, {Ortolani}, \& {Renzini}}]{lecureur07}
{Lecureur}, A., {Hill}, V., {Zoccali}, M., {et~al.} 2007, \aap, 465, 799

\bibitem[{{Leitherer} {et~al.}(2014){Leitherer}, {Ekstr{\"o}m}, {Meynet},
  {Schaerer}, {Agienko}, \& {Levesque}}]{leitherer14}
{Leitherer}, C., {Ekstr{\"o}m}, S., {Meynet}, G., {et~al.} 2014, \apjs, 212, 14

\bibitem[{{Leitherer} {et~al.}(2001){Leitherer}, {Le{\~a}o}, {Heckman},
  {Lennon}, {Pettini}, \& {Robert}}]{leitherer01}
{Leitherer}, C., {Le{\~a}o}, J.~R.~S., {Heckman}, T.~M., {et~al.} 2001, \apj,
  550, 724

\bibitem[{{Leitherer} {et~al.}(2010){Leitherer}, {Ortiz Ot{\'a}lvaro},
  {Bresolin}, {Kudritzki}, {Lo Faro}, {Pauldrach}, {Pettini}, \&
  {Rix}}]{leitherer10}
{Leitherer}, C., {Ortiz Ot{\'a}lvaro}, P.~A., {Bresolin}, F., {et~al.} 2010,
  \apjs, 189, 309

\bibitem[{{Massey} {et~al.}(1988){Massey}, {Strobel}, {Barnes}, \&
  {Anderson}}]{massey88}
{Massey}, P., {Strobel}, K., {Barnes}, J.~V., \& {Anderson}, E. 1988, \apj,
  328, 315

\bibitem[{{Masters} {et~al.}(2014){Masters}, {McCarthy}, {Siana}, {Malkan},
  {Mobasher}, {Atek}, {Henry}, {Martin}, {Rafelski}, {Hathi}, {Scarlata},
  {Ross}, {Bunker}, {Blanc}, {Bedregal}, {Dom{\'{\i}}nguez}, {Colbert},
  {Teplitz}, \& {Dressler}}]{masters14}
{Masters}, D., {McCarthy}, P., {Siana}, B., {et~al.} 2014, \apj, 785, 153

\bibitem[{{McGaugh}(1991)}]{mcgaugh91}
{McGaugh}, S.~S. 1991, \apj, 380, 140

\bibitem[{{Muratov} {et~al.}(2015){Muratov}, {Kere{\v s}},
  {Faucher-Gigu{\`e}re}, {Hopkins}, {Quataert}, \& {Murray}}]{muratov15}
{Muratov}, A.~L., {Kere{\v s}}, D., {Faucher-Gigu{\`e}re}, C.-A., {et~al.}
  2015, \mnras, 454, 2691

\bibitem[{{Murray} {et~al.}(2005){Murray}, {Quataert}, \&
  {Thompson}}]{murray05}
{Murray}, N., {Quataert}, E., \& {Thompson}, T.~A. 2005, \apj, 618, 569

\bibitem[{{Nomoto} {et~al.}(2006){Nomoto}, {Tominaga}, {Umeda}, {Kobayashi}, \&
  {Maeda}}]{nomoto06}
{Nomoto}, K., {Tominaga}, N., {Umeda}, H., {Kobayashi}, C., \& {Maeda}, K.
  2006, Nuclear Physics A, 777, 424

\bibitem[{{Oke} {et~al.}(1995){Oke}, {Cohen}, {Carr}, {Cromer}, {Dingizian},
  {Harris}, {Labrecque}, {Lucinio}, {Schaal}, {Epps}, \& {Miller}}]{oke95}
{Oke}, J.~B., {Cohen}, J.~G., {Carr}, M., {et~al.} 1995, \pasp, 107, 375

\bibitem[{{Osterbrock} \& {Ferland}(2006)}]{osterbrock06}
{Osterbrock}, D.~E., \& {Ferland}, G.~J. 2006, {Astrophysics of gaseous nebulae
  and active galactic nuclei} (University Science Books)

\bibitem[{{Papovich} {et~al.}(2011){Papovich}, {Finkelstein}, {Ferguson},
  {Lotz}, \& {Giavalisco}}]{papovich11}
{Papovich}, C., {Finkelstein}, S.~L., {Ferguson}, H.~C., {Lotz}, J.~M., \&
  {Giavalisco}, M. 2011, \mnras, 412, 1123

\bibitem[{{Peimbert} \& {Peimbert}(2002)}]{peimbert02}
{Peimbert}, M., \& {Peimbert}, A. 2002, in Revista Mexicana de Astronomia y
  Astrofisica, vol. 27, Vol.~14, Revista Mexicana de Astronomia y Astrofisica
  Conference Series, ed. J.~J. {Claria}, D.~{Garcia Lambas}, \& H.~{Levato},
  47--52

\bibitem[{{P{\'e}rez-Montero} \& {Contini}(2009)}]{pmc09}
{P{\'e}rez-Montero}, E., \& {Contini}, T. 2009, \mnras, 398, 949

\bibitem[{{Pettini} \& {Pagel}(2004)}]{pettini04}
{Pettini}, M., \& {Pagel}, B.~E.~J. 2004, \mnras, 348, L59

\bibitem[{{Pettini} {et~al.}(2002){Pettini}, {Rix}, {Steidel}, {Adelberger},
  {Hunt}, \& {Shapley}}]{pettini02}
{Pettini}, M., {Rix}, S.~A., {Steidel}, C.~C., {et~al.} 2002, \apj, 569, 742

\bibitem[{{Pettini} {et~al.}(2000){Pettini}, {Steidel}, {Adelberger},
  {Dickinson}, \& {Giavalisco}}]{pettini00}
{Pettini}, M., {Steidel}, C.~C., {Adelberger}, K.~L., {Dickinson}, M., \&
  {Giavalisco}, M. 2000, \apj, 528, 96

\bibitem[{{Pilyugin} {et~al.}(2012){Pilyugin}, {Grebel}, \&
  {Mattsson}}]{pilyugin12}
{Pilyugin}, L.~S., {Grebel}, E.~K., \& {Mattsson}, L. 2012, \mnras, 424, 2316

\bibitem[{{Pilyugin} {et~al.}(2009){Pilyugin}, {Mattsson}, {V{\'{\i}}lchez}, \&
  {Cedr{\'e}s}}]{pilyugin09}
{Pilyugin}, L.~S., {Mattsson}, L., {V{\'{\i}}lchez}, J.~M., \& {Cedr{\'e}s}, B.
  2009, \mnras, 398, 485

\bibitem[{{Price} {et~al.}(2014){Price}, {Kriek}, {Brammer}, {Conroy},
  {F{\"o}rster Schreiber}, {Franx}, {Fumagalli}, {Lundgren}, {Momcheva},
  {Nelson}, {Skelton}, {van Dokkum}, {Whitaker}, \& {Wuyts}}]{price14}
{Price}, S.~H., {Kriek}, M., {Brammer}, G.~B., {et~al.} 2014, \apj, 788, 86

\bibitem[{{Puls} {et~al.}(2008){Puls}, {Vink}, \& {Najarro}}]{puls08}
{Puls}, J., {Vink}, J.~S., \& {Najarro}, F. 2008, \aapr, 16, 209

\bibitem[{{Quider} {et~al.}(2009){Quider}, {Pettini}, {Shapley}, \&
  {Steidel}}]{quider09}
{Quider}, A.~M., {Pettini}, M., {Shapley}, A.~E., \& {Steidel}, C.~C. 2009,
  \mnras, 398, 1263

\bibitem[{{Quider} {et~al.}(2010){Quider}, {Shapley}, {Pettini}, {Steidel}, \&
  {Stark}}]{quider10}
{Quider}, A.~M., {Shapley}, A.~E., {Pettini}, M., {Steidel}, C.~C., \& {Stark},
  D.~P. 2010, \mnras, 402, 1467

\bibitem[{{Reddy} {et~al.}(2012){Reddy}, {Pettini}, {Steidel}, {Shapley},
  {Erb}, \& {Law}}]{reddy12}
{Reddy}, N.~A., {Pettini}, M., {Steidel}, C.~C., {et~al.} 2012, \apj, 754, 25

\bibitem[{{Reddy} {et~al.}(2008){Reddy}, {Steidel}, {Pettini}, {Adelberger},
  {Shapley}, {Erb}, \& {Dickinson}}]{reddy08a}
{Reddy}, N.~A., {Steidel}, C.~C., {Pettini}, M., {et~al.} 2008, \apjs, 175, 48

\bibitem[{{Reddy} {et~al.}(2015){Reddy}, {Kriek}, {Shapley}, {Freeman},
  {Siana}, {Coil}, {Mobasher}, {Price}, {Sanders}, \& {Shivaei}}]{reddy15}
{Reddy}, N.~A., {Kriek}, M., {Shapley}, A.~E., {et~al.} 2015, \apj, 806, 259

\bibitem[{{Rix} {et~al.}(2004){Rix}, {Pettini}, {Leitherer}, {Bresolin},
  {Kudritzki}, \& {Steidel}}]{rix04}
{Rix}, S.~A., {Pettini}, M., {Leitherer}, C., {et~al.} 2004, \apj, 615, 98

\bibitem[{{Robertson} {et~al.}(2015){Robertson}, {Ellis}, {Furlanetto}, \&
  {Dunlop}}]{robertson15}
{Robertson}, B.~E., {Ellis}, R.~S., {Furlanetto}, S.~R., \& {Dunlop}, J.~S.
  2015, \apjl, 802, L19

\bibitem[{{Rockosi} {et~al.}(2010){Rockosi}, {Stover}, {Kibrick}, {Lockwood},
  {Peck}, {Cowley}, {Bolte}, {Adkins}, {Alcott}, {Allen}, {Brown}, {Cabak},
  {Deich}, {Hilyard}, {Kassis}, {Lanclos}, {Lewis}, {Pfister}, {Phillips},
  {Robinson}, {Saylor}, {Thompson}, {Ward}, {Wei}, \& {Wright}}]{rockosi10}
{Rockosi}, C., {Stover}, R., {Kibrick}, R., {et~al.} 2010, in Society of
  Photo-Optical Instrumentation Engineers (SPIE) Conference Series, Vol. 7735,
  Society of Photo-Optical Instrumentation Engineers (SPIE) Conference Series,
  0

\bibitem[{{Rudie} {et~al.}(2013){Rudie}, {Steidel}, {Shapley}, \&
  {Pettini}}]{rudie13}
{Rudie}, G.~C., {Steidel}, C.~C., {Shapley}, A.~E., \& {Pettini}, M. 2013,
  \apj, 769, 146

\bibitem[{{Sana} {et~al.}(2012){Sana}, {de Mink}, {de Koter}, {Langer},
  {Evans}, {Gieles}, {Gosset}, {Izzard}, {Le Bouquin}, \& {Schneider}}]{sana12}
{Sana}, H., {de Mink}, S.~E., {de Koter}, A., {et~al.} 2012, Science, 337, 444

\bibitem[{{Sanders} {et~al.}(2015){Sanders}, {Shapley}, {Kriek}, {Reddy},
  {Freeman}, {Coil}, {Siana}, {Mobasher}, {Shivaei}, {Price}, \& {de
  Groot}}]{sanders15}
{Sanders}, R.~L., {Shapley}, A.~E., {Kriek}, M., {et~al.} 2015, \apj, 799, 138

\bibitem[{{Sanders} {et~al.}(2016){Sanders}, {Shapley}, {Kriek}, {Reddy},
  {Freeman}, {Coil}, {Siana}, {Mobasher}, {Shivaei}, {Price}, \& {de
  Groot}}]{sanders16}
---. 2016, \apj, 816, 23

\bibitem[{{Shapley} {et~al.}(2003){Shapley}, {Steidel}, {Pettini}, \&
  {Adelberger}}]{shapley03}
{Shapley}, A.~E., {Steidel}, C.~C., {Pettini}, M., \& {Adelberger}, K.~L. 2003,
  \apj, 588, 65

\bibitem[{{Shapley} {et~al.}(2015){Shapley}, {Reddy}, {Kriek}, {Freeman},
  {Sanders}, {Siana}, {Coil}, {Mobasher}, {Shivaei}, {Price}, \& {de
  Groot}}]{shapley15}
{Shapley}, A.~E., {Reddy}, N.~A., {Kriek}, M., {et~al.} 2015, \apj, 801, 88

\bibitem[{{Shirazi} \& {Brinchmann}(2012)}]{shirazi12}
{Shirazi}, M., \& {Brinchmann}, J. 2012, \mnras, 421, 1043

\bibitem[{{Shirazi} {et~al.}(2014){Shirazi}, {Brinchmann}, \&
  {Rahmati}}]{shirazi14}
{Shirazi}, M., {Brinchmann}, J., \& {Rahmati}, A. 2014, \apj, 787, 120

\bibitem[{{Shivaei} {et~al.}(2016){Shivaei}, {Kriek}, {Reddy}, {Shapley},
  {Barro}, {Conroy}, {Coil}, {Freeman}, {Mobasher}, {Siana}, {Sanders},
  {Price}, {Azadi}, {Pasha}, \& {Inami}}]{shivaei16}
{Shivaei}, I., {Kriek}, M., {Reddy}, N.~A., {et~al.} 2016, \apjl, 820, L23

\bibitem[{{Smith}(2014)}]{smith14}
{Smith}, N. 2014, \araa, 52, 487

\bibitem[{{Stanway} {et~al.}(2016){Stanway}, {Eldridge}, \&
  {Becker}}]{stanway16}
{Stanway}, E.~R., {Eldridge}, J.~J., \& {Becker}, G.~D. 2016, \mnras, 456, 485

\bibitem[{{Steidel} {et~al.}(2003){Steidel}, {Adelberger}, {Shapley},
  {Pettini}, {Dickinson}, \& {Giavalisco}}]{steidel03}
{Steidel}, C.~C., {Adelberger}, K.~L., {Shapley}, A.~E., {et~al.} 2003, \apj,
  592, 728

\bibitem[{{Steidel} {et~al.}(2010){Steidel}, {Erb}, {Shapley}, {Pettini},
  {Reddy}, {Bogosavljevi{\'c}}, {Rudie}, \& {Rakic}}]{steidel2010}
{Steidel}, C.~C., {Erb}, D.~K., {Shapley}, A.~E., {et~al.} 2010, \apj, 717, 289

\bibitem[{{Steidel} {et~al.}(2004){Steidel}, {Shapley}, {Pettini},
  {Adelberger}, {Erb}, {Reddy}, \& {Hunt}}]{steidel04}
{Steidel}, C.~C., {Shapley}, A.~E., {Pettini}, M., {et~al.} 2004, \apj, 604,
  534

\bibitem[{{Steidel} {et~al.}(2014){Steidel}, {Rudie}, {Strom}, {Pettini},
  {Reddy}, {Shapley}, {Trainor}, {Erb}, {Turner}, {Konidaris}, {Kulas}, {Mace},
  {Matthews}, \& {McLean}}]{steidel14}
{Steidel}, C.~C., {Rudie}, G.~C., {Strom}, A.~L., {et~al.} 2014, \apj, 795, 165

\bibitem[{{Yoon} \& {Langer}(2005)}]{yoon05}
{Yoon}, S.-C., \& {Langer}, N. 2005, \aap, 443, 643

\end{thebibliography}

\end{document}